\documentclass[prd]{revtex4}

\usepackage{epsf,latexsym}
\usepackage{amsmath,amssymb,amsthm}
\usepackage{graphicx}
\usepackage[all]{xy}
\usepackage{slashed}

\newcommand{\rxy}[1]{{\begin{xy}
0;<2mm,0mm>:<0mm,2mm>::0;0,#1\end{xy}}}

\newtheorem{thm}{Theorem}[section]

\newtheorem{defn}[thm]{Definition}

\newcommand{\bb}{\begin{eqnarray}}

\newcommand{\ee}{\end{eqnarray}}

\newcommand{\eee}{\nonumber\end{eqnarray}}

\newcommand{\pp}[1]{\begin{pmatrix} #1 \end{pmatrix}}

\newcommand{\beq}{\begin{equation}}

\newcommand{\eeq}{\end{equation}}

\newcommand{\bea}{\begin{eqnarray}}

\newcommand{\eea}{\end{eqnarray}}

\newcommand{\T}{{\rm tr}}

\newcommand{\ddfm}{\hbox{$^{\hat{f}}$\hspace{-0.15cm}
$\mathcal{D}$}}

\newcommand{\ddf}{\hbox{$^f$\hspace{-0.15cm} $\mathcal{D}$}}

\def\sv{\left<\sigma v\right>}

\def\eV{\,{\rm eV}}

\def\keV{\,{\rm keV}}

\def\MeV{\,{\rm MeV}}

\def\GeV{\,{\rm GeV}}

\def\s{{\,\rm s}}

\def\cm{{\,\rm cm}}

\def\g{{\,\rm g}}

\def\km{{\,\rm km}}

\def\pc{{\,\rm pc}}

\newcommand{\rxyz}[2]{{\begin{xy} 0;<2mm,0mm>:<0mm,2mm>::0;0,
,(5,-2)*{a}
,(10,-1.8)*{\bar{a}}
,(15,-2)*{b}
,(20,-2)*{c}
,(25,-2)*{d}
,(30,-2)*{e}
,(35,-1.8)*{\bar{e}}
,(40,-2)*{f}
,(2,-5)*{a}
,(2,-10)*{\bar{a}}
,(2,-15)*{b}
,(2,-20)*{c}
,(2,-25)*{d}
,(2,-30)*{e}
,(2,-35)*{\bar{e}}
,(2,-40)*{f}
,(5,-5)*\cir(#1,0){}
,(10,-5)*\cir(#1,0){}
,(15,-5)*\cir(#1,0){}
,(20,-5)*\cir(#1,0){}
,(25,-5)*\cir(#1,0){}
,(30,-5)*\cir(#1,0){}
,(35,-5)*\cir(#1,0){}
,(40,-5)*\cir(#1,0){}
,(5,-10)*\cir(#1,0){}
,(10,-10)*\cir(#1,0){}
,(15,-10)*\cir(#1,0){}
,(20,-10)*\cir(#1,0){}
,(25,-10)*\cir(#1,0){}
,(30,-10)*\cir(#1,0){}
,(35,-10)*\cir(#1,0){}
,(40,-10)*\cir(#1,0){}
,(5,-15)*\cir(#1,0){}
,(10,-15)*\cir(#1,0){}
,(15,-15)*\cir(#1,0){}
,(20,-15)*\cir(#1,0){}
,(25,-15)*\cir(#1,0){}
,(30,-15)*\cir(#1,0){}
,(35,-15)*\cir(#1,0){}
,(40,-15)*\cir(#1,0){}
,(5,-20)*\cir(#1,0){}
,(10,-20)*\cir(#1,0){}
,(15,-20)*\cir(#1,0){}
,(20,-20)*\cir(#1,0){}
,(25,-20)*\cir(#1,0){}
,(30,-20)*\cir(#1,0){}
,(35,-20)*\cir(#1,0){}
,(40,-20)*\cir(#1,0){}
,(5,-25)*\cir(#1,0){}
,(10,-25)*\cir(#1,0){}
,(15,-25)*\cir(#1,0){}
,(20,-25)*\cir(#1,0){}
,(25,-25)*\cir(#1,0){}
,(30,-25)*\cir(#1,0){}
,(35,-25)*\cir(#1,0){}
,(40,-25)*\cir(#1,0){}
,(5,-30)*\cir(#1,0){}
,(10,-30)*\cir(#1,0){}
,(15,-30)*\cir(#1,0){}
,(20,-30)*\cir(#1,0){}
,(25,-30)*\cir(#1,0){}
,(30,-30)*\cir(#1,0){}
,(35,-30)*\cir(#1,0){}
,(40,-30)*\cir(#1,0){}
,(5,-35)*\cir(#1,0){}
,(10,-35)*\cir(#1,0){}
,(15,-35)*\cir(#1,0){}
,(20,-35)*\cir(#1,0){}
,(25,-35)*\cir(#1,0){}
,(30,-35)*\cir(#1,0){}
,(35,-35)*\cir(#1,0){}
,(40,-35)*\cir(#1,0){}
,(5,-40)*\cir(#1,0){}
,(10,-40)*\cir(#1,0){}
,(15,-40)*\cir(#1,0){}
,(20,-40)*\cir(#1,0){}
,(25,-40)*\cir(#1,0){}
,(30,-40)*\cir(#1,0){}
,(35,-40)*\cir(#1,0){}
,(40,-40)*\cir(#1,0){}
#2\end{xy}}}

\begin{document}

\title{Dark matter with invisible light from heavy double charged leptons of almost-commutative geometry? }

\author{M.Yu.~Khlopov}
\email{Maxim.Khlopov@roma1.infn.it}
\address{Centre for CosmoParticle Physics "Cosmion", 125047,
Moscow, Russia\\
Moscow Engineering Physics Institute, 115409 Moscow, Russia
}

\author{C.A.~Stephan}
\email{christoph.stephan@cpt.univ-mrs.fr}
\address{Centre de Physique Theorique, CNRS-Luminy Case 907, \\
13288 Marseille cedex 9, France}
\email{christoph.stephan@cpt.univ-mrs.fr}

\author{D.~Fargion}
\email{daniele.fargion@roma1.infn.it}
\address{Physics department, Universita' degli studi "La Sapienza", \\
         Piazzale Aldo Moro 5,CAP 00185 Roma, Italy and \\
        INFN Roma, Istituto Nazionale di Fisica Nucleare, Italy}

\begin{abstract}

A new candidate of cold dark matter  arises by a novel elementary
particle model:
the \underline{a}lmost-\underline{c}ommutative AC-geometrical
framework. Two heavy leptons are added to the Standard Model, each one sharing
a double opposite electric charge and an own lepton flavor number The novel mathematical theory of almost-commutative geometry \cite{book} wishes to unify gauge
models with gravity. In this scenario two new heavy ($ m_L \geq
100 GeV$), oppositely double charged leptons (A,C),($A$ with
charge -2 and $C$ with charge +2), are born with no twin quark
companions. The model naturally involves a new $U(1)$ gauge
interaction, possessed only by the AC-leptons and providing a
Coulomb-like attraction between them. AC-leptons posses
electro-magnetic as well as $Z$-boson interaction and, according
to the charge chosen for the new $U(1)$ gauge interaction, a new
"invisible light"  interaction.

Their final cosmic relics
are bounded into "neutral" stable atoms $(AC)$ forming the
mysterious cold dark matter, in the spirit of the Glashow's
Sinister model. An $(AC)$ state is reached in the early Universe along a
tail of a few secondary frozen exotic components. They should be
now here somehow hidden in the surrounding matter. The two main
secondary $manifest$ relics are $C^{++}$ (mostly hidden in a
neutral $(C^{++} e^- e^-)$ "anomalous helium" atom, at a $10^{-8}$
ratio) and a corresponding "ion"
 $A^{--}$ bounded with an ordinary helium ion $(^4He)^{++}$ ;
indeed the positive helium ions are able to attract and capture
the free $A^{--}$ fixing them into a neutral relic cage that has
nuclear interaction  $(^4He^{++}A^{--})$. The cage preserves the
leptons to later recombine with neutral $(C^{++} e^- e^-)$ into $(AC)$
evanescent states. In early and late cosmic stages   $(AC)$ gas is
leading to cold dark matter gravity seeds. It can form dense cores
inside dense matter bodies (stars and planets). Binding $(C^{++}
e^- e^-)$ $+$ $(^4He^{++}A^{--})$ into $(AC)$ atoms
results in a steady decrease of the anomalous isotopes and a growing
concentration of AC-gas.  However the $(A He)$ influence on Big Bang
nucleo-synthesis and catalysis of nuclear transformations in
terrestrial matter appear to be a serious problem for the model.
Moreover the zero lepton \emph{OLe-helium} $(AHe)$, $(C^{++} e^-
e^-)$ pollution and its on-going $(AHe)$ catalysis in terrestrial
waters may release frequent tens MeV gamma photons whose pair
production lead to nearly aligned electron pairs; their consequent
expected presence by twin Cherenkov rings poses a crucial test to
the model. Their total absence in Super-Kamiokande or SNO records
might point to the  failure  of the model, while
their  eventual discovery   (above the background) may hint
to  the double charge AC-model to solve the Cold Dark matter
puzzle.

 The new invisible light
attraction allows to stimulate the effective $A-C$ recombination into $(AC)$ atoms
inside dense matter bodies (stars and planets), resulting in a
decrease of anomalous isotopes below the experimental upper
limits. \emph{OLe-helium}
 pollution of terrestrial matter and
$(OHe)$ catalysis of nuclear reactions in it is one of the
dramatic problems (or exciting advantages?) of the present model.

\end{abstract}

\maketitle

\section{Introduction}

The problem of the existence of new leptons is among the most
important in  modern high energy physics. Such new heavy
leptons may be sufficiently long-living to represent a new stable
form of matter. At the present there are in the elementary
particle scenarios at least two main frames for Heavy Stable Quarks
and Leptons: (a) A fourth heavy Quark and a fourth heavy Neutral
Lepton (neutrino) generation (above half the Z-Boson
mass)\cite{Shibaev}, \cite{Sakhenhance}, \cite{Fargion99},
\cite{Grossi}, \cite{Belotsky}, \cite{BKS}, \cite{Khlopov:2005ew}; see also
 \cite{Berezhiani:1995du,Okun,Okun2,Okun3,Okun4,Volovik:2003kh},
\cite{4had}; and (b) A Glashow's  "Sinister"  heavy quark and
heavy Charged Lepton family, whose masses role may be the dominant
dark matter \cite{Glashow,Fargion:2005xz}. Recently another
possibility, based on the approach of almost-commutative geometry
by Alain Connes \cite{book} (the AC-model), was revealed in
\cite{5}. We shall address here our attention only on this latter
option.

 The AC-model \cite{5} extends the fermion content of the
Standard model by two heavy particles with opposite
electromagnetic and Z-boson charges. These particles
(AC-fermions) behave as heavy stable
leptons with charges $+2e$ and $-2e$, to be referred to further as
$A$ and $C$, respectively\footnote{These particles were earlier
called $E^{--}$ and $P^{++}$.
To avoid a
misleading analogy with the single charged electron and the proton
we refer to them as to Anion-Like EXotic Ion of Unknown Matter
(ALEXIUM) $A^{--}$ and to Cathion-Like EXotic Ion of Unknown
Matter (CoLEXIUM) $C^{++}$.}. The mass of AC-fermions has a
"geometrical" origin and is not related to the Higgs mechanism. In
the absence of AC-fermion mixing with light fermions, AC-fermions
can be absolutely stable and we explore this possibility in the
present paper. Furthermore this model admits a new interaction for
the AC-fermions which consists in a new $U(1)$ interaction of
electro-magnetic type, hence the name "invisible light". This new
gauge group will be denoted $U_{AC}(1)$ and the charge will be
denoted by $y$.
 It should be
noted that the only serious reason for the stability of
AC-fermions without $U_{AC}(1)$ interaction which can be given
from the view of almost-commutative
geometry is that no interaction terms with ordinary fermions of
the Standard Model appear in the Lagrangian. Since the
almost-commutative model is purely classical and does not include
quantum gravity or a fully noncommutative geometry, higher order
effects cannot be excluded. This problem is not present when the
AC-leptons posses a $U_{AC}(1)$-charge.

To generate excess,
non-conservation of separate A- and C-numbers is needed, but the
(A-C)-number should be strictly conserved. The excess of A is
taken to be equal to the excess of C, as required in the further
cosmological treatment. This is the main assumption of our
publication, which follows a similar approach of Glashow's sinister
model \cite{Glashow}. The assumption is that the AC-lepton excess saturates
the standard CDM density. The reasoning for this assumption is
that the cosmology for an equal amount of AC-leptons and their
antiparticles is ruled out by observation, as we show in
Appendix 8. If the AC-leptons $A$ and $C$ have equal and opposite sign
of $y$-charges, strict conservation of the $y$-charge does not prevent
the generation of $A$ and $C$ excess.

AC-fermions are sterile relative to the $SU_L(2)$ electro-weak
interaction, and do not contribute to the Standard Model
parameters \cite{Knecht:2006tv}. If their lifetime exceeds the age of the Universe\footnote{If this
lifetime is much less than the age of the Universe, there should
be no primordial AC-fermions, but they can be produced in cosmic
ray interactions and be present in cosmic ray fluxes
 or in their most
recent relics on Earth. The assumed masses for these AC-leptons
make their search a challenge for  present and future
accelerators.}, primordial heavy AC-leptons
should be present in modern matter\footnote{The mechanisms of
production of metastable $Q$ (and $\bar Q$) hadrons and
tera-particles in the early Universe, cosmic rays and accelerators
were analyzed in \cite{4had,Fargion:2005xz} and the possible
signatures of their wide variety and existence were revealed.}.

In the model \cite{5} the properties of heavy AC-fermions are
fixed by the almost-commutative geometry and the physical
postulates given in \cite{1}. The freedom resides in the choice of
their hyper-charge, their $U_{AC}(1)$-charge and their masses. According
to this model, positively charged $C$ and negatively charged $A$
are stable and may form a neutral most probable and stable (while
being evanescent) $(AC)$ atom. The AC-gas of such "atoms" is an
ideal candidate for a very new and fascinating dark matter (like
it was tera-helium gas in \cite{Glashow,Fargion:2005xz}). Because
of their peculiar WIMP-like interaction with matter they may also
rule the  stages of gravitational clustering in early matter
dominated epochs, creating the first gravity seeds for galaxy
formation.

However, in analogy to D, $^3$He and Li relics that are the
intermediate catalyzers of $^4$He formation in Standard Big Bang
Nucleosynthesis (SBBN) and are important cosmological tracers of
this process, the AC-lepton relics from intermediate stages of a
multi-step process towards a final $(AC)$ formation must survive
with high abundance of {\it visible} relics in the present
Universe. We enlist, reveal and classify such tracers, their birth
place and history up to now.

We find that $(eeC^{++})$
 is here to remain among
us and its abundance should be strongly reduced in terrestrial
matter to satisfy known severe bounds on anomalous helium. This
reduction is catalyzed by relic neutral OLe-helium (named so from
\emph{O-Le}pton-\emph{helium}) $(^4He^{++}A^{--})$, because the
primordial component of free AC-leptons $A^{--}$ are mostly
trapped in the first three minutes into this puzzling
\emph{OLe-helium} "atom" $(^4He^{++}A^{--})$ with nuclear
interaction cross section, which provides anywhere eventual later
$(AC)$ binding. This surprising catalyzer with screened Coulomb
barrier can influence the chemical evolution of ordinary matter,
but it turns out that the dominant process of OLe-helium
interaction with nuclei is quasi-elastic and might not result in
the creation of anomalous isotopes. Inside dense matter objects
(stars or planets) its recombination with $(C^{++}ee)$ into $(AC)$
atoms can provide a mechanism for the formation of dense $(AC)$
objects.

We will briefly review the AC-model \cite{5} and the properties of
AC-leptons $A$ and $C$, predicted by it (Sections \ref{flavor} and
\ref{model}). We consider their evolution in the early Universe
and notice (Section \ref{primordial}) that in spite of the assumed
excess of particles ($A^{--}$ and $C^{++}$), the abundance of
frozen out antiparticles ($\bar A^{++}$ and $\bar C^{--}$) is not
negligible, as well as a significant fraction of $A^{--}$ and
$C^{++}$ remains unbound, when $(AC)$ recombination takes place
and most of the AC-leptons form $(AC)$ atoms. This problem of an
unavoidable over-abundance of by-products of "incomplete
combustion" is unresolvable for models assuming dark matter
composed of atoms binding single charged particles, as it was
revealed in \cite{Fargion:2005xz} for the sinister Universe
\cite{Glashow}. As soon as $^4He$ is formed in the Big Bang
nucleosynthesis, it captures all the free negatively charged heavy
particles (Section \ref{catEHe}). If the charge of such particles
is -1 (as it was the case for tera-electron in \cite{Glashow}) the
positively charged ion $(^4He^{++}E^{-})^+$ puts up a Coulomb
barrier for any successive decrease of the abundance of the
species, over-polluting by anomalous isotopes the modern Universe.
A double negative charge of $A^{--}$ in the considered AC-model
provides binding with $^4He^{++}$ in a neutral OLe-helium state
$(^4He^{++}A^{--})^0$, which catalyzes in the first three minutes
an effective binding in $(AC)$ atoms and a complete annihilation
of antiparticles. Products of this annihilation neither cause an
undesirable effect in the CMB spectrum, nor in the light element
abundances. However, Ole-helium influence on reactions of the Big
Bang nucleosynthesis may be dangerous for the considered scenario.
 After the e-print publication of our results such a danger was
claimed in by-passing in the paper \cite{Pospelov:2006sc}, which
developed the aspect of our topic, related with the influence of
unstable massive charged particles on BBN reactions owing to the
decrease of the Coulomb barrier for nuclei, bound with negatively
single charged massive particles. Though the approach of
\cite{Pospelov:2006sc} was not supported by \cite{Kohri:2006cn}
and it does not take into account the back-reaction of cascades of
energetic particles from unstable particle decay on light element
abundance, which implies the theory of non-equilibrium cosmological
nucleosynthesis \cite{Linde3,Linde4,bookKh}, we compared our
results with the constraints, obtained in this paper, and find
that due to the binding in $(AC)$ atoms the amount of $A^{--}$, which
can influence nuclear processes, fits this constraint.
Still, though the CDM in the form of $(AC)$ atoms is successfully
formed, $A^{--}$ (bound in OLe-helium) and $C^{++}$ (forming an
anomalous helium atom $(C^{++}ee)$) should also be present in the
modern Universe and the abundance of primordial $(C^{++}ee)$ is by
more than {\it ten} orders of magnitude higher than the
experimental upper limit on the anomalous helium abundance in
terrestrial matter. This problem can be solved by OLe-helium
catalyzed $(AC)$ binding of $(C^{++}ee)$ (Section
\ref{Interactions}), but different mobilities in matter of the
atomic interacting $(C^{++}ee)$ and the nuclear interacting
$(^4He^{++}A^{--})$ in the absence of $y$-attraction lead to a
fractionation of these species, preventing an effective decrease
of the anomalous helium abundance. The crucial role of long-range
$y$ attraction is to hold these two components together by the
condition of $y$-charge neutrality, thus avoiding their
fractionation and providing strong decrease of their abundance in
dense matter bodies due to effective $(AC)$ binding. Though $(AC)$
binding is not accompanied by strong annihilation effects, as it
was the case for the 4th generation hadrons \cite{4had}, gamma
radiation from it inside large volume detectors might offer
experimental test for the considered model. Another inevitable
element of AC-cosmology is the existence of OLe-helium, causing
nontrivial nuclear transformations and challenging its
experimental search.

A presentation of the mathematical
concepts involved in the AC model is given in Appendices 1-5.
Technical details for the calculations of the primordial
abundances and the recombination rates are given in Appendices 6
and 7. We explain in Appendix 8 that the symmetric case of equal
primordial abundance of AC-leptons and their antiparticles is too
explosive to explain dark matter by an equal amount of $(AC)$
atoms and $(\bar A \bar C)$ anti-atoms without a contradiction
with the observed gamma background.

AC-cosmology, based on the considered model, escapes
most of the troubles revealed for other cosmological scenarios
with stable heavy charged particles \cite{4had,Fargion:2005xz}.
With successive improvements it might provide a realistic scenario
for composite dark matter in the form of evanescent atoms,
composed by heavy stable electrically charged particles with
invisible light.

\section{\label{flavor} A flavor of almost-commutative geometry}

In the last few years several approaches to include the idea of
noncommutative spaces into physics have been established. One of
the most promising and mathematically elaborated is Alain Connes
{\it noncommutative geometry} \cite{book}, where the main idea is
to translate the usual notions of manifolds and differential
calculus into an algebraic language. Here we will mainly focus on
the motivations why noncommutative geometry ss a novel point of
view of space-time, is worthwhile to be considered by theoretical
physics. We will furthermore try to give a glimpse on the main
mathematical notions (for computational details see appendices 1
to 4), but refer to \cite{book} and \cite{costa} for a thorough
mathematical treatment and to \cite{cc} and \cite{schuck} for its
application to the standard model of particle physics.

Noncommutative geometry has its roots in quantum mechanics and
goes back to Heisenberg\footnote{According to Roman Jackiw
\cite{Jackiw} Julius Wess told and documented the following: {\it
Like many interesting quantal ideas, the notion that spatial
coordinates may not commute can be traced to Heisenberg who, in a
letter to Peierls, suggested that a coordinate uncertainty
principle may ameliorate the problem of infinite self-energies.
... Evidently, Peierls also described it to Pauli, who told it to
Oppenheimer, who told it to Snyder, who wrote the first paper on
the subject \cite{Snyder}}} or even Riemann \cite{Riemann}. In the
spirit of quantum mechanics it seems natural that space-time
itself should be equipped with an uncertainty. The coordinate
functions of space-time should be replaced by a suitable set of
operators, acting on some Hilbert space with the dynamics defined
by a Dirac operator. The choice of a relativistic operator is
clear since the theory ought to be Lorentz invariant. As for the
Dirac operator, in favor of the Klein-Gordon operator, matter is
built from Fermions and so the Dirac operator is privileged. This
approach, now known as noncommutative geometry, has been worked
out by Alain Connes \cite{book}. He started out on this field to
find a generalized understanding to cope with mathematical objects
that seemed geometrical, yet escaped the standard approaches. His
work has its predecessors in Gelfand and Naimark \cite{Gelfand},
who stated that the topology of a manifold is encoded in the
algebra of complex valued functions over the manifold. Connes
extended this theorem and translated the whole set of geometric
data into an algebraic language. The points of the manifold are
replaced by the pure states of an algebra, which, inspired by quantum mechanics, acts
on a Hilbert space. With help of a Dirac operator acting as well
on the Hilbert space, Connes formulated a set of axioms which
allows to recover the geometrical data of the manifold. These
three items, the algebra, the Hilbert space and the Dirac operator
are called a spectral triple. But it should be noted that the set
of manifolds, i.e. space-times, which allow to be described by a
spectral triple is limited. These manifolds have to be compact, Riemannian,
i.e. of Euclidean signature, and they have to admit a spin structure,
which is not true for any manifold. The third condition
presents no drawback since
space-time  falls exactly into this class. But asking the manifold
to be Euclidean, whereas special relativity requires a Lorentzian
signature, poses a problem, which is still open. Nevertheless one
can argue, along the line of Euclidean quantum field theory  that
this can be cured by Wick rotations afterwards. This assumes
of course that the Wick rotation is possible on curved space
times, a problem which shall be left aside for the present paper.
The Euclidean treatment also requires fermion doubling \cite{schuck}
and the spurious fermionic degrees of freedom have to be projected
out after rotation to Lorentzian  signature.
There is still no way to describe non-compact pseudo-Riemannian
manifolds by spectral triples, so this shortcoming  has to be
accepted for the moment.

A strong point in favor of the spectral triple approach is,
as the name noncommutative geometry already implies, that the
whole formulation is independent of the commutativity of the
algebra. So even when the algebra is noncommutative it is possible
to define a geometry in a consistent way. But then the geometry
gets equipped with an uncertainty relation, just as in quantum
mechanics.  With this generalization comes a greater freedom to
unify the two basic symmetries of nature, namely the
diffeomorphism invariance (= invariance under general coordinate
transformations) of general relativity and the local gauge
invariance of the standard model of particle physics. In the case
of ordinary manifolds  the theorem of
Mather \cite{Mather} prohibits such a unification (for details see
\cite{book}).

The standard model can be constructed as a classical  gauge theory
which describes the known elementary particles by means of the
symmetries they obey, together with the electro-weak and the
strong force. In contrast to general relativity, this classical
field theory allows to pass over to a quantum field theory. All
elementary particles are fermions and the forces acting between
them are mediated by bosons. The symmetries of the theory are
compact Lie groups, for the standard model of particle physics the
underlying symmetry goup
is $U(1)\times SU(2) \times SU(3)$. Fermions are Dirac spinors,
placed in multiplets which are representations of the symmetry
groups. A peculiar feature of the standard model is that fermions
are chiral. This poses a serious problem, since mass terms mixing
left- and right-handed states would explicitly break the symmetry.
To circumvent this an extra boson, the Higgs boson, has to be
introduced. In the widely used formulation of the standard model
this Higgs mechanism has to be introduced by hand. All the
non-gravitational forces and all known matter is described in a
unified way. But it is not possible to unify it on
the footing of differential geometry with general relativity. The
problem is that no manifold exists, which has general coordinate
transformations and a compact Lie group as its diffeomorphism
group. But here the power of noncommutative geometry comes in.

The first observation is that the general coordinate
transformations of a manifold correspond to the automorphisms\footnote{
An algebra automorphism is a bijective map from the
algebra into itself, which preserves the whole structure of the
algebra (i.e. the addition, the multiplication and, if present,
the involution). The algebra automorphisms form a group.} of the
algebra of complex valued functions over the manifold. Chamseddine
and Connes \cite{cc} discovered that it is possible to define an
action, called the {\it spectral action}, to give space-time in
the setting of spectral triples a dynamics, just as the
Einstein-Hilbert action for general relativity. This spectral
action is given by the number of eigenvalues of the Dirac operator
up to a cut-off. It is most remarkable that this action reproduces
the Einstein-Hilbert action in the limit of high eigenvalues of
the Dirac operator. The crucial observation is now that in
contrast to the diffeomorphisms of a manifold, the automorphisms
of an algebra allow to be extended to include compact Lie groups.
These are the automorphisms of matrix algebras. And since the
whole notion of a spectral triple is independent of the
commutativity of the algebra, it is possible to combine the
algebra of functions over the space-time manifold with an algebra
being the sum of simple matrix algebras by tensorising. These
combined function-matrix geometries are called {\it
almost-commutative} geometries. The part of the spectral triple
based on the matrix algebra is often called the {\it finite} or
{\it internal part}. Indeed, they contain an infinite number of
commutative degrees of freedom plus a finite number of
noncommutative ones. The former are outer, the latter are
inner automorphisms.

To see how the Higgs scalar, gauge potentials and gravity  emerge
one starts out with an almost-commutative spectral triple over a
flat manifold $M$. The corresponding algebra of complex valued
functions over the manifold will be $\mathcal{A}_R$ (where the
subscript $R$ stands for {\it Riemannian} \footnote{These
subscripts will be dropped when no confusion will arise from
it.}), the Hilbert space $\mathcal{H}_R$ is the Hilbert space of
Dirac spinors and the Dirac operator is simply the flat Dirac
operator $\slashed{\partial}$. As mentioned above, the
automorphisms of the algebra $\mathcal{A}_R$ coincide with the
diffeomorphisms, i.e. the general coordinate transformations, of
the underlying manifold, Aut$(\mathcal{A}_R)=$Diff$(M)$. To render
this function algebra noncommutative, a matrix algebra
$\mathcal{A}_f$ (where the subscript $f$ stands for {\it finite})
is chosen. The exact form of this matrix algebra is of no
importance for the moment (as long as its size is at least
two). The Hilbert space $\mathcal{H}_f$ is finite dimensional and
the Dirac operator $\mathcal{D}_f$ is a complex valued matrix. For
the detailed form of the internal Dirac operator see Appendix 1.

It is a pleasant feature of spectral triples that the tensor
product of two spectral triples is again a spectral triple. So
building the tensor product one finds for the algebra and the
Hilbert space of the almost commutative geometry
\begin{equation}
\mathcal{A}_{AC} = \mathcal{A}_R \otimes \mathcal{A}_f, \quad
\mathcal{H}_{AC} = \mathcal{H}_R \otimes \mathcal{H}_f.
\end{equation}
The Dirac operator needs a little bit more care to comply with the
axioms for spectral triples. It is given by
\begin{equation}
\mathcal{D}_{AC} = \slashed{\partial} \otimes 1_f + \gamma^5
\otimes \mathcal{D}_f,
\end{equation}
where $1_f$ is a unity matrix whose size is  the size of the
finite Dirac operator $\mathcal{D}_f$ and $\gamma^5$ is
constructed in the standard way from the Dirac gamma matrices. The
automorphism group of the almost-commutative algebra
$\mathcal{A}_{AC}$ is the semi-direct product of the
diffeomorphisms of the underlying manifolds and the gauged
automorphisms of the matrix algebra. For example, with the matrix
algebra $\mathcal{A}_f=M_2 (\mathbb{C})$ one would have the gauged
unitary group $SU(2)$ in the automorphism group. This is exactly
the desired form for a symmetry group.

Now these automorphisms
\begin{equation}
Aut(\mathcal{A}_{AC}) = Aut(\mathcal{A}_R) \ltimes
Aut(\mathcal{A}_f) \ni (\sigma_R, \sigma_f) \label{auto}
\end{equation}
have to be lifted (=represented) to the Hilbert space
$\mathcal{H}_{AC}$. This is necessary to let them act on the
Fermions as well as to fluctuate or gauge the Dirac operator. It
is achieved by the lift $L(\sigma_R, \sigma_f)$ which is defined
via the representation of the algebra on the Hilbert space. For
details see Appendix 4.

For the moment the Dirac operator $\mathcal{D}_{AC}$ consists of
the Dirac operator on a flat manifold and a complex valued matrix.
Now, to bring in the Higgs scalar, the gauge potentials and
gravity the Dirac operator has to be fluctuated or gauged with the
automorphisms (\ref{auto})
\bb
\ddf_{AC} &&:= L(\sigma_R, \sigma_f) \mathcal{D}_{AC}
L(\sigma_R, \sigma_f)^{-1}
\nonumber \\
&&= L(\sigma_R, \sigma_f) (\slashed{\partial} \otimes 1_f
)L(\sigma_R, \sigma_f)^{-1} +L(\sigma_R, \sigma_f) ( \gamma^5
\otimes \mathcal{D}_f)L(\sigma_R, \sigma_f)^{-1}
\nonumber \\
&&= \slashed{\partial}_{cov.} + \gamma^5 \otimes L(\sigma_f)
\mathcal{D}_f L(\sigma_f)^{-1}=  \slashed{\partial}_{cov.} + \gamma^5 \otimes
\ddf_{f}.
\ee
In the last step it turns out that
$\slashed{\partial}_{cov.}:=L(\sigma_R, \sigma_f)
(\slashed{\partial} \otimes 1_f )L(\sigma_R, \sigma_f)^{-1}$ is
indeed the covariant Dirac operator on a curved space time, when
the appearing gauge potentials have been promoted to arbitrary
functions, i.e. after applying Einstein's equivalence principle
(for details see \cite{cc}). $\slashed{\partial}_{cov.}$ has
automatically the correct representation of the gauge potentials
on the Hilbert space of Fermion multiplets. The gauge potentials
thus emerge from the usual Dirac operator acting on the gauged
automorphisms of the inner algebra.

As for the Higgs scalar, it is identified with $\ddf_{f}:=L(\sigma_f)
\mathcal{D}_f L(\sigma_f)^{-1}$. Here the commutative
automorphisms being the diffeomorphisms $\sigma_R$ of the manifold
drop out, since they commute with the matrix $  \mathcal{D}_f$.
This is not true for the gauged automorphisms $\sigma_f$ since
they are matrices themselves.

>From the gauged Dirac operator $\ddf_{AC}$ the spectral action is calculated
via a heat-kernel expension to be the Einstein-Hilbert action plus the Yang-Mills-Higgs
action. The Higgs potential in its well known quartic form is a
result of this calculation. It should be pointed out that the heat-kernel
expansion is performed up to a cut-off and so the obtained Einstein-Hilbert action
and Yang-Mills-Higgs action should be considered as effective actions.
The details of the calculation of the
spectral action go beyond the scope of this publication and we
refer again to \cite{cc} for a detailed account. For the internal
part $\ddf_{f}$ of the gauged Dirac operator $\ddf_{AC}$ the
spectral action gives exactly the Higgs potential
\bb
V(\ddf_{f} )= \lambda\  \T\!\left[ (\ddf_{f} )^4\right] -\textstyle{\frac{\mu
^2}{2}}\ \T\!\left[ (\ddf_{f}) ^2\right] ,
\ee
where $\lambda $ and $\mu $ are positive constants, as well as
the kinetic term for the Higgs potential.  To determine a sensible
value for the cut-off in the heat-kernel expansion, it is instructive
to note, that at the cut-off the couplings of the non-abelian gauge groups
and the coupling $\lambda$ of the Higgs potential are closely tied together.

Choosing as matrix algebra $\mathcal{A}_f=\mathbb{C} \oplus
\mathbb{H} \oplus M_3(\mathbb{C})$, where $\mathbb{H}$ are the
quaternions, one recovers with a suitable choice for the Hilbert
space that the spectral action reproduces the Einstein-Hilbert
action and the Yang-Mills-Higgs action of the standard model.
The cut-off is then fixed to be at the energy where the coupling
$g_2$ of the weak group $SU(2)_L$ and the coupling $g_3$
of the colour group $SU(3)_C$ become equal ($\sim 10^{17}$GeV).
At the cut-off these two couplings and the Higgs coupling $\lambda$
are related as
\bb
g_3^2=g_2^2=3 \lambda.
\ee
Assuming a great dessert up to the cut-off,  this relation allows
to let the Higgs coupling run back to lower energies and
to calculate the Higgs mass. A detailed calculation
can be found in \cite{schuck} and gives a Higgs mass
of $m_{Higgs}= 175.4 \pm 4.7$ GeV, where the uncertainty
is due to  the uncertainty in the top-quark mass.

Recapitulating, the
Higgs scalar together with its potential emerge naturally as the
"Einstein-Hilbert action" in the noncommutative part of the
algebra. Here it has become possible for the first time to give
the Higgs scalar a geometrical interpretation. In the
almost-commutative setting it plays at the same time the r\^ole of
the metric in the finite part of the geometry as well as that of
the fermionic mass matrix.

One may interpret an almost-commutative geometry as a kind of
Kaluza-Klein theory, where the extra dimensions are discrete.
These extra dimensions are produced by the matrix algebra and they
provide for extra degrees of freedom without being visible.
Furthermore the Yang-Mills-Higgs action can be viewed in the
almost-commutative setting as the gravitational action, or
Einstein-Hilbert analogue for the "discrete part" of space-time.
From this point of view the gauge bosons, i.e. the Higgs boson,
the Yang-Mills bosons and the graviton form a unified
"super-multiplet". But of course space-time in the classical sense
ceases to exist in noncommutative geometry, just as there is no
classical phase space in quantum mechanics. The space-time has
been replaced by operators and extended by discrete extra
dimensions.

\begin{figure}
\begin{center}
\includegraphics[width=11cm]{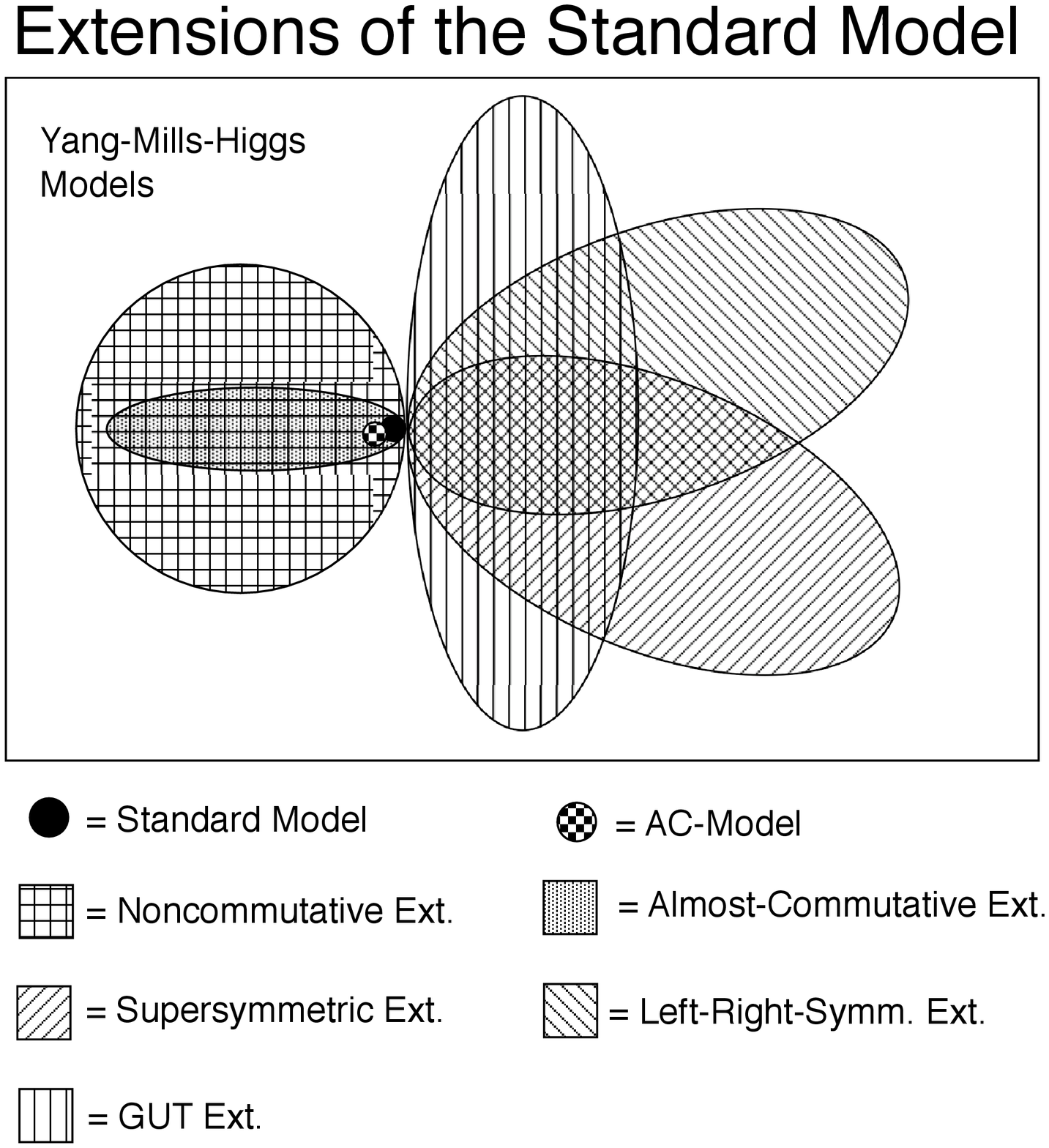}
\caption{Yang-Mills-Higgs models extending the Standard Model}
\label{versus}
\end{center}
\end{figure}

The immediate question that arises is: Which kind of
Yang-Mills-Higgs theory may fit into the frame work of
almost-commutative geometry? The set of all Yang-Mills-Higgs
theories is depicted in figure \ref{versus}. One sees that
left-right symmetric, grand unified and supersymmetric theories do
not belong to the elected group of noncommutative models.
But, as mentioned above, the standard model,
resulting from an almost-commutative geometry,
as well as the AC-model, do.

One of the main tasks of the present research in
almost-commutative geometry is to clarify the  structure of this
restricted sub-set of Yang-Mills-Higgs theories that originate
from  spectral triples. Since this is still an unscalable
challenge it is necessary to adopt a minimal approach. Imposing
certain constraints which are gathered from different areas
reaching from Riemannian geometry over high energy physics to
quantum field theory and starting out with only up to four
summands in the matrix algebra part of the almost commutative
geometry, one can give a classification from the particle
physicist's point of view \cite{1,2,3,4}. As it is custom in particle physics,
space-time curvature  was neglected. Nonetheless the
Riemannian part of the spectral triple plays a crucial role in the
spectral action, introducing derivatives and thus the gauge
bosons. Setting the curvature to zero when the Einstein-Hilbert
and Yang-Mills-Higgs action have been obtained from the spectral
action leaves the Yang-Mills-Higgs action. With respect to this
part of the spectral action the classification has been done. As a
consequence only the finite matrix algebra part of the spectral
triple had to be classified since only the internal Dirac operator
enters into the Higgs scalar, as was shown above. The minimum of
the Higgs potential is the mass matrix of the fermions.

This classification proceeds in two steps. First all the possible
finite spectral triples, with a given number of summands of simple
matrix algebras, have to be found. This classification of finite
spectral triples has been done in the most general setting by
Paschke, Sitarz \cite{pasch} and Krajewski \cite{Kraj}. To
visualize a finite spectral triple Krajewski introduced a
diagrammatic notion, {\it Krajewski diagrams}, which
encode all the algebraic data of a spectral triple. For a
more detailed account see Appendix 3. If one imposes now as a first
condition that the spectral triple be irreducible, i.e. that the
finite Hilbert space be as small as possible, one is led to the
notion of a minimal Krajewski diagram. For a given number of
algebras, the algebra representation on the Hilbert space and the
possible Dirac operators are encoded in these diagrams by arrows
connecting two sub-representations. Finding the minimal diagrams
via this diagrammatic approach is very convenient and quite simple
for up to two summands in the matrix algebra. In this case only a
handful of diagrams exist and it is difficult to miss a diagram.
But with three and more algebras the task quickly becomes
intractable. For three algebras it may be done by hand, but one
risks to overlook some diagrams. It is thus fortunate that the
diagrammatic treatment allows to translate the algebraic problem
of finding spectral triples into the combinatorial problem of
finding minimal Krajewski diagrams. This can then be put into a
computer program. Still the problem is quite involved and the
algorithm to find minimal Krajewski diagrams needs a lot of care.
Furthermore the number of possible Krajewski diagrams increases
rapidly with the number of summands of matrix algebras and reaches
the maximal capacity of an up-to-date personal computer at  five
summands.

If one has found the minimal Krajewski diagrams the second major
step follows.  From each Krajewski diagram all the possible
spectral triples have to be extracted. These are then analyzed
with respect to the following heteroclitic criteria:

\begin{itemize}
\item[$\bullet$] For simplicity and in view of the minimal
approach the spectral triple should be irreducible. This means
simply that the Hilbert space cannot be reduced while still
obeying all the axioms of a spectral triple.
\item[$\bullet$]  The
spectral triple should be non-degenerate, which means that the
fermion masses should be non-degenerate, up to the inevitable
degeneracies which are left and right, particle and antiparticle
and a degeneracy due to a color. This condition has its origin in
perturbative quantum field theory and asserts that the possible
mass equalities are stable under renormalization flow.
\item[$\bullet$] Another criterion also stemming from quantum
field theory is that the Yang-Mills-Higgs models should be free of
Yang-Mills anomalies. In hope of a possible unified quantum theory
of all forces, including gravity, it is also demanded that the
models be free of mixed gravitational anomalies.
\item[$\bullet$]
From particle phenomenology originates the condition that the
representation of the little group has to be complex in each
fermion multiplet, in order to distinguish particles from
antiparticles.
\item[$\bullet$] The last item is the requirement
that massless fermions should be neutral under the little group.
This is of course motivated by the Lorentz force.
\end{itemize}

\noindent Now the Higgs potential has to be minimized and the
resulting models have to be compared with the above list of
criteria. If a model fits all the points of the list it may be
considered  of physical importance, otherwise it will be
discarded.

Nonetheless it is possible to find a Krajewski diagram with six
summands in the matrix algebra which is in concordance with the
physical requirements. It is the aim of this paper
to evaluate its impact on the dark matter problem in cosmology.

\section{\label{model} The particle model}
Among the possible almost-commutative Yang-Mills-Higgs models is
the standard model of particle physics with one generation of quarks
and leptons, for details we refer to
\cite{schuck}. It could furthermore be shown,  \cite{1,2,3,4},
that the standard model takes a most prominent position among
these Yang-Mills-Higgs models.

But the classification of almost-commutative geometries also
allows to go beyond the standard model in a coherent way. Here
heavy use is made of Krajewski diagrams \cite{Kraj}, which allow
to visualize the structure of almost-commutative geometries. The
particle model analyzed in the present publication is an extension
of the AC-lepton model presented in \cite{5}.
It is remarkable to note that the almost-commutative geometry of
the basic AC-lepton model, which builds on the internal algebra
\bb \mathcal{A}= \mathbb{C} \oplus \mathbb{H} \oplus
M_3(\mathbb{C}) \oplus \mathbb{C} \oplus \mathbb{C}  \oplus
\mathbb{C}, \ee
allows to be enlarged to
\bb \mathcal{A}= \mathbb{C} \oplus M_2(\mathbb{C}) \oplus
M_3(\mathbb{C})  \oplus \mathbb{C} \oplus \mathbb{C}  \oplus
\mathbb{C} \ee
which produces through the so called centrally extended lift, for
details see \cite{farewell}, a second $U(1)$ gauge group in
addition to the standard model hypercharge group $U_Y(1)$. In
spirit with the previous nomenclature this group will be called
$U_{AC}(1)$, where AC stands again for
\underline{a}lmost-\underline{c}ommutative. We will call the
corresponding interaction $y$-interaction.
For a detailed derivation from the corresponding
Krajewski diagram to the Lagrangian of the model we refer to
Appendix 5.

Choosing the central
charges as in Appendix 5 reproduces the  standard model and the AC-particles with
the desired electric charges.
Furthermore the central extension determines whether the AC-particles
couple to the $U_{AC}(1)$ gauge group or not. The
simplest model without a $y$-coupling is achieved by setting the
corresponding central charge to zero, and thus decoupling the $U_{AC}(1)$
gauge group completely from the particles. This situation is completely
equivalent to the model presented in \cite{5}. By choosing a non-zero
central charge for the novel interaction the AC-fermions will
experience a $y$-interaction.

The AC-particles do not participate in the Higgs
mechanism and consequently the AC-gauge group stays unbroken:
\bb U_Y(1)\times SU_L(2)\times SU_c(3)  \times U_{AC}(1)
\longrightarrow U_{em}(1)\times SU_c(3)  \times U_{AC}(1)
\nonumber \ee
If the central charge of the $y$-coupling is set to zero the $U_{AC}(1)$
disappears completely from the model.
The Lagrangian of the model consists of the usual standard model
Lagrangian, the  Lagrangian for the AC-particles and the new term
for the AC-gauge potential. We shall only give the two new parts
of the Lagrangian for the AC-fermion spinors $\psi_A$ and $\psi_C$
and the AC-gauge curvature $\tilde B_{\mu \nu}$:
\bb \mathcal{L}_{AC} &=& i \psi_{A L}^\ast D_A \psi_{A L} + i
\psi_{A R}^\ast D_A \psi_{A R} + m_A \psi_{A L}^\ast \psi_{A R} +
 m_A \psi_{A R}^\ast \psi_{A L}
\nonumber  \\
&&+ \; i \psi_{C L}^\ast D_C \psi_{C L} + i \psi_{C R}^\ast D_C
\psi_{C R} + m_C \psi_{C L}^\ast \psi_{C R} + m_C \psi_{C R}^\ast
\psi_{C L}
\nonumber \\
&& \; - \frac{1}{4} \tilde B_{\mu \nu} \tilde B^{\mu \nu}.
\nonumber \ee
The covariant derivatives $D_{A/C}$ and the gauge curvature are
given by
\bb D_{A/C} &=& \gamma^\mu \partial_\mu + \frac{i}{2} \, g' \,
Y_{A/C} \gamma^\mu B_\mu +  \frac{i}{2} \, g_{AC} \, \tilde
Y_{A/C}  \gamma^\mu \tilde B_\mu
\nonumber \\
&=&\gamma^\mu  \partial_\mu + \frac{i}{2} \, e \, Y_{A/C}
\gamma^\mu  A_\mu - \frac{i}{2} \, g' \, \sin \theta_w Y_{A/C}
\gamma^\mu  Z_\mu +  \frac{i}{2} \, g_{AC} \, \tilde Y_{A/C}
\gamma^\mu \tilde B_\mu, \nonumber \ee
and
\bb \tilde B_{\mu \nu} = \partial_\mu \tilde B_{\nu} -
\partial_\nu \tilde B_{\mu} \ee
Again, if the central charge for the $y$-interaction is set to zero the
$U_{AC}(1)$ gauge group disappears and so does its generator
$\tilde B_\mu$.
Setting the central charge to zero leads of course to $\tilde Y_{A/C}=0$.

In the considerations above the Wick rotation has already been performed and the spurious
degrees of freedom from fermion doubling have been projected out.
There is no difference in the treatment of AC-particles and standard
model particles.
On cosmological reasons the electric charge
of the AC-leptons has to be $Q_{em} = \pm 2 e$, where $e$ is the
electric charge of the electron. Otherwise unwanted forms of
OLe-Helium ions would appear.
This requires $Y_{A/C} = \mp 2$. For
simplicity if the AC-model with $y$-interaction is considered,
the AC-hyper charge is also chosen to be $\tilde
Y_{A/C} = \mp 2$, but it cannot be fixed by almost-commutative
geometry. This  also applies to the AC-coupling $g_{AC}$ which
has to be fixed by experiment. If the coupling is chosen small
enough, the AC-fermions will exhibit a supplementary long range
force with a Coulomb behavior. The corresponding necessarily massless
"photons"
will be called $y$-photons. Implications and effects on the high
energy physics of the standard model will not be considered in
this paper, but there may be detectable effects due to
interactions between AC-fermions and standard model particles on
loop level.

Indeed, loop diagrams with virtual $A$ and $C$ pairs induce mixing
between $y$-photon and ordinary gauge bosons ($y-\gamma$ and
$y-Z$). Due to this mixing ordinary particles acquire new long
range ($y$) interaction, which, however, can be masked in the
electro-neutral matter.
 Such mixing can also result in a small electric charge for
electrically neutral particles \cite{Holdom:1986eq}. But this
effect and the strong experimental constraints on such milli-charged
particles \cite{Davidson:2000hf} would be important, if the AC-leptons
possess only $y$-interaction. This is not appropriate in the
considered case, where AC-leptons are electrically charged.

The masses of the standard model fermions are obtained by
minimizing the Higgs potential. As in the standard model, in the
first generation the neutrino appears left handed only and
massless. But this is not in conflict with the fact that neutrino
masses have been observed through neutrino oscillations. In these
experiments only mass differences can be measured, and
almost-commutative geometry allows to introduce masses for the
second and the third generation.

Furthermore it turns out that the masses $m_A$
and $m_C$ of the new fermions do not feel the fluctuations of the
Dirac operator and are thus Dirac masses which do not stem from
the Higgs mechanism but have a purely geometrical origin. This is
due to the necessarily vector like coupling of the gauge group induced by the
lift of the automorphisms. Consequently these Dirac masses do not
break gauge invariance. The mass scale will later be fixed on
cosmological grounds.

\section{\label{primordial} Primordial AC-particles from the Big Bang Universe}

The AC-model admits that in the early Universe a charge
asymmetry of AC-fermions can be generated so that an $A$ and a $C$
excess saturates the modern dark matter density, dominantly in the
form of $(AC)$ atoms. For light baryon excess $\eta_b=
n_{b\,mod}/n_{\gamma \,mod} = 6 \cdot 10^{-10}$ it gives an
AC-excess
 \begin{eqnarray}
 \eta_{A}&=&n_{A\,mod}/n_{\gamma \,mod} = \eta_{C}=n_{C\,mod}/n_{\gamma \,mod}
 \nonumber \\
 &=& 3 \cdot 10^{-11}
(\frac{100{\GeV}}{M}),
\label{excess}
\end{eqnarray}
where $M=m_A+m_C$ is the sum of the masses of $A$ and $C$. For
future use, following \cite{Glashow,Fargion:2005xz}, it is
convenient to relate the baryon density $\Omega_b=0.044$ and the
AC-lepton density $\Omega_{CDM}=0.224$ with the entropy density
$s$, and to introduce $r_b = n_b/s$ and
$r_{A}=r_{C}=n_{A}/s=n_{C}/s$. Taking into account that
$s_{mod}=7.04\cdot n_{\gamma\,mod},$ one obtains $r_b \sim 8 \cdot
10^{-11}$ and \beq r_{A} =r_{C} = 4 \cdot 10^{-12}
(\frac{100{\GeV}}{M}). \label{sexcess} \eeq We'll further assume
that $m_A=m_C=M/2=m$, so the AC -fermion excess Eq.(\ref{sexcess})
is given by
\begin{eqnarray}
\kappa_{A} &=&\kappa_{C} =r_{A} -r_{\bar A} =r_{C} -r_{\bar C}
\nonumber \\
&=& 2 \cdot 10^{-12} (\frac{100{\GeV}}{m}) = 2 \cdot
10^{-12}/S_2, \label{Eexcess}
\end{eqnarray}
where $S_2 = m/100{\GeV}$.

\subsection{\label{Chronology} Chronological cornerstones of the AC-Universe}
After the generation of AC-lepton asymmetry, the thermal history
of AC-matter in chronological order looks as follows for $m_A =
m_C = m = 100 S_2{\GeV} $:

1) $10^{-10}S_2^{-2}\s \le t \le 6 \cdot10^{-8}S_2^{-2}\s$ at $m
\ge T \ge T_f=m/31 \approx 3 S_2 \GeV.$  AC-lepton pair $A \bar A$
and $C \bar C$ annihilation and freezing out (Subsection
\ref{Efreezing} and Appendix 6). For large $m$ the abundance of
frozen out AC-lepton pairs is not suppressed in spite of an
AC-lepton excess.

2)$1.5 \cdot 10^{-4}S_2^{-2}k_y^{-4}\s \le t \le 1.3 \cdot 10^{-1}
S_2^{-2}k_y^{-4}\s$ at $I_{AC} \approx 80 S_2 k_y^{2}\MeV \ge T
\ge I_{AC}/30 .$ In this period recombination of negatively
charged AC-leptons $A^{--}$ with positively charged
$C^{++}$-leptons can lead to the formation of AC-lepton atoms
$(AC)$ with potential energy $I_{AC}= Z_A^2 Z_C^2\alpha^2 k_y^{2}m
\approx 80 S_2 k_y^{2}\MeV $ ($Z_A=Z_C=2$ and $k_y = (1 +
\alpha_y/(Z_AZ_C\alpha))/2 \approx 1$ for $\alpha_y \sim 1/30$).
Together with neutral $(AC)$ atoms free charged $A^{--}$ and
$C^{++}$ are also left, being the dominant form of AC-matter at
$S_2 > 6$.

3)$ t \sim 1.5 \cdot 10^{-4}S_2^{-2}k_y^{-4}\s$  at $T \sim I_{A}
= I_{C} = 80 S_2 k_y^2\MeV.$ The temperature corresponds to the
binding energy $I_{A} = I_{C} = Z_A^4 \alpha^2 k_y^2 m = Z_C^4
\alpha^2 k_y^2 m \approx 80 S_2 k_y^2 \MeV$ ($Z_A=Z_C=2$) of twin
AC-positronium "atoms" $(A^{--} \bar A^{++})$ and $(C^{++} \bar
C^{--})$, in which $\bar A^{++}$ and $\bar C^{--}$ annihilate. At
large $m$ this annihilation is not at all effective to reduce the
$A \bar A$ and $C \bar C$ pairs abundance. These pairs are
eliminated in the course of the successive evolution of AC-matter.

4)$100\s \le t \le 300\s$  at $100 \keV\ge T \ge I_{OHe}/27
\approx 60 \keV,$ where $I_{OHe}= Z_{He}^2 Z_{A}^2 \alpha^2
m_{He}/2 = 1.6 \MeV$ is the ionization potential of a
$(^4He^{++}A^{--})$ "atom". Helium $^4$He is formed in the
Standard Big Bang Nucleosynthesis and virtually all free $A^{--}$
are trapped by $^4$He in OLe-helium $(^4He^{++}A^{--})$. Note that
in the period $100 \keV \le T \le 1.6 \MeV$ helium $^4$He is not
formed, therefore it is only after {\it the first three minutes},
when $(^4He^{++}A^{--})$ trapping of $A^{--}$ can take place.
Being formed, OLe-helium catalyzes the binding of free $C^{++}$
with its constituent $A^{--}$ into $(AC)$ atoms. In this period
free $\bar C^{--}$ are also captured by $^4$He. At large $m$
effects of $(A^{--} \bar A^{++})$ and $(C^{++} \bar C^{--})$
annihilation, catalyzed by OLe-helium, do not cause any
contradictions with observations.

The presence of new relativistic species - a gas of primordial
$y$-photons - does not influence the light element abundances,
since the $y$-photons decouple at $T< T_f (Z^2 \alpha/\alpha_y)$
from the cosmological plasma after AC-lepton pairs are frozen out
at $T_f=m/30 \approx 3 S_2 \GeV.$ Here $\alpha_y$ is the fine
structure constant of the $y$-interaction and $Z=2$. Therefore the
contribution of $y$-photons into the total number of relativistic
species in the period of SBBN is suppressed.

5) $t \sim 2.5 \cdot 10^{11}\s$  at $T \sim I_{He}/30 \approx 2
eV.$ Here $I_{He}= Z^2 \alpha^2 m_{e}/2 = 54.4 \eV$ is the
potential energy of an ordinary He atom.  Free $C^{++}$ with
charge $Z=+2$ recombine with $e^-$ and form anomalous helium atoms
$(eeC^{++})$.

6) $t \sim 10^{12}\s$  at $T \sim T_{RM} \approx 1 \eV.$ AC-matter
dominance starts with $(AC)$ atoms, playing the role of CDM in
the formation of Large Scale structures.

7) $z \sim 20.$ The formation of galaxies starts, triggering
$(AC)$ recombination in dense matter bodies.

All these species should be present around us and we turn now to
the stages of their formation.

\subsection{\label{Efreezing} Freezing out of $AC$-leptons}

In the early Universe at temperatures highly above their masses,
AC-fermions  were in thermodynamical equilibrium with the
relativistic plasma. It means that at $T>m$ the excessive $A$ and
$C$ were accompanied by $A \bar A$ and $C \bar C$ pairs.

When in the course of the expansion the temperature $T$ falls down
\footnote{This picture assumes that the reheating temperature
$T_r$ after inflation exceeds $m$. A wide variety of inflationary
models involve a long pre-heating stage, after which the reheating
temperature does not exceed $T_r < 4\cdot 10^6 GeV$. This upper
limit appears, in particular as a necessary condition to suppress
an over-abundance of primordial gravitinos (see e.g.
\cite{Linde,Linde2,Linde3}, for review and Refs. \cite{bookKh}).
Therefore the successive description of the freezing out of
AC-fermions may not be strictly applicable for very large $S_2 >
10^4$, when nonequilibrium mechanisms of AC-particle creation can
become important. However, even the out-of-equilibrium mechanisms
of AC-particle creation in the early  Universe can hardly avoid
the appearance of AC-lepton pairs.} below the mass of
AC-particles the concentration of particles and
antiparticles is given by the equilibrium. The equilibrium
concentration of $A \bar A$ and $C \bar C$ pairs starts to
decrease at $T<m_A=m_C=m=100 S_2 \GeV$. At the freezing out
temperature $T_f$ the rate of expansion exceeds the rate of
annihilation to photons $A \bar A  \rightarrow \gamma \gamma$, to
$y$-photons $A \bar A  \rightarrow y y$ or to pairs of light
fermions $f$ (quarks and charged leptons) $A \bar A \rightarrow
\bar f f$ (and of the same processes for $C \bar C$ pairs). Then
$A$ (and $C$) leptons and their antiparticles $\bar A$ ($\bar C$)
are frozen out.

Simultaneously, since the  freezing out of $AC$ leptons results in
decrease of their number density, they cannot provide effective
energy transfer between $y$-photon gas and other species of
equilibrium plasma. Therefore $y$-photons decouple from
cosmological plasma and their successive contribution into the
entropy and density of relativistic plasma (being determined by
the fraction of entropy, corresponding to them) is suppressed due
to decrease of effective number of degrees of freedom, which
corresponds to $g_s \sim 100$ in the period of decoupling and
becomes an order of magnitude smaller after QCD phase transition.

The $S_2$-dependence of the frozen out abundances (in units of the entropy
density) of AC-leptons and their antiparticles
\begin{eqnarray}
r_{AC}= f_{AC}(S_2)\nonumber\\
r_{\bar{AC}}=f_{\bar{AC}}(S_2) \label{rEfr}
\end{eqnarray}
is given by Eq.(\ref{rUpm}) in Appendix 6. For growing $S_2 \gg 1$
the solution Eq.(\ref{rUpm}) approaches the values
\begin{eqnarray}
r_{AC} \approx \sqrt{f_2(x_f)} + \kappa/2 \approx\nonumber\\
\approx 1.5 \cdot 10^{-13}S_2\cdot (1 - \ln{(S_2)}/30) + 10^{-12}/S_2\nonumber\\
r_{\bar{AC}} \approx \sqrt{f_2(x_f)} - \kappa/2 \approx \nonumber\\
\approx 1.5 \cdot 10^{-13}S_2\cdot (1 - \ln{(S_2)}/30) -
10^{-12}/S_2. \label{rESfr}
\end{eqnarray}
At $S_2 < 3.6$ there is a exponential suppression of
the $\bar{AC}$ abundance. For $S_2$, close to 1, one has
\begin{eqnarray}
r_{AC}=f_{AC}(S_2) \approx \kappa = 2 \cdot 10^{-12}/S_2\nonumber\\
r_{\bar{AC}}=f_{\bar{AC}}(S_2) \approx 5 \cdot 10^{-3}\kappa S_2^4
\exp\left( -13/S_2^2 \right). \label{rE1fr}
\end{eqnarray}
At $S_2=1$ the solution Eq.(\ref{rE1fr}) gives
$$r_{AC} \approx \kappa_A = \kappa_C = 2 \cdot 10^{-12}$$ and
$$r_{\bar{AC}} \approx 3 \cdot 10^{-20}.$$
On the other hand, at $S_2 > 5$ the concentration of frozen out
AC-lepton pairs exceeds the one of the AC-lepton excess, given by
Eq.(\ref{Eexcess}) and this effect grows with $S_2$ as $\propto
S_2^2$ at large $S_2$. So in this moment, in spite of an assumed
AC-lepton asymmetry, the frozen out concentration of antiparticles
$\bar A$ and  $\bar C$ is not strongly suppressed and they cannot
be neglected in the cosmological evolution of AC-matter.

Antiparticles $\bar A$ and $\bar C$ should be effectively
annihilated in the successive processes of $A \bar A$ and $C \bar
C$ recombination in bound $(A \bar A)$ and $(C \bar C)$
AC-positronium states.

\subsection{\label{anE} $\bar A$ and $\bar C$ annihilation in AC-positronium states}
The frozen out antiparticles ($\bar A$ and $\bar C$) can bind at
$T< I_{A} = I_{C} = Z_A^4 \alpha^2 k_y^2 m = Z_C^4 \alpha^2 k_y^2
m \approx 80 S_2 k_y^2 \MeV$ with the corresponding particles ($A$
and $C$) into positronium-like systems and annihilate. The binding
is provided both by Coulomb interaction of electromagnetic charges
$Z_A=Z_C=2$ as well as by Coulomb-like $y$ attraction. Both
effects are taken into account by the factor $k_y = (1 +
\alpha_y/(Z_AZ_C\alpha))/2 \approx 1$ for $\alpha_y \sim 1/30$.
Since the lifetime of these positronium-like systems is much less
than the timescale of their disruption by energetic photons, the
direct channel of $\bar A$ and $\bar C$ binding in $(A \bar A)$
and $(C \bar C)$, followed by a rapid annihilation, cannot be
compensated by an inverse reaction of photo-ionization. That is
why $\bar A$ ($\bar C$) begin to bind with $A$ ($C$) and
annihilate as soon as the temperature becomes less than $I_{A} =
I_{C} = Z_A^4 \alpha^2 k_y^2 m = Z_C^4 \alpha^2 k_y^2 m \approx 80
S_2 k_y^2 \MeV$, where $m=m_A=m_C=100 S_2 \GeV$ and $Z=Z_A=Z_C=2$.
The decrease of the $\bar A$ abundance owing to $A \bar A$
recombination is governed by the equation \beq \frac{dr_{\bar
A}}{dt} = -r_A r_{\bar A} \cdot s \cdot \sv, \label{hadrecombE}
\eeq where $s$ is the entropy density and (see Appendix 6)
$$\sv=  (\frac{16\pi}{3^{3/2}}) \cdot \frac{\bar \alpha}{T^{1/2}\cdot m^{3/2}}.$$
Here $\bar \alpha = Z^2 \alpha + \alpha_y$. For $m_A=m_C=m$ the
same equation describes the decrease of the $\bar C$ abundance due
to $C \bar C$ recombination. Therefore all the successive results
for $\bar A$ are also valid for $\bar C$.

Using the formalism of Appendix 6 we can re-write
Eq.(\ref{hadrecombE}) as \beq \frac{dr_{\bar A}}{dx}=f_{1 \bar
A}\left<\sigma v\right> r_{\bar A}(r_{\bar A}+\kappa_A),
\label{Esrecomb} \eeq where $x=T/I_{A}$, the asymmetry $\kappa_A =
r_A-r_{\bar A} = 2 \cdot 10^{-12}/S_2$ is given by
Eq.(\ref{Eexcess}) and
$$f_{1 \bar A}=\sqrt{\frac{\pi g_s^2}{45g_{\epsilon}}}m_{Pl}I_{A} \approx m_{Pl}I_{A}.$$
The concentration of the remaining $\bar A$ is given by
Eq.(\ref{rrr}) of Appendix 6 \beq r_{\bar A}=\frac{\kappa_A\cdot
r_{f\bar A}}{(\kappa_A+r_{f\bar A}) \exp\left( \kappa_A J_{A}
\right) - r_{f\bar A}}, \label{rEbar} \eeq where from
Eq.(\ref{rEfr})
$$r_{f \bar A}= r_{\bar{AC}}$$
and
$$J_{A}=\int_0^{x_{f\bar A}} f_{1\bar A}\left<\sigma v\right>dx =$$
\beq = m_{Pl}I_{A} 4 \pi (\frac{2}{3^{3/2}}) \cdot \frac{ \bar
\alpha^2}{I_{A}\cdot m} \cdot 2 \cdot x_{f\bar A}^{1/2}\approx 3.2
\cdot 10^{15}k_y^2/S_2. \label{JEbar} \eeq In the evaluation of
Eq.(\ref{JEbar}) we took into account that the decrease of $\bar
A$ starts at $T\sim I_{A}$, so that $x_{f\bar A} \sim 1$. At $S_2
< 80k_y$ the abundance of $\bar A$ is suppressed exponentially.

Indeed, one has $\kappa_A J_{A} \approx 6400 k_y^2/S_2^2$ in the
exponent of Eq.(\ref{rEbar}). It differs significantly from the
situation revealed in \cite{Fargion:2005xz} for tera-positrons in
Glashow's sinister model \cite{Glashow}. Though in both cases a
decrease of antiparticles due to the formation of positronium like
systems is induced by electromagnetic interaction and the factor
in the exponent is determined by the square of the fine structure
constant $\alpha$, in the case of AC-leptons, this factor is
enhanced by $4Z^4k_y^2=64k_y^2$ times owing to the $4Z^4k_y^2$
dependence of $\bar \alpha^2$. It results in a much wider mass
interval for AC-leptons, in which the primordial pair abundance is
exponentially suppressed.

At $S_2$ close to 1  the condition $r_{f\bar A} \ll \kappa_A$ in
the solution Eq.(\ref{rEbar}) provides the approximate solution
$$r_{\bar A}= r_{f\bar A} \cdot  \exp\left(- \kappa_A J_{A} \right) \approx
10^{-14}S_2^3 \exp\left( -6413k_y^2/S_2^2 \right).$$
For $S_2>5$ the condition $r_{f\bar A} \gg \kappa_A$ is valid.
Therefore the solution Eq.(\ref{rEbar}) has the form \beq r_{\bar
A} \approx \frac{\kappa_A}{ \exp\left( \kappa_A J_{A} \right) -
1}, \label{rEbars} \eeq which gives for $S_2 < 80k_y$
$$r_{\bar A}= \kappa_A \cdot  \exp\left(- \kappa_A J_{A} \right)
\approx \left(\frac{2\cdot 10^{-12}}{S_2}\right) \exp\left(
-6400k_y^2/S_2^2\right).$$ At large $S_2 > 80k_y$ the approximate
solution is given by
$$r_{\bar A} \approx \frac{1}{ J_{A} }  -  \frac{\kappa_A}{ 2 } \approx 3 \cdot 10^{-16} S_2/k_y^2 - 10^{-12}/S_2.$$
In the result the residual amount of $\bar A$ remains at $S_2 >
80k_y$ enormously high, being  for  $S_2 > 98k_y$  larger than the
AC-lepton excess. This effect grows with $S_2 > 98k_y$ as $\propto
S_2^2.$

The general expression for the particle ($A$ and $C$) abundance
$r_A$ after twin AC-positronium annihilation has the form (see
Eq.(\ref{rrr}) of Appendix 6)
$$r_A=\frac{\kappa_A\cdot r_{Af}}{r_{Af}-(r_{Af}-\kappa_A) \exp\left(-\kappa_A J_{A}\right)},$$
where $J_{A}$ is given by Eq.(\ref{JEbar}) and from
Eq.(\ref{rEfr})
$$r_{Af}= r_{AC}.$$
With the account for $r_{Af} > \kappa_A$ for all $S_2$ one obtains
\beq r_A=\frac{\kappa_A}{1-\exp\left(-\kappa_A J_{A}\right)}.
\label{rEtpan} \eeq It tends to $r_{A} \approx 1/J_{A}  +
\kappa_A/2 \approx 3 \cdot 10^{-16} S_2 + 10^{-12}/S_2$ at large
$S_2$ and to $\kappa_A$ for $S_2 < 80k_y$.

\subsection{\label{recE} A-C recombination}
At the temperature $T< I_{AC} =Z_A^2 Z_C^2 \alpha^2 k_y^2 m
\approx 80 k_y^2 S_2 MeV$ (where the electric charge of $A$ and
$C$ is $Z_A=Z_C=2$ and additional binding due to $y$-attraction is
taken into account) $A$ and $C$ can form atom-like bound systems.
Reactions \beq A + C \rightarrow (AC) + \gamma \label{EUUUbind}
\eeq are balanced by inverse reactions of photo-destruction. In
the direct channel the reaction \beq A + C \rightarrow (AC) + y
\label{EUUUybind} \eeq is also essential, while the inverse
process is effective at $S_2 > 2$, when the contribution of
$y$-photons is not suppressed. According to the Saha-type equation
\beq \frac{n_{A} n_C}{n_{\gamma} n_{(AC)}} =
\exp{(-\frac{I_{AC}}{T})} \label{recEPSaha1} \eeq  an effective
formation of $(AC)$ systems is delayed.

In this period composite $(AC)$ cold dark matter is formed.
However, though at $S_2 \sim 1$ most of $A$ bind with $C$ into
$(AC)$, a significant fraction of free $A$ and $C$ remains
unbound, what we show below. Moreover, such a binding is not
efficient at large $S_2 > 6$.

In the considered case $m_A=m_C=m$ the frozen out concentrations
of $A$ and $C$ are equal, being $r_{A}=r_{C}=\kappa_A = 2 \cdot
10^{-12}/S_2$ for $S_2 < 80k_y$ and $r_{A}=r_{C} = 3 \cdot
10^{-16} S_2 + 10^{-12}/S_2$ for larger $S_2$.

Defining the fraction of free $A$ and $C$ as $r_{Af}=r_{Cf}=p
\cdot r_{A}=p \cdot r_{C}$, we have for the fraction of bound
$(AC)$ states $r_{(AC)}=(1-p) \cdot r_{A}=(1-p) \cdot r_{C}$,
where $y$ is determined by the equilibrium and we can re-write the
Saha equation Eq.(\ref{recEPSaha1}) as $$\frac{p^2}{1-p} W=1,$$
with
$$W=\exp{(\frac{I_{AC}}{T} + \ln{\eta_A})}.$$ Here $\eta_A =
\eta_C$ is defined by Eq.(\ref{excess}) and is given by $\eta_A =
\eta_C = 1.5 \cdot 10^{-11}/S_2$.

The equilibrium value of $p$ is equal to $p_{eq}\approx 1-W$ for
$W \ll 1$ (at $x=T/I_{AC}>\frac{1}{26(1+\ln{S_2}/26)}$) and
$p_{eq}\approx 1/\sqrt{W}$ for $W \gg 1$ (at $x <
\frac{1}{26(1+\ln{S_2}/26)}$). The equilibrium ratio $p_{eq}$ is
frozen out at $T = T_f$, when the rate of recombination
$R(T_f)=n_{eq}(T_f)\left<\sigma_{ann}v(T_f)\right>$ becomes less
than the rate of expansion $R(T_f)=H(T_f)$. With the use of
formulae of Appendix 6 the freezing out temperature
$x_f=T_f/I_{AC}$ is determined by the relationship \beq
p_{eq}(T_f)r_A f_{1AC} \left<\sigma_{ann}v(T_f)\right> x_f = 1,
\label{EPfrcomb} \eeq where
$$f_{1AC}=\sqrt{\frac{\pi g_s^2}{45g_{\epsilon}}}m_{Pl}I_{AC} \approx m_{Pl}I_{AC}$$
and $$f_{1AC} \left<\sigma_{ann}v(T_f)\right> x_f \approx J_A
\sqrt{x_f} = 3.2 \cdot 10^{14}k_y^2\sqrt{x_f}/S_2.$$ Here $J_A$ is
given by Eq.(\ref{JEbar}). At $S_2 < 80k_y$ $$p_{eq}(x_f)r_A
f_{1AC} \left<\sigma_{ann}v(T_f)\right> x_f \approx
p_{eq}(x_f)\frac{640k_y^2}{S_2^2}\sqrt{x_f}$$ and $x_f > 1/26$ for
$S_2>6$. It leads to $W = \exp{(1/x_f - 26)}/S_2 \ll 1$ for
$S_2>6$ and only  a small fraction of $A$ and $C$ $$W \approx
\exp{(- 26[1-100/S_2^4 ])}/S_2$$ binds in evanescent $(AC)$
states. At larger $S_2 > 80k_y^2$ binding virtually does not take
place. Only for $S_2 < 6$ the value of $W$ exceeds 1 and the
abundance of free $A$ and $C$ decreases owing to their binding.
However, even for $S_2 \sim 1$ a fraction of free AC-particles
remains significant, being only 30 times less, than the abundance
of bound $(AC)$ states.

\subsection{\label{briefSum} Brief summary of AC-traces at $t \sim 1\s$}
The AC-matter content of the Universe to the end of MeV-era is:

1. Free $A$ and $C$ with $r_{A}=r_{C}= 3 \cdot 10^{-16} S_2 +
10^{-12}/S_2$ at $S_2 > 80k_y$. Their abundance tends to
$r_{A}=r_{C}= \kappa_A = \kappa_C= 2 \cdot 10^{-12}/S_2$ at $6 <
S_2 < 80k_y$ and decreases down to $7 \cdot 10^{-14}/S_2$ at $S_2
\rightarrow 1.$

2. Free $\bar A$ and $\bar C$ with $r_{\bar A}=r_{\bar C}= 3 \cdot
10^{-16} S_2 - 1/S_2 10^{-12}$ at $S_2 > 80k_y$ decreasing
$\propto \exp\left( -6400k_y^2/S_2^2\right)$ at smaller $S_2 <
80k_y$.

3. Neutral AC-atoms $(AC)$ with an abundance $r_{(AC)} \approx
\kappa_A = \kappa_C= 2 \cdot 10^{-12}/S_2$ at $ S_2 < 6$ and
decreasing exponentially at larger $ S_2 > 6$.

4. An exponentially small ($\ll 10^{-30}$) amount of AC-anti-atoms
$(\bar A \bar C)$. At small $ S_2$ it is small owing to the
exponential suppression of frozen out antiparticles, while at
large $ S_2$ the antiparticles do not form bound states.

For $S_2 \ge 100$ pairs of free AC-leptons $A \bar A$ and $C \bar
C$ dominate among these relics from MeV era. The abundance of
these pairs relative to the AC-lepton excess grows with $S_2 >100$
as $\propto S_2^2$. Since the mass of an AC-lepton is $\propto
S_2$, the contribution of AC-lepton pairs to the total density
grows as $\propto S_2^3$ relative to the AC-excess. If they
survive, AC-lepton pairs can over-close the Universe even at
modest $S_2 >100$. However they do not survive due to
annihilation, and such annihilation in the first three minutes
does not lead to contradictions with the observed light element
abundance and the CMB spectrum.

In the Big Bang nucleosynthesis $^4$He is formed with an abundance
$r_{He} = 0.1 r_b = 8 \cdot 10^{-12}$ and, being in excess, bind
all the negatively charged AC-species into atom-like systems.
Since the electric charge of $A$ (and $\bar C$) is $-2$, neutral
"atoms" are formed, catalyzing effective $(AC)$ binding and
anti-particle annihilation. It turns out that the electromagnetic
cascades from this annihilation cannot influence the light element
abundance and the energy release of this annihilation takes place
so early that it does not distort the CMB spectrum.

\section{\label{catEHe} Helium-4 cage for free negative charges}

At $T<I_{o} = Z_E^2 Z_{He}^2 \alpha^2 m_{He}/2 \approx 1.6 \MeV$
the reaction \beq A^{--}+^4He^{++}\rightarrow \gamma +(^4HeA)
\label{EHeIg} \eeq can take place. In this reaction a neutral
OLe-helium $(EHe)$ "atom" is produced. The size of this "atom" is
\beq R_{o} \sim 1/(Z_A Z_{He}\alpha m_{He}) \approx 2 \cdot
10^{-13} \cm \label{REHe} \eeq and it can play a nontrivial
catalyzing role in nuclear transformations. This aspect needs
special thorough study, but some arguments, which we present
below, favor no significant influence on the SBBN picture.

For our problem another aspect is also important.  The reaction
Eq.(\ref{EHeIg}) can start only after $^4$He is formed, what
happens at $T<100 \keV$. Then inverse reactions of ionization by
thermal photons support Saha-type relationships between the
abundances of these "atoms", free $A^{--}$, $^4$He and $\gamma$:
\beq \frac{n_{He} n_A}{n_{\gamma} n_{(HeA)}} =
\exp{(-\frac{I_{o}}{T})}. \label{recHeSaha1} \eeq When $T$ falls
down below $T_{rHe} \sim I_{o}/\ln{\left(n_{\gamma}/n_{He}\right)}
\approx I_{o}/27 \approx 60 \keV$ free $A^{--}$ are effectively
bound with helium in the reaction Eq.(\ref{EHeIg}). The fraction
of free $A$, which forms neutral $(^4He^{++}A^{--})$ depends on
the ratio of $A^{--}$ and the $^4$He abundances. For $S_2<6.4
\cdot 10^3$ this ratio is less than 1. Therefore owing to $^4$He
excess virtually all $A^{--}$ form $(^4He^{++}A^{--})^+$ "atoms"
in reaction Eq.(\ref{EHeIg}).

As soon as neutral OLe-helium $(^4HeA)$ is formed it catalyzes the
reaction of $AC$ binding \beq C+(HeA) \rightarrow (AC) +^4He.
\label{EHeUUU} \eeq In these reactions a heavy $C$-ion penetrates
the neutral $(HeA)$ "atom", expels $^4He$ and binds with $A$ into
the AC-atom $(AC)$.

At $S_2 > 80k_y$ the concentration of primordial antiparticles
$\bar A$ and $\bar C$ is not negligible. At $T_{rHe} \sim
I_{o}/\log{\left(n_{\gamma}/n_{He}\right)} \approx I_{o}/27
\approx 60 \keV$ $\bar C$ form in the reaction \beq \bar
C^{--}+^4He\rightarrow \gamma +(^4He \bar C) \label{EHePg} \eeq
P-OLe-helium "atoms" $(^4He \bar C)$.  As soon as OLe-helium and
P-OLe-helium are formed heavy antiparticles can penetrate them,
expelling $^4He$ and forming twin AC-positronium states $(A \bar
A)$ and $(C \bar C)$, in which the antiparticles annihilate.
Therefore antiparticle ($\bar A$ and $\bar C$) annihilation
through twin AC-positronium formation, such as \beq (HeA)+\bar A
\rightarrow (A \bar A \,annihilation\,products ) +^4He
\label{EHeP} \eeq takes place.

\subsection{\label{Hetrap} $(^4He)$ trapping of free negative charges}

At $T \le T_{rHe} \sim I_{o}/\log{\left(n_{\gamma}/n_{He}\right)}
\approx I_{o}/27 \approx 60 \keV$, when the inverse reaction of
photo-destruction cannot prevent reaction (\ref{EHeIg}), the
decrease of the free $A$ abundance owing to $A He$ binding is
governed by the equation \beq
\frac{dr_{E}}{dx}=f_{1He}\left<\sigma v\right> r_{A}r_{He},
\label{Esrecomb} \eeq where $x=T/I_{o}$, $r_{He} = 8 \cdot
10^{-12}$, $\sv$ according to Appendix 7 is given by
$$\sv=  (\frac{4\pi}{3^{3/2}}) \cdot \frac{\bar \alpha^2}{I_{o}\cdot m_{He}}\frac{1}{x^{1/2}},$$
$\bar \alpha = Z_A Z_{He} \alpha$ and
$$f_{1He}=\sqrt{\frac{\pi g_s^2}{45g_{\epsilon}}}m_{Pl}I_{o} \approx m_{Pl}I_{o}.$$
The solution of Eq.(\ref{Esrecomb}) is given by
$$r_{A}=r_{A0} \exp\left(-r_{He} J_{He}\right)=
r_{A0}\exp\left(-1.28 \cdot 10^{4} \right).$$ Here
$$J_{He}=\int_0^{x_{fHe}} f_{1He}\left<\sigma v\right>dx =$$
\beq = m_{Pl} (\frac{4 \pi}{3^{3/2}}) \cdot \frac{Z_A^2 Z_{He}^2
\alpha^2}{m_{He}} \cdot 2 \cdot \sqrt{x_{fHe}}\approx 1.6 \cdot
10^{15} \label{JHe} \eeq and $x_{fHe}=1/27.$ Thus, virtually all
the free $A$ are trapped by helium and their remaining abundance
becomes exponentially small.

For particles $Q^-$ with charge $-1$, as for tera-electrons in the
sinister model \cite{Glashow}, $^4He$ trapping results in the
formation of  a positively charged ion $(^4He^{++} Q^-)^+$,
putting up a Coulomb barrier for any successive process of
recombination \cite{Fargion:2005xz}. Therefore, only the choice of
$\pm 2$ electric charge for the AC-leptons makes it possible to
avoid this problem. In this case $(^4He)$ trapping leads to the
formation of neutral {\it OLe-helium} "atoms" $^4He^{++}Q^{--}$,
which can catalyze an effective binding of $A$ and $C$ into an
$(AC)$ dark matter species.

\subsection{\label{OleHecat} OLe-helium catalysis of (AC) binding}
The process of $C^{++}$ capture by the $(HeA)$ atom looks as
follows \cite{leptons}. Being in thermal equilibrium with the
plasma, free $C^{++}$ have momentum $k = \sqrt{2T m_C}.$ If their
wavelength is much smaller than the size of the $(HeA)$ atom, they
can penetrate the atom and bind with $A^{--}$, expelling $He$ from
it. The rate of this process is determined by the size of the
$(HeA)$ atoms  and is given (without $y$-interaction)
by $$\sv_0 \sim \pi R_{o} ^2 \sim \frac{\pi}{(\bar \alpha
m_{He})^2} = \frac{\pi}{2 I_{o} m_{He}} \approx 3 \cdot
10^{-15}\frac{cm^3}{s}. $$ Here $\bar \alpha = Z_A Z_{He} \alpha.$
At $T< T_a = \bar \alpha^2 m_{He} \frac{m_{He}}{2m_{C}} =
\frac{I_{o}m_{He}}{m_{C}} = 4 \cdot 10^{-2} I_{o}/S_2$ the
wavelength of $C^{++}$, $\lambda$, exceeds the size of $(HeA)$ and
the rate of $(HeA)$ catalysis is suppressed by a factor
$(R_{o}/\lambda)^3 = (T/T_a)^{3/2}$ and is given by
$$\sv_{cat}(T<T_a)= \sv_0 \cdot (T/T_a)^{3/2}.
$$

In the presence of the $y$-interaction both OLe-helium and
$C^{++}$ are y-charged and for slow charged particles a
Coulomb-like factor of the "Sakharov enhancement"
\cite{Sakhenhance} should be added in these expressions
$$C_y=\frac{2 \pi \alpha_y/v}{1 - \exp{(-2 \pi \alpha_y/v)}},$$
where $v=\sqrt{2T/m}$ is relative velocity. It results in \beq
\sv_0 = \pi R_{o} ^2 \cdot 2 \pi\alpha_y \cdot (m/2T)^{1/2} =
\frac{\alpha_y}{\bar \alpha} \frac{\pi^2}{I_{o} m_{He}} \cdot
(\frac{m}{ m_{He}})^{1/2}\frac{1}{x^{1/2}} \approx
10^{-13}\frac{\alpha_y}{1/30}(\frac{S_2}{x})^{1/2}\frac{cm^3}{s},
\label{Epsv} \eeq where $x=T/I_{o}.$ At $T < T_a$ the rate of
OLe-helium catalysis is given by \beq \sv_{cat}(T<T_a)= \sv_0
\cdot (T/T_a)^{3/2}=\frac{\alpha_y}{\bar \alpha} \frac{\pi^2}{T_a
m_{He}} \cdot (\frac{T}{T_a})\approx 2 \cdot 10^{-19}
\frac{\alpha_y}{1/30}S_2^2 (\frac{T}{300K})\frac{cm^3}{s}.
\label{catETa} \eeq

At modest values of $S_2$ the abundance of primordial
antiparticles is suppressed and the abundance of free $C$, $r_C$,
is equal to the abundance of $A^{--}$, trapped in the $(HeA)$
atoms, $r_A=r_{HeA}$. Therefore a decrease of their concentration
due to the OLe-helium catalysis of $(AC)$ binding is determined by
the equation \beq \frac{dr_{C}}{dx}=f_{1He}\left<\sigma v\right>
r_{C}r_{HeA}, \label{Esrecomb} \eeq where $x=T/I_{o}$,
$r_{HeA}=r_C$, $\sv$ is given by Eqs.(\ref{Epsv}) at $T>T_a$ and
(\ref{catETa}) at $T<T_a$, $\bar \alpha = Z_A Z_{He} \alpha$ and
$$f_{1He}=\sqrt{\frac{\pi g_s^2}{45g_{\epsilon}}}m_{Pl}I_{o} \approx m_{Pl}I_{o}.$$
The solution of Eq.(\ref{Esrecomb}) is given by
$$r_C=r_{HeA}=\frac{r_{C0}}{1+r_{C0}J_{o}} \approx \frac{1}{J_{o}} \approx
7 \cdot 10^{-20}(\frac{1/30}{\alpha_y})/f(S_2).$$ Here
$$J_{o}=\int_0^{x_{fHe}} f_{1He}\left<\sigma v\right>dx =$$ \beq
= \pi^2 \frac{\alpha_y}{\bar \alpha}(\frac{ m_{Pl}}{2 m_{He}})
f(S_2) \approx 1.4 \cdot 10^{19}(\frac{\alpha_y}{1/30})\cdot
f(S_2), \label{JEHe} \eeq $x_{fHe}=1/27$ and the dependence on
$S_2$ is described by the function $f(S_2) =
4(\frac{S_2}{1.08})^{1/2} - 3$ for $S_2> 1.08$; $f(S_2) =
(\frac{S_2}{1.08})^{2}$ for $S_2 < 1.08.$

At large $S_2 > 80k_y$ the primordial abundance of antiparticles
($\bar A$ and $\bar C$) is not suppressed. OLe-helium and
C-OLe-helium catalyze in this case the annihilation of these
antiparticles through the formation of AC-positronium. Similar to
the case of tera-particles, considered in \cite{Fargion:2005xz},
it can be shown that the products of the annihilation cannot cause
a back-reaction, ionizing OLe-helium and C-OLe-helium and
suppressing the catalysis.

Indeed, energetic particles, created in $A \bar A$ and $C \bar C$
annihilation, interact with the cosmological plasma. In the
development of the  electromagnetic cascade the creation of
electron-positron pairs in the reaction $\gamma + \gamma
\rightarrow e^+ + e^-$ plays an important role in astrophysical
conditions (see \cite{BL,AV,bookKh} for review). The threshold of
this reaction puts an upper limit on the energy of the
nonequilibrium photon spectrum in the cascade \beq E_{max} =
a\frac{m_{e}^2}{25T}, \label{Emax} \eeq where the factor $a =
\ln{(15 \Omega_b + 1)} \approx 0.5$.

At $T>T_{rbHe} = a m^2_e/(25 I_{o}) \approx 1 \keV$ in the
spectrum of the electromagnetic cascade from $A \bar A$ and $C
\bar C$ annihilation the maximal energy $E_{max} <I_{o}$ and $\bar
A$ ($ \bar C$) annihilation products cannot ionize $(^4HeA^{--})$.
So, there is no back reaction of the $\bar A$ ($ \bar C$)
annihilation until $T \sim T_{rbHe}$ and in this period
practically all free $A^{--}$ and $ \bar C^{--}$ are bound into
$(^4HeA^{--})$ and $(^4He \bar C^{--})$ atoms.

On the same reason electromagnetic showers, induced by annihilation
products and having a maximal  energy below the binding energies of the SBBN
nuclei, cannot initiate reactions of non-equilibrium
nucleosynthesis and influence the light element abundance.

\subsection{\label{ANcat} Complete elimination of antiparticles by OLe-helium catalysis}
In the absence of a back-reaction of annihilation products nothing
prevents the complete elimination of antiparticles ($\bar A^{++}$
and $\bar C^{--}$) by (C-)OLe-helium catalysis.

In the case of $\bar C^{--}$, C-OLe-helium $(^4He \bar C^{--})$
binds virtually all $\bar C^{--}$ and $r_{\bar{AC}} =r_{\bar
C}=r_{\bar CHe}$. Then free $C^{++}$ with primordial abundance
$r_{C} =r_{\bar C} + \kappa_{C}= r_{\bar{AC}} + \kappa$ collide
with $(^4He \bar C^{--})$, expel $^4He$ and form twin
C-positronium $(C^{++} \bar C^{--})$, which rapidly annihilates.

$\bar A^{++}$ with primordial abundance $r_{\bar{AC}} = r_{\bar
A}$, colliding with OLe-helium with abundance $r_{EHe} = r_{A} =
 r_{\bar A} + \kappa_{A}= r_{\bar{AC}} + \kappa$, expel $^4He$
from $(^4HeA^{--})$ and annihilate in twin A-positronium $(\bar
A^{++} A^{--})$.

In both cases the particle excess $\kappa= \kappa_{A}= \kappa_{C}
= 2 \cdot 10^{-12}/S_2$ is given by Eq.(\ref{Eexcess}) and the
decrease of antiparticle abundance $r_{\bar{AC}}$ is described by
the same equation \beq
\frac{dr_{\bar{AC}}}{dx}=f_{1He}\left<\sigma v\right>
r_{\bar{AC}}(r_{\bar{AC}} + \kappa), \label{AEsrecomb} \eeq where
for large $S_2$, when the primordial abundance of antiparticles is
not suppressed, $\sv$ is given by Eq.(\ref{Epsv}). The solution of
this equation is given by Eq.(\ref{rrr}) of Appendix 7, which in
the considered case has the form \beq
r_{\bar{AC}}=\frac{\kappa\cdot
r_{f\bar{AC}}}{(\kappa_A+r_{f\bar{AC}}) \exp\left( \kappa J_{o}
\right) - r_{f\bar{AC}}}, \label{rHeEbar} \eeq where
$r_{f\bar{AC}}$ is given by Eq.(\ref{rEbars}) and $J_{o} = 1.4
\cdot 10^{19}(\frac{\alpha_y}{1/30})\cdot f(S_2)$ is given by
Eq.(\ref{JEHe}) with $f(S_2) \approx 1$ at large $S_2 \gg 1$. The
factor in the exponent is $\kappa J_{o} = 2.8 \cdot 10^{7}/S_2$.
It leads to a huge exponential suppression of the antiparticles at
$S_2 \ll 10^{7}$.

\subsection{\label{EHESBBN} OLe-helium in the  SBBN}
The question of the participation of OLe-helium in nuclear
transformations and its direct influence on the light element
abundance is less evident. Indeed, OLe-helium looks like an $\alpha$
particle with a shielded electric charge. It can closely approach
nuclei due to the absence of a Coulomb barrier. On that reason it
seems that in the presence of OLe-helium the character of SBBN
processes should change drastically. However, it might not be the
case.

The following simple argument can be used to indicate that
OLe-helium influence on SBBN transformations might not lead to
binding of $A^{--}$ with nuclei, heavier, than $^4He$.

In fact, the size of OLe-helium is of the order of the size of
$^4He$ and for a nucleus $^A_ZQ$ with electric charge $Z>2$ the
size of the Bohr orbit for an $Q A^{--}$ ion is less than the size
of the nucleus $^A_ZQ$. This means that while binding with a heavy
nucleus alexium $A^{--}$ penetrates it and effectively interacts
with a part of the nucleus with a size less than the corresponding
Bohr orbit. This size corresponds to the size of $^4He$, making
OLe-helium the most bound $Q A^{--}$-atomic state. It favors a
picture, according to which an OLe-helium collision with a
nucleus, results in the formation of OLe-helium and the whole
process looks like an elastic collision.

Interaction of the $^4He$ component of $HeA$ with a $^A_ZQ$
nucleus can lead to a nuclear transformation due to the reaction
\beq ^A_ZQ+(HeA) \rightarrow ^{A+4}_{Z+2}Q +A^{--}, \label{EHeAZ}
\eeq provided that the masses of the initial and final nuclei
satisfy the energy condition \beq M(A,Z) + M(4,2) - I_{o}>
M(A+4,Z+2), \label{MEHeAZ} \eeq where $I_{o} = 1.6 \MeV$ is the
binding energy of OLe-helium and $M(4,2)$ is the mass of the
helium-4 nucleus.

This condition is not valid for stable nuclei participating in
reactions of the SBBN. However, tritium $^3H$, which is also
formed in SBBN with abundance $^3H/H \sim 10^{-7}$ satisfies this
condition and can react with OLe-helium, forming $^7Li$ and
opening the path of successive OLe-helium catalyzed
transformations to heavy nuclei. This effect might strongly
influence the chemical evolution of matter on the pre-galactic
stage and needs self-consistent consideration within Big Bang
nucleosynthesis network. The following arguments show, however,
that this effect is not as dramatic as it might appear.
\begin{itemize}
\item[$\bullet$] On the path of reactions (\ref{EHeAZ}) the final
nucleus can be formed in the excited $[\alpha, M(A,Z)]$ state,
which can rapidly experience an $\alpha$- decay, giving rise to
OLe-helium regeneration and to an effective quasi-elastic process of
$(HeA)$-nucleus scattering. It leads to a possible suppression of
the OLe-helium catalysis of nuclear transformations.
\item[$\bullet$]
The path of reactions (\ref{EHeAZ}) does not stop on $^7Li$ but
goes further through $^{11}B$, $^{15}N$, $^{19}F$, ... along the
table of nuclides.
\item[$\bullet$] The cross section of reactions
(\ref{EHeAZ}) grows with the mass of the nucleus, making a
formation of heavier elements more probable and moving the main output
away from a
potentially dangerous Li and B overproduction.
\end{itemize}

E-print publication of these results gave rise to the development
of another aspect of the problem - to CBBN - Charged massive
particles BBN, studying influence of unstable negatively charged
massive particles on BBN \cite{Pospelov:2006sc,Kohri:2006cn}. The
important difference of CBBN, considered in
\cite{Pospelov:2006sc}, from our approach is that singly charged
particles $X^-$ with charge $-1$ do not screen the $+2$ charge of He
in $(HeX)$ ion-like bound system, and the Coulomb barrier of
$(HeX)^+$ ion can strongly hamper the path to creation of
isotopes, heavier than $^6Li$. Therefore, $^6Li$ created in the
$D+(HeX)$ reaction cannot dominantly transform into heavier
elements and if not destructed by $X$-decay products should retain
in the primordial chemical content. It makes the $^6Li$ overproduction
found in \cite{Pospelov:2006sc} a really serious trouble for a
wide range of parameters for unstable particles $X$.

It should be noted that the approach of \cite{Pospelov:2006sc} is
not supported by \cite{Kohri:2006cn}. Moreover, we can mention the
following effects, missed in its solution for the $^7Li$ problem:
(i) the competitive process of $^7Li$ creation by similar
mechanism in reaction $^3H+(HeX)^+$  with tritium and (ii) the
effects of non-equilibrium nucleosynthesis reactions, induced by
hadronic and electromagnetic cascades from products of $X$ decays.
The latter effect, which was discussed in \cite{Kohri:2006cn},
implies a self-consistent treatment based on the theory of
non-equilibrium cosmological nucleosynthesis
\cite{Linde3,Linde4,bookKh} (see also
\cite{Kawasaki:2004yh,Kawasaki:2004qu,Kohri:2005wn,Jedamzik:2006xz}). Both effects
(i) and (ii) were not studied in \cite{Pospelov:2006sc}.

The main role of neutral Ole-helium in our approach is its
catalyzing role in binding free $C^{++}$ and bound in neutral
$(HeA^{--})$ OLe-helium  $A^{--}$ leptons into weakly interacting
$(AC)$ atoms. It leaves the amount of Ole-helium, giving rise to
nuclear transformations, at the level of the solution of
Eq.(\ref{Esrecomb}), $n_{AC}/s \le 10^{-19}$ which is formally by
two orders of magnitude less than the constraint $n_X/s \le
10^{-17}$, derived in Eq.(10) of \cite{Pospelov:2006sc}. It should
be also noted that this constraint is not valid for our case if the
binding energy $I_o= 1589 \keV$ of OLe-helium is taken,
corresponding to the Bohr orbit of the $(HeA)$ atom. According to
\cite{Kohri:2006cn} this approximation is valid for $0<ZZ_A \alpha
M_Z R_Z<1$, where $R_Z \sim 1.2 A^{1/3}/200 \MeV^{-1}$ is the size
of nucleus, which is the case for $(HeA)$ atom. Then $D+(HeA) ->
^6Li + A$ reaction, on which the constraint is based, does not go.
This reaction can take place only for a reduced binding energy of
$(HeA)$ ($E= 1200 \keV$ and $E= 1150 \keV$) calculated in
\cite{Pospelov:2006sc} with the account for the charge distribution in
the He nucleus. Then this channel becomes possible, but similar to the
case of tritium the chain of OLe-helium transformations
(\ref{EHeAZ}), started from deuterium does not stop on $^6Li$, but
goes further through $^{10}B$, $^{14}N$, $^{18}F$, ... along the
table of nuclides. Such a qualitative change of the physical picture
appeals to necessity in a detailed nuclear physics treatment of
($A^{--}$+ nucleus) systems and in the account for the whole set
of transformations induced by OLe-helium, including an analysis of a
possibility of fast conversion of helium to carbon and of the
formation  of a $(^8BeA^{--})$ system, mentioned in
\cite{Pospelov:2006sc} as possible dangers for our approach.
Though the above arguments do not seem to make these dangers
immediate and obvious, a detailed study of this complicated
problem is needed.


Colexium $C^{++}$ ions, which remain free after OLe-helium
catalysis, are in thermal equilibrium due to their Coulomb
scattering with the matter plasma. At $T < T_{od} \approx 1 \keV$
energy and momentum transfer due to the nuclear interaction from
baryons to OLe-helium is not effective $n_b \sv (m_p/m) t < 1$.
Here
\beq \sigma \approx \sigma_{o} \sim \pi R_{o}^2 \approx
10^{-25}\cm^2.\label{sigOHe}
\eeq
and $v = \sqrt{2T/m_p}$ is the
baryon thermal velocity. Then OLe-helium gas decouples from the plasma
and the radiation and must behave like a sparse component of dark
matter. However, for a small window of parameters $1 \le S_2 \le
2$ at $T<(\frac{\alpha_y}{1/30})\frac{10 \eV}{S_2 f(S_2)^2}$
Coulomb-like scattering due to $y$ interaction with $C^{++}$ ions
returns OLe-helium to the thermal equilibrium with the plasma and supports
an effective energy and momentum exchange between the $A$ and the $C$
components during all the pre-galactic stage.

\section{\label{Interactions} Forms of AC-Matter in the modern Universe}

The development of gravitational instabilities of AC-atomic gas
follows the general path of the CDM scenario, but the composite
nature of $(AC)$ atoms leads to some specific difference. In
particular, one might expect that particles with a mass $m_{AC} =
200 S_2 \GeV$
 should form gravitationally bound objects
with the minimal mass \beq M = m_{Pl} (\frac{m_{Pl}}{m_{AC}})^2
\approx 5\cdot 10^{28}/S_2^2 \g. \label{MHEm} \eeq However, this
estimation is not valid for  composite CDM particles, which $(AC)$
atoms are.

For $S_2<6$ the bulk of $(AC)$ bound states appear in the Universe
at $T_{fAC} = 0.7 S_2 \MeV$ and the minimal mass of their
gravitationally bound systems is given by the total mass of $(AC)$
within the cosmological horizon in this period, which is of the
order of \beq M = \frac{T_{RM}}{T_{fAC}} m_{Pl}
(\frac{m_{Pl}}{T_{fAC}})^2 \approx 6\cdot 10^{33}/S_2^3 \g,
\label{MHEm} \eeq where $T_{RM}=1 \eV$ corresponds to the
beginning of the AC-matter dominated stage.

These objects, containing $N = 2 \cdot 10^{55} \cdot S_2^{-4}$
 $(AC)$ atoms, separated by the cosmological expansion at $z_s \sim
20$,  have the internal number density
$$n \approx 6 \cdot 10^{-5}\cdot S_2^{-1} (\frac{1+z_s}{1+20})^3 \cm^{-3}$$
and the size \beq R = (\frac{N}{4 \pi n/3})^{1/3}\approx 3 \cdot
10^{19}\cdot S_2^{-1} (\frac{1+20}{1+z_s}) \cm. \label{RHEgb} \eeq
At $S_2>6$ the bulk of $(AC)$ atoms is formed only at $T_{o} =
60 \keV$ due to OLe-helium catalysis. Therefore at $S_2>6$ the
minimal mass is independent of $S_2$ and is given by \beq M =
\frac{T_{RM}}{T_{o}} m_{Pl} (\frac{m_{Pl}}{T_{o}})^2 \approx
10^{37} \g. \label{MEPm} \eeq
The size of $(AC)$ atoms is ($Z_A=Z_C=2$) $$R_{AC} \sim 1/(Z_A
Z_C \alpha k_y m) \sim 0.7 \cdot 10^{-14}\cdot S_2^{-1}k_y^{-1}
\cm$$ and their mutual collision cross section is about \beq
\sigma_{AC} \sim \pi R_{AC}^2 \approx 1.5 \cdot 10^{-28}\cdot
S_2^{-2}k_y^{-2}\cm^2. \label{sigHEHE} \eeq $(AC)$ atoms can be
considered as collision-less gas in clouds with a number density
$n_{AC}$ and size $R$, if $n_{AC}R < 1/\sigma_{AC}$.

At a small energy transfer $\Delta E \ll m$ the cross section for
the interaction of $(AC)$ atoms with matter is suppressed by the
factor $\sim Z^2 (\Delta E/m)^2$, being for scattering on nuclei
with charge $Z$ and atomic weight $A$ of the order of
$\sigma_{ACZ} \sim Z^2/\pi (\Delta E/m)^2 \sigma_{AC} \sim 0.25
Z^2 A^2 10^{-43} \cm^2 /S^2_2.$ Here we take $\Delta E \sim 2 A
m_p v^2$ and $v/c \sim 10^{-3}$ and find that even for heavy
nuclei with $Z \sim 100$ and $A \sim 200$ this cross section does
not exceed $ 10^{-35} \cm^2 /S^2_2.$ It proves the WIMP-like
behavior of $(AC)$ atoms in the ordinary matter.

Mutual collisions of $(AC)$ atoms determine the evolution
timescale for a gravitationally bound system of collision-less
$(AC)$-gas \beq t_{ev} = \frac{1}{n \sigma_{AC} v} \approx 4 \cdot
10^{23} S_2^{17/6} (\frac{1 \cm^{-3}}{n})^{7/6}\s,
\label{evHEcoll} \eeq where the relative velocity $v = \sqrt{G
M/R}$ is taken at $S_2<6$ for a cloud of mass Eq.(\ref{MHEm}) and
an internal number density $n$. The timescale Eq.(\ref{evHEcoll})
exceeds substantially the age of the Universe even at $S_2<6$.
Therefore the internal evolution of AC-atomic self-gravitating
clouds cannot lead to the formation of dense objects.

\subsection{\label{Matter} The problem of Anomalous Helium}
The main possible danger for the considered cosmological scenario
is the over-production of primordial anomalous isotopes. The
pregalactic abundance of anomalous helium (of $C$-lepton atoms
$(eeC^{++})$) exceeds by more than 12 orders of magnitude the
experimental upper limits on its content in terrestrial matter.
The best upper limits on the anomalous helium were obtained in
\cite{exp3}. It was found, by searching with the use of laser
spectroscopy for a heavy helium isotope in the Earth's atmosphere,
that in the mass range 5 GeV - 10000 GeV the terrestrial abundance
(the ratio of anomalous helium number to the total number of atoms
in the Earth) of anomalous helium is less than $3 \cdot 10^{-19}$
- $2 \cdot 10^{-19}$. Therefore spectrometric effects of the
$(eeC^{++})$) atom are just the way, by which its search in the
terrestrial atmosphere is possible as well as the present
experimental constraints on anomalous helium were obtained.  As
concerns the spectra of dense astrophysical objects, both due to
the $(AC)$ recombination and ionization of $(eeC^{++})$), as we
show below, $(eeC^{++})$)-atomic concentration in them should be
strongly suppressed and can hardly lead to observable effects.

It may be shown, however, that a purely gravitational
concentration in stars is not effective to reduce the pregalactic
abundance of anomalous helium. Indeed, for a number density $n_s$
of stars with mass $M_s$ and radius $R_s$ the decrease of the
number density $n_i$ of free particles, moving with the relative
velocity $v$, is given by
\begin{eqnarray}
\frac{dn_i}{dt} &=& -n_s n_i \pi R_s(R_s + \frac{2 G M_s}{v^2})v
\nonumber \\
&\approx& -n_s n_i 2\pi \frac{2 G M_s}{v}.
\end{eqnarray}
Therefore, to be
effective (i.e. to achieve substantial decrease of number density
$n_i=n_{i0} \exp(-t/\tau)$) the timescale of capture
$$\tau = \frac{1}{n_s 2 \pi R_s G M_s/v}$$
should be much less than the age of the Universe $\tau \ll t_U=
4\cdot 10^{17}\s$, whereas for $n_s \sim 1 \pc^{-3}$, $M_s =
M_{\odot} \approx 2 \cdot 10^{33}\g$, $R_s = R_{\odot} \approx 7
\cdot 10^{10}\cm$ and $v\sim10^{6}\cm/\s$, $\tau \sim 5 \cdot
10^{23} \s \gg t_U$. Even for supergiants with $M_s \sim 20
M_{\odot}$ and $R_s \sim 10^4 R_{\odot}$ (and even without account
for a smaller number density of these stars) we still obtain $\tau
\sim 3 \cdot 10^{18} \s \gg t_U$.

Moreover, the mechanisms of the above mentioned kind cannot in
principle suppress the abundance of remnants in interstellar gas
by more than factor $f_g \sim 10^{-2}$, since at least $1\%$ of
this gas has never passed through stars or dense regions, in which
such mechanisms are viable.

Though the binding energy of basic DM composites - $(AC)$ atoms -
is rather low ($20 MeV S_2$), owing to their small size and
corresponding WIMP-like interaction, they can be hardly broken by
cosmic rays
and electromagnetic transitions in these "atoms" seem hardly
observable. It does not exclude the possibility of some peculiar
features in the electromagnetic background, which will be a subject of
further studies. Due to very small abundance of anomalous helium and
OLe-helium in dense matter bodies (e.g. in stars) their
contribution to the electromagnetic radiation from such bodies is
negligible.
 Nuclear interaction of OLe-helium ($(HeA)$ state) and
atomic interaction of $(Cee)$ with cosmic rays gives rise to an
ionization of these bound states in the interstellar gas and to an
acceleration of free charged AC leptons in the Galaxy. It may lead
to the presence of a $C^{++}$ (and $A^{--}$) component in cosmic
rays at a level $\sim f_g \xi_i$. Therefore based on the AC-model
one can expect the anomalous helium fractions in cosmic rays
\begin{eqnarray}
 \frac{C^{++}}{He} \ge 10^{-10}/f(S_2),\nonumber\\
\frac{A^{--}}{He} \ge 10^{-10}/f(S_2).
\end{eqnarray}
These predictions are maximal at $S_2 \sim 1$ and decrease with
$S_2$ as $\propto S_2^{-2}$ for $S_2 < 1.08$ and as $\propto
S_2^{-1/2}$ for $S_2 > 1.08$. These fluxes may be within the reach
for future cosmic ray experiments, in particular, for AMS.

The only way to solve the problem of anomalous isotopes is to find
a possible reason for their low abundance inside the Earth and
a solution to this problem implies a mechanism of effective
suppression of anomalous helium in dense matter bodies (in
particular the Earth).
The idea of such a suppression, first proposed in \cite{fractons}
and recently realized in \cite{4had} is as follows.

If anomalous species have an initial abundance relative to baryons
$\xi_{i0}$, their recombination with the rate $\sv$ in a body with
baryonic number density $n$ reduces their abundance during the age
of the body $t_b$ down to \beq \xi_{i} = \frac{\xi_{i0}}{1 +
\xi_{i0}n \sv t_{b}}. \label{matsym} \eeq If $\xi_{i0} \gg 1/(n
\sv t_{b})$ in the result, the abundance is suppressed down to \beq
\xi_{i} = \frac{1}{n \sv t_{b}}. \label{supsym} \eeq To apply this
idea to the case of the AC-model, OLe-helium catalysis can be
considered as the mechanism of anomalous isotope suppression.

The crucial role of the $y$-attraction comes into the realization
of this mechanism. The condition of $y$-charge neutrality makes
Ole-helium to follow anomalous helium atoms $(C^{++}e^-e^-)$ in
their condensation in ordinary matter objects. Due to this
condition OLe-helium and anomalous helium can not separate and
$AC$ recombination goes on much more effectively, since its rate
is given now by Eq.(\ref{sigimpact}) of Appendix 7 $$\sv \approx 2
\cdot 10^{-12}
(\frac{\alpha_y}{1/30})^{2}(\frac{300K}{T})^{9/10}S_2^{-11/10}
\frac{cm^3}{s}. $$ This increase of the recombination rate reduces
the primeval anomalous helium (and OLe-helium) terrestrial content
down to $r \le 5 \cdot 10^{-30}.$

In the framework of our consideration, interstellar gas contains a
significant ($\sim f_g \xi_{C}$) fraction of $(eeC^{++})$. When
the interstellar gas approaches Solar System, it might be stopped
by the Solar wind in the heliopause at a distance $R_h \sim
10^{15} \cm$ from the Sun. In the absence of detailed experimental
information about the properties of this region we can assume for
our estimation following \cite{leptons} that all the incoming
ordinary interstellar gas, containing dominantly ordinary
hydrogen, is concentrated in the heliopause and the fraction of
this gas, penetrating this region towards the Sun, is pushed back
to the heliopause by the Solar wind. In the result, to the present
time during the age of the Solar system $t_E$ a semisphere of
width $L\sim R_h$ is formed in the distance $R_h$, being filled by
a gas with density $n_{hel} \sim (2 \pi R_h^2 v_g t_E n_g)/(2 \pi
R_h^2 L) \sim 10^8 \cm^{-3}.$ The estimations of Appendix 9 show
that this region is transparent for $(HeA)$, but opaque for atomic
size remnants, in particular, for $(eeC^{++})$. Owing to the
$y$-interaction both components can thus be stopped in heliopause.
Though the Solar wind cannot directly stop heavy $(eeC^{++})$, the
gas shield in the heliopause slows down their income to Earth and
suppresses the incoming flux $I_{C}$ by a factor $S_h \sim
1/(n_{hel} R_h \sigma_{tra})$, where  $\sigma_{tra} \approx
10^{-18} S_2 \cm^2$. So the incoming flux, reaching the Earth, can
be estimated as \cite{4had,leptons} \beq I_{C} = \frac{\xi_{C} f_g
n_g v_g}{8\pi}S_h \approx \frac{10^{-10}}{f(S_2)}\frac{S_h}{5
\cdot 10^{-5}}(cm^2 \cdot s \cdot ster)^{-1}. \label{IincP} \eeq
Here $n_g \sim 1 \cm^{-3}$ and $v_g \sim 2 \cdot 10^{4} \cm/\s.$

Kinetic equilibrium between interstellar AC-gas pollution and $AC$
recombination in Earth holds \cite{4had} their concentration in
terrestrial matter at the level \beq n = \sqrt{\frac{j}{\sv}},
\label{statsol} \eeq where \beq j_{A}=j_{C}=j \sim \frac{2 \pi
I_C}{L} = 2.5 \cdot 10^{-11} S_h/f(S_2) \cm^{-3}\s^{-1},
\label{statin} \eeq within the water-circulating surface layer of
thickness $L \approx 4 \cdot 10^5 \cm$.  Here $I_{C} \approx 2
\cdot 10^{-6} S_h (\cm^2 \cdot \s \cdot ster)^{-1}$ is given by
Eq.(\ref{IincP}), factor $S_h$ of incoming flux suppression in
heliopause can be as small as $S_h \approx 5 \cdot 10^{-5}$ and
$\sv$ is given by the Eq.(\ref{sigimpact}). For these values of
$j$ and $\sv$ one obtains in water \beq n \le 3.5
\sqrt{S_h/f(S_2)} (\frac{1/30}{\alpha_y})S_2^{11/20}
\cm^{-3}.\label{ncPE} \eeq It corresponds to a terrestrial
anomalous helium abundance
$$ r \le 3.5 \cdot 10^{-23}\sqrt{S_h/f(S_2)}(\frac{1/30}{\alpha_y})S_2^{11/20},$$
being below the above mentioned experimental upper limits for
anomalous helium ($ r < 10^{-19}$).

The reduction of the anomalous helium abundance due to $(AC)$
recombination in dense matter objects is not accompanied by an
annihilation, which was the case for $U$-hadrons in \cite{4had},
therefore the AC-model escapes the severe constraints \cite{4had}
on the products of such an annihilation, which follow from the
observed gamma background and the data on neutrino and upward muon
fluxes.

\subsection{\label{EpMeffects} Effects of $(OHe)$ catalyzed processes in the Earth}
The first evident consequence of the proposed excess is the
inevitable presence of $(OHe)$ in terrestrial matter. $(HeA)$
concentration in the Earth can reach the value (\ref{ncPE}) for
the incoming $(HeA)$ flux, given by Eq.(\ref{IincP}). The
relatively small size of neutral $(HeA)$ may provide a catalysis
of cold nuclear reactions in ordinary matter (much more
effectively, than muon catalysis). This effect needs special and
thorough nuclear physical treatment. On the other hand, if
$A^{--}$ capture by nuclei, heavier than helium, is not effective
and does not lead to a copious production of anomalous isotopes,
$(HeA)$ diffusion in matter is determined by an elastic collision
cross section (\ref{sigpEpcap}) of Appendix 9 and may effectively
hide OLe-helium from observations.

One can give the following argument for an effective regeneration
of OLe-helium in terrestrial matter. OLe-helium can be destroyed
in reactions (\ref{EHeAZ}). Then free $A^{--}$ are released and
owing to a hybrid Auger effect (capture of $A^{--}$ and ejection
of ordinary $e$ from the atom with atomic number $A$ and charge of
$Z$ of the nucleus) $A^{--}$-atoms are formed, in which $A^{--}$
occupies highly an excited level of the $(Z-A^{--})$ system, which
is still much deeper than the lowest electronic shell of the
considered atom. $(Z-A^{--})$-atomic transitions to lower-lying
states cause radiation in the range intermediate between atomic
and nuclear transitions. In course of this falling down to the
center of the $(Z-A^{--})$ system, the nucleus approaches
$A^{--}$. For $A>3$ the energy of the lowest state $n$ (given by
$E_n=\frac{M \bar \alpha^2}{2 n^2} = \frac{2 A m_p Z^2
\alpha^2}{n^2}$)  of the $(Z-A^{--})$ system (having reduced mass
$M \approx A m_p$) with a Bohr orbit, $r_n=\frac{n}{M \bar
\alpha}= \frac{n}{2 A Z m_p \alpha}$, exceeding the size of the
nucleus, $r_A \sim A^{1/3}m^{-1}_{\pi}$, is less, than the binding
energy of $(OHe)$. Therefore the regeneration of OLe-helium in a
reaction, inverse to (\ref{EHeAZ}), takes place. An additional
reason for dominantly elastic channel for reactions (\ref{EHeAZ})
is that the final state nucleus is created in the excited state
and its de-excitation via $\alpha$-decay can also result in
OLe-helium regeneration.

Another effect is the energy release from OLe-helium catalysis of
$(AC)$ binding. The consequences of the latter process are not as
pronounced as those discussed in \cite{4had,leptons} for the
annihilation of  4th generation hadrons in terrestrial matter, but
it might lead to a possible test for the considered model.

In our mechanism the terrestrial abundance of anomalous
$(C^{++}ee)$ is suppressed due to the $(HeA)$ catalyzed binding of
most of the $C$ from the incoming flux $I_{C}$, reaching the
Earth. $AC$ binding is accompanied by de-excitation of the
initially formed bound $(AC)$ state. To expel $^4He$ from
OLe-helium, this state should have binding energy exceeding
$I_{He}=1.6 MeV$, therefore MeV range $\gamma$ transitions from
the lowest excited levels to the ground state of $(AC)$ with
$I_{AC}=80 S_2 k_y MeV$ should take place. The danger of gamma
radiation from these transitions is determined by the actual
magnitude of the incoming flux, which was estimated in subsection
\ref{Matter} as Eq.(\ref{IincP}).

The stationary regime of $(HeA)$ catalyzed recombination of these
incoming $C^{++}$ in the Earth should be accompanied by gamma
radiation with the flux $F(E) = N(E) I_C l_{\gamma}/R_E$, where
$N(E)$ is the energy dependence of the multiplicity  of $\gamma$
quanta with energy $E$ in $(AC)$-atomic transitions, $R_E$ is the
radius of the Earth and $l_{\gamma}$ is the mean free path of such
$\gamma$ quanta. At $E
> 10 MeV$ one can roughly estimate the flux $F(E> 10 MeV) \sim
\cdot \frac{10^{-16}}{f(S_2)}\frac{S_h}{5 \cdot 10^{-5}} (cm^2
\cdot s \cdot ster)^{-1}$, coming from the atmosphere and the
surface layer $l_{\gamma} \sim 10^3 cm$. Even without the
heliopause suppression (namely, taking $S_h = 1$) $\gamma$
radiation from $AC$ binding seems to be hardly detectable.

In the course of $(AC)$ atom formation electromagnetic transitions
with $\Delta E > 1 \MeV$ can be a source of $e^+e^-$ pairs, either
directly with probability $\sim 10^{-2}$ or due to development of
electromagnetic cascade. If $AC$ recombination goes on
homogeneously in Earth within the water-circulating surface layer
of the depth $L \sim 4 \cdot 10^5 \cm$ inside the volume of Super
Kamiokande with size $l_{K} \sim 3 \cdot 10^3 \cm$ equilibrium
$AC$ recombination should result in a flux of $e^+e^-$ pairs $F_e
= N_e I_C l_{K}/L$, which for $N_e \sim 1$ can be as large as $F_e
\sim \cdot \frac{10^{-12}}{f(S_2)}\frac{S_h}{5 \cdot 10^{-5}}
(cm^2 \cdot s \cdot ster)^{-1}.$

Such an internal source of electromagnetic showers in large volume
detectors inevitably accompanies the reduction of the anomalous
helium abundance due to $AC$ recombination and might give an
advantage of experimental tests for the considered model. Their
signal might be easily  disentangled \cite{leptons} (above a few
MeV range) with respect to common charged
    current neutrino interactions and single electron tracks,
     because the tens MeV gamma lead, by pair productions, to twin
    electron  tracks,  nearly aligned along their Cerenkov rings.
    The signal is piling  the energy in windows where
   few atmospheric neutrino and  cosmic Super-Novae  radiate. The same gamma flux produced is comparable
    to expected secondaries of tau decay secondaries  while showering  in air at
    the horizons edges(\cite{FarSolar03},\cite{FarTau02},\cite{FarTau03}). The predicted signal
strongly depends, however, on the uncertain astrophysical
parameters (concentration OLe-helium and anomalous helium in the
interstellar gas, their flux coming to Earth etc) as well as on
the geophysical details of the actual distribution of OLe-helium
and anomalous helium in the terrestrial matter, surrounding large
volume detectors.

It should be noted that OLe-helium represents a tiny fraction of
dark matter and thus escapes the severe constraints
\cite{McGuire:2001qj,McGuire:2001qj2,McGuire:2001qj3} on strongly
interacting dark matter particles (SIMPs)
\cite{Starkman,Starkman2,Starkman3,Starkman31,Starkman4,Starkman5,McGuire:2001qj,McGuire:2001qj2,McGuire:2001qj3}
imposed by the XQC experiment \cite{XQC,XQC2}.

Direct search for OLe-helium from dark matter halo is not possible
in underground detectors due to OLe-helium slowing down in
terrestrial matter. Therefore special strategy of such search is
needed, which can exploit sensitive dark matter detectors before
they are installed under ground. In particular, future superfluid
$^3He$ detector \cite{Winkelmann:2005un} and, as it was revealed
in \cite{Belotsky:2006fa}, even its existing few $\g$ laboratory
prototype can be used to put constraints on the in-falling
OLe-helium flux from galactic halo.

\subsection{\label{EpMeffects} $(HeA)$ catalyzed formation of $(AC)$-matter
objects inside ordinary matter stars and planets}

$(AC)$ atoms  from the halo interact weakly with ordinary matter
and can be hardly captured in large amounts by a matter object.
However the following mechanism can provide the existence of a
significant amount of $(AC)$ atoms in matter bodies and even the
formation of gravitationally bound dense $(AC)$- bodies inside
them.

Inside a dense matter body $(OHe)$ catalyzes $C$ aggregation into
$(AC)$ atoms in the reaction \beq (eeC^{++}) + (A^{--}He)
\rightarrow (AC)+He+2e. \label{EpUUUEe} \eeq In the result of this
reaction $(OHe)$, interacting with matter with a nuclear cross
section given by Eq.(\ref{sigpEpcap}) of Appendix 9
$$\sigma_{trAb} = \pi R_{o}^2 \frac{m_{p}}{m_A} \approx
10^{-27}/S_2 \cm^{2},$$ and $(eeC^{++})$, having a nearly atomic
cross section of that interaction (\ref{sigpUEcap})$$\sigma_{tra}=
\sigma_a (m_p/m_C) \approx 10^{-18} S_2 \cm^2,$$ bind into weakly
interacting $(AC)$ atom, which decouples from the surrounding
matter.

In this process "products of incomplete AC-matter combustion"
(OLe-helium and anomalous helium), which were coupled to the
ordinary matter by hadronic and atomic interactions, convert into
$(AC)$ atoms, which immediately sinks down to the center of the
body.

The  amount of $(AC)$ atoms produced inside matter object by the
above mechanism is determined by the initial concentrations of
OLe-helium $(A^{--}He)$ and anomalous helium atoms $(eeC^{++})$.
This amount $N$ defines the number density of $(AC)$-matter inside
the object, being initially $n \sim N/R_s^3$, where $R_s$ is the
size of body. At the collision timescale $t \sim (n \sigma_{AC}
R_s)^{-1}$, where the $(AC)$ atom collision cross section is given
by Eq.(\ref{sigHEHE}), in the central part of body a dense and
opaque $(AC)$-atomic core is formed. This core is surrounded by a
cloud of free $(AC)$ atoms, distributed as $\propto R^{-2}$.
Growth and evolution of this $(AC)$-atomic conglomeration may lead
to the formation of a dense self-gravitating $(AC)$-matter object,
which can survive after the star, inside which it was formed,
exploded.

The relatively small mass fraction of AC-matter inside matter
bodies corresponds to the mass of the $(AC)$-atomic core $\ll
10^{-4} S_2 M_{\odot}$ and this mass of AC-matter can be hardly
put within its gravitational radius in the result  of the
$(AC)$-atomic core evolution. Therefore it is highly improbable
that such an evolution could lead to the formation of black holes
inside matter bodies.

\section{\label{Discussion} Discussion}
In the present paper we explored the cosmological implications of
the AC-model presented in \cite{5} with an additional
Coulomb like interaction, mediated by the $y$-photon. This new
$U_{AC}(1)$ interaction appears naturally in the almost-commutative
framework. For the standard model particles the $y$-photons are
invisible, the only source of this invisible light are the
AC-particles. Due to this new strict gauge symmetry the AC-leptons
acquire stability, similar to the case of 4th generation hadrons
\cite{4had} and fractons \cite{fractons}.

The AC-particles are lepton like, coupling apart from  the
$y$-photons only to the ordinary photon and the $Z$-boson. Their
electric charge is taken to be $-2e$ for the $A^{--}$-lepton and
$+2e$ for the $C^{++}$-lepton. They may form atom like bound
states $(A^{--}C^{++})$ with WIMP like cross section which can
play the role of evanescent Cold Dark Matter in the modern
Universe. The AC-model escapes most of the drastic problem of the
Sinister Universe \cite{Glashow}, related with the primordial
$^4He$ cage for $-1$ charge particles and a consequent
overproduction of anomalous hydrogen \cite{Fargion:2005xz}. These
charged $^4He$ cages pose a serious problems to CDM models with
single charged particles, since their Coulomb barrier prevents
successful recombination of positively and negatively charged
particles. The doubly charged $A^{--}$-leptons bind with helium in
the neutral OLe-helium catalyzers of $AC$ binding and AC-leptons
may thus escape this trap.

Nonetheless the binding of AC-leptons into $(AC)$ atoms is a
multi step process, which, due to the expansion of the Universe,
produces necessarily exotic combinations of AC-matter and ordinary
matter, as well as free charged AC-ions. A mechanism to suppress
these unwanted remnants is given by the OLe helium catalysis
$(AHe)+C \rightarrow (AC)+He$. This process is enhanced by the
long-range interaction between the AC-leptons due to the
$y$-photons. It prevents the fractionating of  AC-particles and in
this way enhances also the binding of AC-particles in dense matter
bodies today. This process is necessary to efficiently suppress
exotic atoms to avoid the strong observational bounds.

The AC-model with $y$-interaction may thus solve the serious
problem of anomalous atoms, such as anomalous helium, which
appeared in the AC-cosmology presented in \cite{leptons} as well
as the question of the stability of the AC-leptons. However the
AC-cosmology, even with $y$-interaction, can only be viewed as  an
illustration of the possible solution for the Dark Matter problem
since the following problems remain open:

1. The reason for particle-antiparticle asymmetry.

The AC-model cannot provide a mechanism to explain the necessary
particle-antiparticle asymmetry. Such a mechanism may arise from
further extensions of the AC-model within noncommutative geometry
or due to phenomena from quantum gravity.

2. Possibly observable nuclear processes due to OLe helium.

A challenging problem is the possible existence of OLe helium
$(AHe)$ and of nuclear transformations, catalyzed by $(AHe)$. The
question about its consistency with observations remains open,
since special nuclear physics analysis is needed to reveal what
are the actual OLe-helium effects in SBBN and in terrestrial
matter.

3. Recombination of AC-particles in dense matter objects.

  The recombination into (AC)-atoms and the consequent release
 of gamma energy at tens MeV, at the edge of detection in Super Kamiokande underground
  detector, (at rate comparable to cosmic neutrino Supernovae
  noise or Solar Flare thresholds \cite{FarSolar03}).
    Their signal might be easily  disentangled \cite{leptons} (above a few MeVs ) respect common charged
    current neutrino interactions and single electron tracks
     because the tens MeV gamma lead, by pair productions, to twin
    electron  tracks,  nearly aligned along their Cerenkov rings.
    The signal is piling  the energy in windows where
   few atmospheric neutrino and  cosmic Super-Novae  radiate.

4. Mixing of $y$-photons with neutral gauge bosons.

Due to the interaction of AC-leptons with photons and $Z$-bosons
the invisible $y$-photons will appear in fermionic AC-loops. Thus
standard model fermions may acquire a weak long-range
$y$-interaction. Furthermore it may be necessary to take the
AC-lepton loops into account for high precision calculations of
QED parameters. It has recently been shown \cite{Knecht:2006tv} that the
anomalous magnetic moment of the muon provides a lower
bound for the masses of the AC-leptons of $\sim 10$GeV.

In the context of AC-cosmology search for AC leptons at
accelerators acquires the meaning of crucial test for the
existence of basic components of the composite dark matter. One
can hardly overestimate the significance of positive results of
such searches, if AC leptons really exist and possess new long
range interaction.

To conclude, in the presence of the $y$-interaction AC-cosmology
can naturally resolve the problem of anomalous helium, avoiding
all the observational constraints on the effects, accompanying
reduction of its concentration. Therefore the AC-model with
invisible light for its dark matter components might provide a
realistic model of composite dark matter.

\begin{acknowledgments}
We are grateful to K.M. Belotsky, M. Buenard, A.A.Starobinsky,
I.I. Tkachev, T. Sch\"ucker and E. B\"ohm for fruitful
discussions. M.Kh. is grateful to CRTBT-CNRS and LPSC, Grenoble,
France for hospitality and to D.Rouable for help.
\end{acknowledgments}

\section*{Appendix 1: Basic definitions of noncommutative geometry}

In this section we will give the necessary basic definitions for an almost-commutative
geometries from a particle physics point of view. As mentioned above only the matrix part
will be taken into account, so we restrict ourselves to real, $S^0$-real, finite spectral triples
($\mathcal{A},\mathcal{H},\mathcal{D}, $ $J,\epsilon,\chi$). The algebra $\mathcal{A}$ is
a finite sum of matrix algebras
$\mathcal{A}= \oplus_{i=1}^{N} M_{n_i}(\mathbb{K}_i)$ with $\mathbb{K}_i=\mathbb{R},\mathbb{C},\mathbb{H}$ where $\mathbb{H}$
denotes the quaternions.
A faithful representation $\rho$ of $\mathcal{A}$ is given on the finite dimensional Hilbert space $\mathcal{H}$.
The Dirac operator $\mathcal{D}$ is a selfadjoint operator on $\mathcal{H}$ and plays the role of the fermionic mass matrix.
$J$ is an antiunitary involution, $J^2=1$, and is interpreted as the charge conjugation
operator of particle physics.
The $S^0$-real structure $\epsilon$ is a unitary involution, $\epsilon^2=1$. Its eigenstates with
eigenvalue $+1$ are the particle states, eigenvalue $-1$ indicates antiparticle states.
The chirality $\chi$ is as well a unitary involution, $\chi^2=1$, whose eigenstates with eigenvalue
$+1$ $(-1)$ are interpreted as right (left) particle states.
These operators are required to fulfill Connes' axioms for spectral triples:

\begin{itemize}
\item  $[J,\mathcal{D}]=[J,\chi]=[\epsilon,\chi]=[\epsilon,\mathcal{D}]=0, \quad \epsilon
J=-J \epsilon,\quad\mathcal{D}\chi =-\chi \mathcal{D}$,

$[\chi,\rho(a)]=[\epsilon,\rho(a)]=[\rho(a),J\rho(b)J^{-1}]=
[[\mathcal{D},\rho(a)],J\rho(b)J^{-1}]=0, \forall a,b \in \mathcal{A}$.
\item The chirality can be written as a finite sum $\chi =\sum_i\rho(a_i)J\rho(b_i)J^{-1}.$
This condition is called {\it orientability}.
\item The intersection form
$\cap_{ij}:=\T(\chi \,\rho (p_i) J \rho (p_j) J^{-1})$ is non-degenerate,
$\rm{det}\,\cap\not=0$. The
$p_i$ are minimal rank projections in $\mathcal{A}$. This condition is called
{\it Poincar\'e duality}.
\end{itemize}
Now the Hilbert space $\mathcal{H}$ and the representation $\rho$ decompose with respect to the
eigenvalues of $\epsilon$ and $\chi$ into left and right, particle and antiparticle spinors
and representations:
\begin{eqnarray}
\mathcal{H}=\mathcal{H}_L\oplus\mathcal{H}_R\oplus\mathcal{H}_L^c\oplus\mathcal{H}_R^c
\nonumber
\end{eqnarray}
\begin{eqnarray}
\rho = \rho_L \oplus \rho_R \oplus \overline{ \rho_L^c} \oplus \overline{ \rho_R^c}
\label{representation}
\end{eqnarray}
In this representation the Dirac operator has the form
\begin{eqnarray}
\mathcal{D}=\pp{0&\mathcal{M}&0&0\\
\mathcal{M}^*&0&0&0\\ 0&0&0&\overline{\mathcal{M}}\\
0&0&\overline{\mathcal{M}^*}&0}, \label{opdirac}
\end{eqnarray}
where $\mathcal{M}$ is the fermionic mass matrix connecting the left and the right handed Fermions.

Since the individual matrix algebras have only one irreducible representation for $\mathbb{K}=
\mathbb{R},\mathbb{H}$ and two for $\mathbb{K}=\mathbb{C}$ (the fundamental one and its complex
conjugate), $\rho$ may be written as a direct sum of these fundamental representations with
mulitiplicities
\begin{eqnarray}
\rho(\oplus_{i=1}^N a_i):=(\oplus_{i,j=1}^N
a_i \otimes
1_{m_{ji}} \otimes 1_{(n_j)})\
\oplus\ ( \oplus_{i,j=1}^N 1_{(n_i)} \otimes 1_{m_{ji}} \otimes
\overline{a_j} ).
\nonumber
\end{eqnarray}
There arise certain subtleties which are described in detail in
\cite{1,Kraj,pasch} and will be treated in a later extension of our work.

The first summand denotes the particle sector and the second the antiparticle sector. For the dimensions
of the unity matrices we have $(n)=n$ for $\mathbb{K}=\mathbb{R},\mathbb{C}$ and $(n)=2n$ for
$\mathbb{K}=\mathbb{H}$ and the convention $1_0=0$.
The multiplicities $m_{ji}$ are non-negative integers. Acting with the real structure
$J$ on $\rho$ permutes the main summands and complex conjugates them. It is also possible to write
the chirality as a direct sum
\begin{eqnarray}
\chi=(\oplus_{i,j=1}^N 1_{(n_i)} \otimes \chi_{ji}1_{m_{ji}} \otimes
1_{(n_j)})\
\oplus\
(\oplus_{i,j=1}^N 1_{(n_i)} \otimes \chi_{ji}1_{m_{ji}} \otimes 1_{(n_j)}),
\nonumber
\end{eqnarray}
where $\chi_{ji}=\pm 1$ according to our previous convention on left-(right-)handed spinors.
One can now define the multiplicity matrix $\mu \in M_N(\mathbb{Z})$ such that
$\mu _{ji}:=\chi _ {ji}\, m_{ji}$. This matrix is symmetric and decomposes into a particle and an antiparticle matrix, the second being just the particle matrix transposed, $\mu= \mu_P + \mu_A = \mu_P + \mu_P^T$. The intersection form of the Poincar\'e duality is now simply $\cap = \mu + \mu^T$, see \cite{Kraj,pasch}.

The mass matrix $\mathcal{M}$ of the Dirac operator connects the left and the right handed Fermions. Using
the decomposition of the representation $\rho$ and the corresponding decomposition of the Hilbert
space $\mathcal{H}$ we find two types of submatrices in $\mathcal{M}$, namely $M\otimes 1_{(n_k)}$ and
$1_{(n_k)}\otimes M$. $M$ is a complex $(n_i)\times(n_j)$ matrix connecting the i-th and the j-th sub-Hilbert
space and its eigenvalues give the masses of the fermion multiplet. We will call the k-th algebra the
colour algebra.

\section*{Appendix 2: Irreducibility, Non-Degeneracy}

We will impose
some extra conditions as in \cite{1}.  The
spectral triples are required to be irreducible and non-degenerate according to
the following definitions:

\begin{defn}
 i) A spectral triple $(\mathcal{A},\mathcal{H},\mathcal{D})$ is {\it degenerate} if the kernel of
$\mathcal{D}$ contains a non-trivial subspace of the complex Hilbert space $\mathcal{H}$
invariant under the representation $\rho$ on $\mathcal{H}$ of the real algebra
 $\mathcal{A}$.  \\
 ii) A non-degenerate spectral triple $(\mathcal{A},\mathcal{H},\mathcal{D})$ is {\it reducible} if
there is a proper subspace
$\mathcal{H}_0\subset\mathcal{H}$ invariant under the algebra $\rho(\mathcal{A})$  such that
$(\mathcal{A},\mathcal{H}_0,\mathcal{D}|_{\mathcal{H}_0})$ is a non-degenerate spectral triple. If the
triple is real, $S^0$-real and even, we require  the subspace
$\mathcal{H}_0$ to be also invariant under the real structure $J$, the $ S^0$-real
structure $\epsilon $ and under the chirality
$\chi $ such that the triple $(\mathcal{A},\mathcal{H}_0,\mathcal{D}|_{\mathcal{H}_0})$ is again real,
$S^0$-real and even.
\end{defn}

\begin{defn} The irreducible
 spectral triple $(\mathcal{A},\mathcal{H},\mathcal{D})$ is {\it dynamically non-degenerate} if all
minima $\ddfm$ of the action $V(\ddf)$ define a non-degenerate spectral
triple $(\mathcal{A},\mathcal{H},\ddfm )$ and if the spectra of all minima  have no
degeneracies other than the three kinematical degeneracies: left-right,
particle-antiparticle and colour. Of course in the massless case there is no
left-right degeneracy. We also suppose that the colour degeneracies are
protected by the little group. By this we mean that all eigenvectors of
$\ddfm$ corresponding to the same eigenvalue are in a common orbit of
the little group (and scalar multiplication and charge conjugation).
\label{irred}
\end{defn}

In physicists' language non-degeneracy excludes all models with
pairwise equal fermion masses in the left handed particle sector
up to colour degeneracy. Irreducibility means that we restrict
ourselves to one fermion generation and wish to keep the number of
fermions as small as allowed by the axioms for spectral triples.
The last requirement of definition \ref{irred} means
noncommutative colour groups are unbroken. It ensures that the
corresponding mass degeneracies are protected from quantum
corrections. It should be noted that the standard model of
particle physics meets all these requirements.

\section*{Appendix 3: Krajewski Diagrams}
Connes' axioms, the decomposition of the Hilbert space, the
representation and the Dirac operator allow a diagrammatic
dipiction. As was shown in \cite{Kraj} and \cite{1} this can be
boiled down to simple arrows, which encode the multiplicity matrix
and the fermionic mass matrix. From this information all the
ingredients of the spectral triple can be recovered. For our
purpose a simple arrow and connections of arrows at one point
(i.e. double arrows, edges, etc) are sufficient. The arrows always
point from right fermions (positive chirality) to left fermions
(negative chirality). We may also restrict ourselves to the
particle part, since the information of the antiparticle part is
included by transposing the particle part. We will adopt the
conventions of \cite{1} so that algebra elements tensorised with
$1_{m_{ij}}$ will be written as a direct sum of $m_{ij}$ summands.

$\bullet$ The Dirac operator: The components of the (internal) Dirac
operator are represented by horizontal or vertical lines connecting two
nonvanishing entries of opposite signs in the multiplicity matrix $\mu $
and we will orient them from plus to minus. Each arrow represents a
nonvanishing, complex submatrix in the Dirac operator: For instance
$\mu_{ij}$ can be linked to $\mu_{ik}$ or $\mu_{kj}$ by
\begin{center}\begin{tabular}{cc}
\rxy{
,(0,0)*\cir(0.7,0){}
,(5,0)*\cir(0.7,0){}
,(5,0);(0,0)**\dir{-}?(.6)*\dir{>}
,(0,-3)*{\mu_{ij}}
,(5,-3)*{\mu_{ik}}
}
&\;\;\;\;\;
\rxy{
,(0,0)*\cir(0.7,0){}
,(0,-5)*\cir(0.7,0){}
,(0,0);(0,-5)**\dir{-}?(.6)*\dir{>}
,(-3,0)*{\mu_{kj}}
,(-3,-5)*{\mu_{ij}}
}
\end{tabular} \end{center}
and these arrows represent respectively submatrices of $\mathcal{M}$ in $\mathcal{D}$ of
type $M\otimes 1_{(n_i)}$ with $M$ a complex $(n_j)\times(n_k)$ matrix
and $1_{(n_j)}\otimes M$  with $M$ a complex $(n_i)\times(n_k)$ matrix.

\noindent The requirement of
non-degeneracy of a spectral triple means that every nonvanishing
entry in the multiplicity matrix
$\mu $ is touched by at least one arrow.

$\bullet$ Convention for the diagrams: We will see that (for sums of up to
three simple algebras) irreducibility implies that most entries of $\mu $
have an absolute value less than or equal to two. So we will use a {\it simple
arrow} to connect plus one to minus one and {\it double arrows} to
connect plus one to minus two or plus two to minus one:)
\begin{center}
\begin{tabular}{ccc}
\rxy{
,(0,0)*\cir(0.7,0){}
,(5,0)*\cir(0.7,0){}
,(5,0);(0,0)**\dir{-}?(.6)*\dir{>}
,(0,-3)*{-1}
,(5,-3)*{+1}
}
&
\;\;\;\;\;
\rxy{
,(20,0)*\cir(0.7,0){}
,(20,0)*\cir(0.4,0){}*\frm{*}
,(15,0)*\cir(0.7,0){}
,(20,0);(15,0)**\dir2{-}?(.6)*\dir2{>}
,(20,-3)*{+1}
,(15,-3)*{-2}
}
&
\;\;\;\;\;
\rxy{
,(35,0)*\cir(0.7,0){}
,(30,0)*\cir(0.7,0){}
,(30,0)*\cir(0.4,0){}*\frm{*}
,(35,0);(30,0)**\dir2{-}?(.6)*\dir2{>}
,(35,-3)*{+2}
,(30,-3)*{-1}
}
\\
\\
\end{tabular}
\end{center}
Multiple arrows beginning or ending at one point are with or without edges are built
in an obvious way iterating the procedure above. We will give examples below that will
clarify these constructions.

Our arrows always point from plus, that is right chirality, to minus, that is
left chirality.
For a given algebra, every spectral triple is encoded in its
multiplicity matrix which itself is encoded in its Krajewski diagram, a field
of arrows. In our conventions, for particles, $\epsilon =1$, the column label
of the multiplicity matrix indicates the representation, the row label
indicates the multiplicity. For antiparticles, the row label
of the multiplicity matrix indicates the representation, the column label
indicates the multiplicity.

\noindent Every arrow comes with three algebras:
Two algebras that localize its end
points, let us call them {\it right and left algebras}
and a third algebra that localizes the arrow, let us call it {\it colour
algebra}.  For example for the arrow
\bb\rxy{
,(0,0)*\cir(0.7,0){}
,(5,0)*\cir(0.7,0){}
,(5,0);(0,0)**\dir{-}?(.6)*\dir{>}
,(0,-3)*{\mu _{ij}}
,(5,-3)*{\mu _{ik}}}\eee
the left algebra is $\mathcal{A} _j$, the right algebra is $\mathcal{A}_k$ and the colour
algebra is $\mathcal{A}_i$.

\noindent The {\it circles} in the diagrams only intend to guide the
eye. A {\it black disk} on a multiple arrow indicates that the coefficient of the
multiplicity matrix is plus or minus one at this location, ``the arrows
are joined at this location''.  For example the following arrows
\begin{center}
\begin{tabular}{cc}
\rxy{
,(20,0)*\cir(0.7,0){}
,(20,0)*\cir(0.4,0){}*\frm{*}
,(15,0)*\cir(0.7,0){}
,(20,0);(15,0)**\dir2{-}?(.6)*\dir2{>}
,(20,-3)*{\mu_{ik}}
,(15,-3)*{\mu_{ij}}
}
&\quad
\rxy{
,(35,0)*\cir(0.7,0){}
,(30,0)*\cir(0.7,0){}
,(30,0)*\cir(0.4,0){}*\frm{*}
,(35,0);(30,0)**\dir2{-}?(.6)*\dir2{>}
,(35,-3)*{\mu_{ik}}
,(30,-3)*{\mu_{ij}}
}
\end{tabular}
\end{center}
represent respectively submatrices of
$\mathcal{M}$ of type
$$\pp{M_1\cr M_2} \otimes 1_{(n_i)}\quad
{\rm and} \quad
\pp{M_1&M_2} \otimes 1_{(n_i)}
$$
with $M_1,M_2$ of size $(n_j)\times(n_k)$ or
in the third case, a matrix of type $\pp{M_1\otimes 1_{(n_i)}&1_{(n_j)}
\otimes M_2}$  where $M_1$ and $M_2$ are of size $(n_j)\times(n_k)$ and
$(n_i)\times(n_\ell)$.
According to these rules, we can omit the number
$\pm1,\pm2$ under the arrows.

\section*{Appendix 4: Obtaining the Yang-Mills-Higgs theory}

To complete our short survey on the almost-commutative standard model, we will give a brief glimpse
on how to construct the actual Yang-Mills-Higgs theory.
We started out with the fixed (for convenience flat) Dirac operator of a 4-dimensional spacetime with a fixed fermionic
mass matrix. To generate curvature we have to perform a general coordinate transformation and then
fluctuate the Dirac operator. This can be achieved by lifting the automorphisms of the algebra to
the Hilbert space, unitarily transforming the Dirac operator with the lifted automorphisms
and then building linear combinations. Again we restrict ourselves to the finite case.
Except for complex conjugation in $M_n(\mathbb{C})$ and permutations of
identical summands in the algebra $\mathcal{A}=\mathcal{A}_1\oplus\mathcal{A}_2\oplus ...\oplus\mathcal{A}_N$,
every algebra automorphism
$\sigma
$  is inner, $\sigma (a)=uau^{-1}$ for a unitary $ u\in U(\mathcal{A})$. Therefore
the connected component of the automorphism group is
Aut$(\mathcal{A})^e=U(\mathcal{A})/(U(\mathcal{A})\cap{\rm Center}(\mathcal{A}))$. Its lift to the Hilbert
space \cite{real}
\bb L(\sigma )=\rho (u)J\rho (u)J^{-1}\eee is multi-valued. To avoid the multi-valuedness in the fluctuations, we allow  a central extension of the automorphism group.

The {\it fluctuation $\ddf$} of the Dirac operator $\mathcal{D}$ is given by a
finite collection $f$ of real numbers
$r_j$ and algebra automorphisms $\sigma _j\in{\rm Aut}(\mathcal{A})^e$ such
that
\bb
\ddf :=\sum_j r_j\,L(\sigma _j) \, \mathcal{D} \, L(\sigma_j)^{-1},\quad r_j\in\mathbb{R},\
\sigma _j\in{\rm Aut}(\mathcal{A})^e.
\eee
The fluctuated Dirac operator $\ddf$ is often denoted by $\varphi $, the
`Higgs scalar', in the physics literature.  We consider only fluctuations
with real coefficients since
$\ddf$ must remain selfadjoint.

The sub-matrix of the fluctuated Dirac operator $\ddf$ which is equivalent to
the mass matrix $\mathcal{M}$,  is often denoted by $\varphi $, the
`Higgs scalar', in physics literature. But one has to be careful, as will be shown
below explicitly. It may happen that the lifted automorphisms commute with
the initial Dirac operator and one finds $\ddf =\sum_i r_i \mathcal{D}$ for the finite
part of the spectral triple. This behaviour appeared for the first time in
the electro-strong model in \cite{4}, where the fermions couple vectorially
to all gauge groups and no Higgs field appears. In the model presented
below, the spectral triple can be decomposed into a direct sum consisting
of the standard model and two new particles. The initial Dirac operator
of the new particles commutes with the corresponding part of the lift and
thus does not participate in the Higgs mechanism.

As mentioned in the introduction an almost commutative geometry is the tensor product of a finite
noncommutative triple with an infinite, commutative spectral triple. By
Connes' reconstruction theorem \cite{grav} we know that the latter comes
from a Riemannian spin manifold, which we will take to be any
4-dimensional, compact, flat manifold like the flat 4-torus.  The spectral
action of this almost commutative spectral triple reduced to the finite part
is a functional on the vector space of all fluctuated, finite Dirac operators:
\bb V(\ddf )= \lambda\  \T\!\left[ (\ddf )^4\right] -\frac{\mu
^2}{2}\
\T\!\left[
(\ddf) ^2\right] ,\eee where $\lambda $ and $\mu $ are positive constants
\cite{cc}.
The spectral action is invariant under lifted automorphisms and even
under the unitary group $U(\mathcal{A})\owns u$,
\bb V( [\rho (u)J\rho (u)J^{-1}] \, \ddf \, [\rho (u)J\rho
(u)J^{-1}]^{-1})=V(\ddf),\eee and it is bounded from below.
To obtain the physical content of a diagram and its associated spectral triple one has
to find the minima $\ddfm $ of this action with respect to the lifted automorphisms.

\section*{Appendix 5: Deriving the spectral triple of AC-fermions}

We will start out with the general case of AC-fermions with $y$-interaction
since it boils down to the simpler case without a new gauge group as in \cite{5}, by
setting certain central charges to zero.
The Krajewski diagram of the particle model under consideration encodes an almost-commutative spectral triple with six summands in the internal algebra:

\begin{center}
\begin{tabular}{c}
\rxyz{0.4}{
,(10,-10);(15,-10)**\dir{-}?(.6)*\dir{>}
,(5,-20);(15,-20)**\crv{(10,-17)}?(.6)*\dir{>}
,(10,-20);(15,-20)**\dir{-}?(.6)*\dir{>}
,(15,-20)*\cir(0.2,0){}*\frm{*}
,(30,-25);(25,-25)**\dir{-}?(.6)*\dir{>}
,(40,-35);(35,-35)**\dir{-}?(.6)*\dir{>}
} \\
\\
\end{tabular}
\end{center}
All the necessary translation rules between Krajewski
diagrams and the corresponding spectral triples can be found in \cite{Kraj} and \cite{1}.
The matrix is already {\it blown up} in the sense that the
representations of the complex parts of the matrix algebra have been fixed.
It is the same diagram from which the AC-fermions model of \cite{5} was derived.
One can clearly see the sub-diagram of the standard model in the upper $4\times 4$ corner.

To incorporate a new interaction for the AC-fermions the simplest possible extension
of the almost-commutative spectral triple \cite{5} will be to extend the quaternion
algebra $\mathbb{H}$ to the algebra of complex $2 \times 2$-matrices,
$M_2 (\mathbb{C})$. The notation for the algebra and its elements will be
the following:
\bb
\mathcal{A}= \mathbb{C} \oplus M_2(\mathbb{C}) \oplus M_3(\mathbb{C}) \oplus \mathbb{C} \oplus \mathbb{C} \oplus \mathbb{C} \ni (a,b,c,d,e,f),
\nonumber
\ee
which has as its representation
\bb
\rho_L (a,b,c,d,e,f)&=&  \pp{ b \otimes 1_3 & 0 & 0 & 0 \\ 0 & b & 0 & 0 \\ 0 & 0 & d & 0 \\ 0& 0& 0& \bar{e} },
\quad
\rho_R (a,b,c,d,e,f)=  \pp{ a 1_3 & 0 & 0 & 0 & 0 \\ 0 & \bar{a} 1_3 & 0 & 0 & 0\\ 0 & 0 & \bar{a}  & 0 & 0 \\ 0& 0& 0& e & 0 \\
0&0&0&0&f },
\nonumber \\ \nonumber
\ee

\bb
\rho_L^c (a,b,c,d,e,f)&=&  \pp{  1_2 \otimes c & 0 & 0 & 0 \\ 0 & \bar{a} 1_2 & 0 & 0 \\ 0 & 0 & d & 0 \\ 0& 0& 0& e },
\quad
\rho_R^c (a,b,c,d,e,f)=  \pp{ c & 0 & 0 & 0 & 0 \\ 0 & c & 0 & 0 & 0\\ 0 & 0 & \bar{a}  & 0 & 0 \\ 0& 0& 0& e & 0 \\
0&0&0&0& \bar{e}} .
\nonumber
\ee
These representations are faithful on the Hilbert space given below and serve as well to construct the lift of the automorphism group.
Roughly spoken each diagonal entry of the
representation of the algebra can be associated to fermion multiplet. For example the first entry of $\rho_L$, $b \otimes 1_3$,
is the representation of the algebra on the up and down quark doublet, where each quark is again a colour triplet.

It is now straight forward that the algebra, as well as its representation split
into direct sums of the usual standard model algebra
\bb
\mathcal{A}_{SM}= \mathbb{C} \oplus M_2(\mathbb{C}) \oplus M_3(\mathbb{C}) \ni (a,b,c)
\ee
and its  representation
\bb
\rho_{L,SM} (a,b,c,d,e,f)&=  \pp{ b \otimes 1_3 & 0  \\ 0 & b  },
\quad
\rho_{R,SM} (a,b,c,d,e,f)&=  \pp{ a 1_3 & 0 & 0  \\ 0 & \bar{a} 1_3 & 0 \\ 0 & 0 & \bar{a}   },
\nonumber \\ \nonumber \\
\rho_{L,SM}^c (a,b,c,d,e,f)&=  \pp{  1_2 \otimes c & 0 \\ 0 & \bar{a} 1_2  },
\quad
\rho_{R,SM}^c (a,b,c,d,e,f)&=  \pp{ c & 0 & 0 \\ 0 & c & 0 \\ 0 & 0 & \bar{a}}.
\label{rep}
\ee
as well as  an algebra and representation for the AC-particles:
\bb
\mathcal{A}_{AC} =  \mathbb{C} \oplus \mathbb{C} \oplus \mathbb{C}  \ni (d,e,f)
\ee
and
\bb
\rho_{L,AC} (d,e,f)&=  \pp{d & 0 \\  0& \bar{e} },
\quad
\rho_{R,AC} (d,e,f)&=  \pp{ e & 0 \\0&f },
\nonumber \\ \nonumber \\
\rho_{L,AC}^c (d,e,f)&=  \pp{   d & 0 \\  0&\bar{e} },
\quad
\rho_{R,AC}^c (d,e,f)&=  \pp{  e & 0 \\ 0& \bar{e}}.
\label{repnew}
\ee
The explicit form of the direct sum of the representation is then
\bb
\rho= \rho_{L,SM} \oplus \rho_{L,AC} \oplus \rho_{R,SM} \oplus \rho_{R,AC} \oplus \overline \rho^c_{L,SM} \oplus \overline \rho^c_{L,AC} \oplus  \overline \rho^c_{R,SM} \oplus \overline \rho^c_{R,AC}.
\ee
As pointed out in \cite{farewell} the commutative sub-algebras of $\mathcal{A}$ which are equivalent to the complex numbers, serve
as receptacles for the $U(1)$ subgroups embedded in the automorphism group $U(2)\times U(3)$ of the $M_2(\mathbb(C) \oplus M_3(\mathbb{C})$ matrix algebra. One
can easily see that in contrast to the matrix algebra considered in \cite{5} there is now
a second $U(1)$ embedded in the automorphism group of the algebra. This new
$U(1)$ will be coupled to the AC-fermions only, as a minimal extension of the
standard model gauge group.
The extended lift is defined by
\bb
L(v,w) := \rho(\hat u , \hat v, \hat w, \hat x, \hat y, \hat z) J \rho(\hat u , \hat v, \hat w, \hat x, \hat y, \hat z) J^{-1},
\label{Lift}
\ee
with
\bb
\rho (a,b,c,d,e,f)  := \rho_L (a,b,c,d,e,f) \oplus \rho_R (a,b,c,d,e,f)
\oplus \rho_L^c (a,b,c,d,e,f) \oplus \rho_R^c (a,b,c,d,e,f),
\nonumber
\ee
where $J$ in (\ref{Lift}) is the real structure of the spectral triple, an anti-unitary
operator which coincides with the charge conjugation operator.
For the central extension the unitary entries in $L(u,v)$ are defined as
\bb
\hat u &:=&  (\det v )^{p_1} (\det w)^{q_1}  \nonumber \\
\hat v &:=&  v (\det v )^{p_2} (\det w)^{q_2}  \nonumber \\
\hat w &:=&  w (\det v )^{p_3} (\det w)^{q_3}   \nonumber \\
\hat x &:=&  (\det v )^{p_4} (\det w)^{q_4}   \nonumber \\
\hat y &:=&  (\det v )^{p_5} (\det w)^{q_5}  \nonumber \\
\hat z &:=&  (\det v )^{p_6} (\det w)^{q_6}  \nonumber
\ee
and the unitaries $(v,w) \in U(M_2 (\mathbb{C}) \oplus M_3(\mathbb{C}))$.
The exponents, or central charges, of the determinants will constitute the hypercharges corresponding
to the $U(1)$ subgroups of the gauge group. To ensure the absence of harmful
anomalies a rather cumbersome calculation, \cite{Anomaly,ZouZou}, results in the following values
for the central charges:
\bb
&p_1& \in \mathbb{Q}, \;\; q_1 \in \mathbb{Q},
\nonumber  \\
&p_2& = -\frac{1}{2}, \;\; q_2 = 0,
\nonumber  \\
&p_3& = \frac{p_1}{3}, \;\; q_3 = \frac{q_1 -1}{3},
\nonumber  \\
&p_4& \in \mathbb{Q}, \;\; q_4 \in \mathbb{Q},
\nonumber  \\
&p_5& = p_4, \; \; q_5 = q_4,
\nonumber  \\
&p_6& = - p_4, \;\; q_6 = - q_4
\nonumber
\ee
In the spirit of a minimal extension of the standard model with
AC-fermions as presented in \cite{leptons} the particle content of
the model should stay unchanged. Furthermore the standard model
fermions should not acquire any new interactions on tree-level. To
obtain the standard model hypercharge $U_Y(1)$ one can choose the
relevant central charges to be $p_1 := 0$ and $q_1 := - 1/2$.
Setting $q_4 := -1$ will produce electro-magnetic charges of $\mp
2$ for the AC-fermions $A^{--}$ and $C^{++}$, as required by
\cite{leptons}.

Now $p_4$ governs the existence of the $U_{AC}(1)$ gauge group
and the coupling of the AC-particles to it.  If $p_4$ is chosen to be
zero the AC-model without $y$ interaction as presented in \cite{5}
is recovered. This case is completely equivalent to the
choice of $\mathbb{H}$ instead of $M_2(\mathbb{C})$ as summand in
the internal algebra.

It is also possible to choose $p_4$ to be any
rational number. If $p_4$ is taken to be different from zero, the AC-fermions
will be furnished with a new interaction generated by the second $U_{AC}(1)$ sub-group,
in the group of unitaries of the algebra. Since the AC-fermions do not couple
to the Higgs scalar, as will be explained later,
this new gauge group will not be
effected by the Higgs mechanism and stays thus unbroken.
In the model considered in this paper $p_4 := -1$ for simplicity and the
whole gauge group, before and after symmetry breaking is given by
\bb
U_Y(1)\times SU_w(2)\times SU_c(3) ( \times U_{AC}(1)) \longrightarrow
U_{em}(1)\times SU_c(3) ( \times U_{AC}(1)) \nonumber
\ee
Here the brackets shall indicate the possible absence of the $U_{AC}(1)$
gauge group in the case of $p_4=0$.
The Hilbert space is  the direct sum of the standard model Hilbert
space, for details see \cite{schuck}, and the Hilbert space
containing the AC-fermions $A^{--}$ and $C^{++}$, see \cite{5}:
\bb \mathcal{H} = \mathcal{H}_{SM} \oplus \mathcal{H}_{AC},
\nonumber \ee where \bb \mathcal{H}_{AC} \ni \pp{\psi_{A L} \\
\psi_{C L}} \oplus \pp{\psi_{A R} \\ \psi_{C R}} \oplus
\pp{\psi_{A L}^c \\ \psi_{C L}^c} \oplus \pp{\psi_{A R}^c \\
\psi_{C R}^c}. \nonumber \ee
The wave functions $\psi_{A L}$, $\psi_{C L}$, $\psi_{A R}$ and
$\psi_{C R}$ are the respective left and right handed Dirac
4-spinors. The initial internal Dirac operator,  which is to be
fluctuated with the lifted automorphisms is chosen to be the mass
matrix
\bb
\mathcal{M}= \pp{ \pp{m_u & 0 \\ 0 & m_d} \otimes 1_3 &0&0&0 &0\\ 0&0&m_e&0&0 \\ 0&0&0&m_1&0 \\ 0&0&0&0&m_2 } =  \mathcal{M}_{SM} \oplus \mathcal{M}_{AC},
\nonumber
\ee
with $m_u,m_d,m_e,m_1,m_2 \in \mathbb{C}$,
\bb
\mathcal{M}_{SM}= \pp{ \pp{m_u & 0 \\ 0 & m_d} \otimes 1_3 &0&0\\ 0&0&m_e}
\quad {\rm and} \quad \mathcal{M}_{AC}= \pp{m_1&0 \\0&m_2 }.
\ee
Here again the general structure of a direct sum appears. But the physical model
cannot be considered as a direct sum of the standard model and the AC-particles,
since the gauge group for the new particles, which is generated by the lift,
has its origin in the $M_2(\mathbb{C})\oplus M_3(\mathbb{C})$ summand of the
algebra.
It should be pointed out that the above choice of the summands of the matrix algebra,
the Hilbert space and the Dirac operator
is rather unique, if one
requires the Hilbert space to be minimal and the fermion masses to be non-degenerate.

Fluctuating the Dirac operator leads to a curious new feature which has first been
observed in \cite{5}. The lift \ref{Lift} (here only the particle part will be given,
since this is sufficient for the following considerations) splits as the representation of the
algebra into a direct sum
\bb
L_P = L_{L,SM} \oplus L_{L,AC} \oplus L_{R,SM} \oplus L_{R,AC}
\ee
where $L_{L,SM}$ and $L_{R,SM}$ are the left and right handed parts
of the well known lift appearing in the treatment of the pure standard model
\cite{farewell}. The left and right handed part of the AC-particle lift are given
by
\bb
L_{L,AC}= \rho_{L,AC} \rho^c_{L,AC} \quad {\rm and} \quad L_{R,AC}= \rho_{R,AC} \rho^c_{R,AC}
\ee
which are, as can be easily seen from the representations \ref{repnew}, diagonal
matrices with complex entries.

The fluctuations of the Dirac operator lead for the standard model part
of the mass matrix $\mathcal{M}_{SM}$ to the usual Higgs potential
\bb
\varphi = \sum_i r_i\, (L_{L,SM})_i \mathcal{M}_{SM}
(L_{R,SM})^{-1}_i,
\label{Higgs}
\ee
where the sub-script $i$ on the lifts indicates for the summation over
automorphisms.
For the AC-particles on the other hand the lift commutes with the corresponding
diagonal mass matrix. And for the choice of
vectorial couplings one finds
\bb
L_{L,AC}  \mathcal{M}_{AC}
(L_{R,AC} )^{-1}
= L_{L,AC}
(L_{R,AC} )^{-1}\mathcal{M}_{AC}=\mathcal{M}_{AC}.
\ee
Therefore the AC-particles do not participate in the Higgs mechanism but
acquire their masses through the Dirac operator directly from the geometry.
These masses do not break gauge invariance, since they couple vectorially to
the Abelian $U_Y(1) \times U_{AC}(1)$ subgroup of the standard model, only.

Calculating the spectral action gives the usual Einstein-Hilbert action,
the Yang-Mills-Higgs
action of the standard model and a new part in the Lagrangian for the two AC-fermions
as well as a term for the standard gauge potential $\tilde B_{\mu \nu}$ of the new
$U_{AC}(1)$ sub-group:
\bb
\mathcal{L}_{AC} &=& i \psi_{A L}^\ast D_A \psi_{A L} + i
\psi_{A R}^\ast D_A \psi_{A R} + m_A \psi_{A L}^\ast \psi_{A R} +
 m_A \psi_{A R}^\ast \psi_{A L}
\nonumber  \\
&&+ \; i \psi_{C L}^\ast D_C \psi_{C L} + i \psi_{C R}^\ast D_C
\psi_{C R} + m_C \psi_{C L}^\ast \psi_{C R} + m_C \psi_{C R}^\ast
\psi_{C L}
\nonumber \\
&& \; - \frac{1}{4} \tilde B_{\mu \nu} \tilde B^{\mu \nu}.
\nonumber
\ee
The covariant derivative couples the AC-fermions to the $U(1)_Y$ sub-group of the standard model gauge group
and to the $U_{AC}(1)$ sub-group,
\bb D_{A/C} &=& \gamma^\mu \partial_\mu + \frac{i}{2} \, g' \,
Y_{A/C} \gamma^\mu B_\mu +  \frac{i}{2} \, g_{AC} \, \tilde
Y_{A/C}  \gamma^\mu \tilde B_\mu
\nonumber \\
&=&\gamma^\mu  \partial_\mu + \frac{i}{2} \, e \, Y_{A/C}
\gamma^\mu  A_\mu - \frac{i}{2} \, g' \, \sin \theta_w Y_{A/C}
\gamma^\mu  Z_\mu +  \frac{i}{2} \, g_{AC} \, \tilde Y_{A/C}
\gamma^\mu \tilde B_\mu, \nonumber \ee
where $\tilde B$ is the gauge field corresponding to $U(1)_{AC}$,
$g_{AC}$ is the corresponding coupling and $ \tilde Y_{A/C} $ the
almost-commutative hypercharge. From $\tilde Y_A = -\tilde Y_C= 2
p_4$ follows with the choice $p_4=-1$ that $\tilde Y_A = - \tilde
Y_C = -2$. The possible range of the coupling $g_{AC}$ cannot be
given by almost-commutative geometry, but has to be fixed by
experiment. Again it is straight forward to see that the $y$-interaction may be
switched off by setting the central charge $p_4$ to zero.

Furthermore $B$ is the gauge field corresponding to
$U(1)_Y$, $A$ and $Z$ are the photon and the Z-boson fields, $e$
is the electro-magnetic coupling and $\theta_w$ is the weak angle.
The hyper-charge $Y_{A/C}= 2 q_4$ of the AC-fermions can be any
non-zero fractional number with $Y_A = - Y_C$ so that $\psi_A$ and
$\psi_C$ have opposite electrical charge. To reproduce the
AC-model $q_4=-1$ was chosen, as stated above, and so  $Y_A=-2$
which results in opposite electro-magnetic charges $\mp 2 e$ for
the AC-fermions $A$ and $C$.

\section*{ Appendix 6. Charge asymmetry in the freezing out of particles and antiparticles
} The frozen number density of cosmic relics, which were in
equilibrium with the primordial plasma, is conventionally deduced
\footnote{We follow here the results of \cite{Fargion:2005xz}}
from the equation \cite{ZeldNov}
\begin{equation}
\dot{n}+3Hn=\left<\sigma_{ann}v\right>(n_{eq}^2-n^2). \label{sym}
\end{equation}
This equation is written for the case of a charge symmetry of the
particles in question, i.e. for the case when number densities of
particles ($X$) and antiparticles ($\bar{X}$) are equal
$n_X=n_{\bar{X}}=n$. The value $n_{eq}$ corresponds to their
equilibrium number density and is given by the Boltzmann distribution
\begin{equation}
n_{eq}=g_S \frac{mT}{2\pi}^{3/2}\exp \left(-\frac{m}{T}\right).
\label{neq}
\end{equation}
Here $g_S$ and $m$ are the number of spin states and the mass of
the given particle.

In course of the cooling down, $n_{eq}$ decreases exponentially and
becomes, below the freezing out temperature $T_f$, much less then the real
density $n$, so the term $\left<\sigma_{ann}v\right>n_{eq}^2$,
describing the creation of $X\bar{X}$ from the plasma, can be neglected
\cite{Turner}. It allows to obtain an approximate solution of
Eq.(\ref{sym}).

In case of a charge asymmetry one needs to split Eq.(\ref{sym}) in
two: for $n_X$ and $n_{\bar{X}}$, which are not equal now.
\begin{eqnarray}
\dot{n}_X+3Hn_X=\left<\sigma_{ann}v\right>(n_{eq\,X}n_{eq\,\bar{X}}-n_{X}n_{\bar{X}}),\nonumber\\
\dot{n}_{\bar{X}}+3Hn_{\bar{X}}=\left<\sigma_{ann}v\right>(n_{eq\,X}n_{eq\,\bar{X}}-n_{X}n_{\bar{X}}).
\label{asym}
\end{eqnarray}
The values $n_{eq\,X}$ and $n_{eq\,\bar{X}}$ are given by
Eq.(\ref{neq}) with inclusion of the chemical potential, which for $X$
and for $\bar{X}$ are related as $\mu_X=-\mu_{\bar{X}}=\mu$
(see, e.g., \cite{Dolgov}). So
\begin{equation}
n_{eq\,X,\bar{X}}=\exp\left(\pm\frac{\mu}{T}\right) n_{eq},
\label{nmueq}
\end{equation}
where upper and lower signs are for $X$ and $\bar{X}$
respectively. So
\begin{equation}
n_{eq\,X}n_{eq\,\bar{X}}= n_{eq}^2. \label{neq2}
\end{equation}
A degree of asymmetry will be described in the conventional manner (as
for baryons) by the ratio of the difference between $n_{X}$ and
$n_{\bar{X}}$ to the number density of relic photons at the modern
period
\begin{equation}
\kappa_{\gamma\,mod}=\frac{n_{X\,mod}-n_{\bar{X}\,mod}}{n_{\gamma\,mod}}.
\label{kappagamma}
\end{equation}
However, for practical purposes it is more suitable to use the
ratio to the entropy density which, unlike Eq.(\ref{kappagamma}), does
not change in time provided entropy conservation. Photon number
density $n_{\gamma}$ and entropy density $s$ are given by
\begin{equation}
n_{\gamma}=\frac{2\zeta(3)}{\pi^2}T^3,\;\;\; s=\frac{2\pi^2
g_s}{45}T^3=1.80g_sn_{\gamma}, \label{ngammas}
\end{equation}
where
\begin{equation}
g_s=\sum_{bos}
g_S(\frac{T_{bos}}{T})^3+\frac{7}{8}\sum_{ferm}g_S(\frac{T_{ferm}}{T})^3.
\label{gs}
\end{equation}
The sums in Eq.(\ref{gs}) are over ultrarelativistic bosons and
fermions. So
\begin{equation}
\kappa=\frac{n_{X}-n_{\bar{X}}}{s},\;\;\;
\kappa=\frac{\kappa_{\gamma\,mod}}{1.8g_{s\,mod}}, \label{kappa}
\end{equation}
$g_{s\,mod}\approx 3.93$.
Equation (\ref{kappa}) provides a connection between $n_{X}$ and
$n_{\bar{X}}$. Let us pass to the variables
\begin{equation}
r_+=\frac{n_X}{s},\;\;\; r_-=\frac{n_{\bar{X}}}{s},\;\;\;
r=\frac{n_X+n_{\bar{X}}}{s},\;\;\; x=\frac{T}{m}. \label{rx}
\end{equation}
The apparent relations between the $r_i$ are
\begin{equation}
r_+-r_-=\kappa,\;\;\; r_++r_-=r. \label{r-r-r}
\end{equation}
Provided that the essential entropy redistribution does not take
place ($g_s=const.$) during the period of freezing out,
a transformation to the variable $x$ is possible $$-Hdt=dT/T=dx/x.$$ On
the RD stage the Hubble parameter depends on $T$ as
\begin{equation}
H=\frac{2\pi}{3} \sqrt{\frac{\pi g_{\epsilon}}{5}}
\frac{T^2}{m_{Pl}}, \label{Heps}
\end{equation}
where $g_{\epsilon}$ is given by
\begin{equation}
g_{\epsilon}=\sum_{bos}
g_S(\frac{T_{bos}}{T})^4+\frac{7}{8}\sum_{ferm}g_S(\frac{T_{ferm}}{T})^4.
\label{geps}
\end{equation}
For $r_+$, $r_-$ and $r$ from Eqs.(\ref{asym}) one obtains
the equations
\begin{eqnarray}
\frac{dr_+}{dx}=f_1\left<\sigma_{ann}v\right>\left( r_+(r_+-\kappa)-f_2(x)\right)\nonumber\\
\frac{dr_-}{dx}=f_1\left<\sigma_{ann}v\right>\left( r_-(r_-+\kappa)-f_2(x)\right)\nonumber\\
\frac{dr}{dx}=\frac{1}{2}f_1\left<\sigma_{ann}v\right>\left(
r^2-\kappa^2-4f_2(x)\right). \label{drrr}
\end{eqnarray}
Here
\begin{eqnarray}
f_1=\frac{s}{Hx}\nonumber\\
f_2(x)=\frac{n_{eq}^2}{s^2}=\frac{45^2 g_S^2}{2^5\pi^7g_s^2
x^3}\exp\left(-\frac{2}{x}\right). \label{f12}
\end{eqnarray}
With the use of Eqs.(\ref{ngammas}) and Eq.(\ref{Heps}) one finds
that on the RD stage $f_1$ is equal to
$$f_1=\sqrt{\frac{\pi g_s^2}{45g_{\epsilon}}}m_{Pl}m$$
and independent of $x$.

To solve Eqs.(\ref{drrr}) analogously to Eq.(\ref{sym}), namely
neglecting $f_2(x)$ in them, starting with some $x=x_f$, it would
not be more difficult  to define the moment $x=x_f$.
Nonetheless, if one supposes that such a moment is defined then,
say, $r_i$ will be
\begin{eqnarray}
r_+(x\approx 0)=\frac{\kappa\cdot r_{+f}}{r_{+f}-(r_{+f}-\kappa) \exp\left(-\kappa J\right)}\nonumber\\
r_-(x\approx 0)=\frac{\kappa\cdot r_{-f}}{(\kappa+r_{-f}) \exp\left( \kappa J \right)-r_{-f}}\\
r(x\approx 0)=\kappa \frac{(\kappa+r_{f})\exp\left( \kappa J
\right)+r_f-\kappa} {(\kappa+r_{f})\exp\left( \kappa J
\right)-(r_f-\kappa)}.\nonumber \label{rrr}
\end{eqnarray}
Here $r_{i\,f}=r_i(x=x_f)$,
$$J=\int_0^{x_f} f_1\left<\sigma_{ann}v\right>dx.$$
All $r_i$ (at any moment) are related with the help of
Eqs.(\ref{r-r-r}). Taking into account Eq.(\ref{nmueq}) or
Eq.(\ref{neq2}) for $r_{i\,f}$ one obtains
\begin{eqnarray}
r_{\pm\,f}&=&\frac{1}{2}\left(\sqrt{4f_2(x_f)+\kappa^2}\pm \kappa
\right),
\nonumber \\
r_{f}&=&\sqrt{4f_2(x_f)+\kappa^2}. \label{rpm}
\end{eqnarray}
For $\left<\sigma_{ann}v\right>$ independent of $x$ on the RD stage,
when $f_1$ is also independent of $x$, with  account for the
definition of $x_f$ from the condition  $R(T_f)=H(T_f)$ for the
reaction rate $R(T_f)=n_{eq}(T_f)\left<\sigma_{ann}v(T_f)\right>$,
leading to
$$n_{eq}(T_f)\left<\sigma_{ann}v(T_f)\right>/H(T_f)=\frac{n_{eq}}{s}\cdot \frac{s}{H x_f}\cdot \left<\sigma_{ann}v(x_f)\right> \cdot x_f =$$
\beq = \sqrt{f_2(x_f)} f_1 \left<\sigma_{ann}v(x_f)\right> \cdot
x_f =1, \label{f2f1H} \eeq one obtains \beq \sqrt{f_2(x_f)} =
\frac{1}{f_1 \left<\sigma_{ann}v\right> \cdot x_f} = \frac{1}{J}.
\label{f2f1J} \eeq
If (a) $\left<\sigma_{ann}v\right>=\alpha^2/m^2$ or (b)
$\left<\sigma_{ann}v\right>=C\alpha/\sqrt{Tm^3}$ and one assumes
$f_1=const$ then
\begin{eqnarray}
J_a=\sqrt{\frac{\pi g_s^2}{45g_{\epsilon}}}m_{Pl}\frac{\alpha^2}{m}x_f, \nonumber\\
J_b=\sqrt{\frac{\pi
g_s^2}{45g_{\epsilon}}}m_{Pl}C\frac{\alpha}{m}2\sqrt{x_f}.
\end{eqnarray}
In the case of freezing out of $AC$-leptons one has
$$f_{1AC}=\sqrt{\frac{\pi g_s^2}{45g_{\epsilon}}}m_{Pl}m \approx 2.5 m_{Pl}m,$$
$\left<\sigma_{ann}v\right> =\frac{\bar \alpha^2}{m^2}$ and \beq
J_{AC}=\sqrt{\frac{\pi g_s^2}{45g_{\epsilon}}}m_{Pl}\frac{\bar
\alpha^2}{m}x_f, \label{JU} \eeq where $\bar \alpha=Z^2\alpha +
\alpha_y$ and $Z=Z_E=Z_P=2$ is the charge of AC-leptons. Putting
in Eq.(\ref{f12}) $g_S=2$, $g_s \sim 100$, one obtains the
solution of the transcendent equation (\ref{f2f1J})
$$x_f \approx \left(\ln{\left(\frac{45 g_S}{2^{5/2}\pi^{7/2}g_s} \cdot f_{1AC} \left<\sigma_{ann}v\right>\right)}\right)^{-1} \approx$$
$$ \approx \frac{1}{30}\cdot \frac{1}{(1 - \ln{(S_2)}/30)}.$$
Taking $g_s \approx g_{\epsilon} \sim 100$ one finds from
Eq.(\ref{JU}) $J_U = 6.5 \cdot 10^{13}/S_2 (1 -
\ln{(S_6)}/30)^{-1}$ and from Eq.(\ref{f2f1J}) $\sqrt{4f_2(x_f)} =
2/J_{AC} =3 \cdot 10^{-13} S_2\cdot (1 - \ln{(S_2)}/30)$. For
$\kappa = r_E=r_P= 2 \cdot 10^{-12}/S_2$ one has $\kappa J_{AC} =
13/S_2^2$. At $S_2<2.7$ $4f_2(x_f) < \kappa^2$ and $r_{\pm\,f}$ is
given by Eq.(\ref{rpm}). Since $4f_2(x_f) \gg \kappa^2$ for $S_2
\gg 1$ one obtains from Eq.(\ref{rpm}) \beq
r_{\pm\,f}=\frac{1}{2}\left(\sqrt{4f_2(x_f)}\pm \kappa \right).
\label{rpmf} \eeq The frozen out abundances of AC-leptons and
their antiparticles is given by
\begin{eqnarray}
r_{AC}=\frac{\kappa\cdot r_{+f}}{r_{+f}-(r_{+f}-\kappa) \exp\left(-\kappa J_{AC}\right)}=f_{AC}(S_2)\nonumber\\
r_{\bar{AC}}=\frac{\kappa\cdot r_{-f}}{(\kappa+r_{-f}) \exp\left(
\kappa J_{AC} \right)-r_{-f}} = f_{\bar{AC}}(S_2). \label{rUpm}
\end{eqnarray}
For growing $S_2 \gg 1$ the solution Eq.(\ref{rUpm}) approaches
the values
\begin{eqnarray}
r_{AC} \approx \sqrt{f_2(x_f)} + \kappa/2 \approx\nonumber\\
\approx 1.5 \cdot 10^{-13}S_2\cdot (1 - \ln{(S_2)}/30) + 10^{-12}/S_2\nonumber\\
r_{\bar{AC}} \approx \sqrt{f_2(x_f)} - \kappa/2 \approx \nonumber\\
\approx 1.5 \cdot 10^{-13}S_2\cdot (1 - \ln{(S_2)}/30) -
10^{-12}/S_2. \label{rUSpm}
\end{eqnarray}
At $S_2 < 3.6$ the factor in the exponent $\kappa J_{AC}$ exceeds 1,
and some suppression of the $(\overline{AC})$-abundance takes place. For
$S_2$, close to 1, one has
\begin{eqnarray}
r_{AC}=f_{AC}(S_2) \approx \kappa = 2 \cdot 10^{-12}/S_2\nonumber\\
r_{\bar{AC}}=f_{\bar{AC}}(S_2) \approx 5 \cdot 10^{-3}\kappa S_2^4
\exp\left( -13/S_2^2 \right). \label{rU1pm}
\end{eqnarray}
At $S_2=1$ the factor in the exponent reaches the value $\kappa
J_{AC}=13$ and the solution Eq.(\ref{rUpm}) gives $r_{AC} \approx
\kappa = 2 \cdot 10^{-12}$, $r_{-f} \approx 10^{-14}$ from
Eq.(\ref{rpm}) and
\begin{eqnarray}
r_{\bar{AC}} &\approx& \frac{\kappa\cdot r_{-f}}{\kappa+r_{-f}}\exp\left(- \kappa J_{AC} \right) \approx  r_{-f} \exp\left(- \kappa J_{AC} \right)
\nonumber \\
&\approx& 10^{-14} \exp\left( -13 \right) \approx 3 \cdot 10^{-20}.
\nonumber
\end{eqnarray}

\section*{\label{radiative} Appendix 7. Recombination and binding of heavy charged particles.}
In the analysis of various recombination processes we can use the
interpolation formula for the recombination cross section, deduced in
\cite{4had,Fargion:2005xz}: \beq
 \sigma_r=(\frac{2\pi}{3^{3/2}}) \cdot \frac{\bar \alpha ^3}{T\cdot I_1} \cdot \log{(\frac{I_1}{T})}\eeq
and the recombination rate given by \cite{4had,Fargion:2005xz}\beq
 \sv=(\frac{2\pi}{3^{5/2}}) \cdot \frac{\bar \alpha ^3}{T\cdot I_1} \cdot \log{(\frac{I_1}{T})} \cdot \frac{k_{in}}{M}
\label{recdisc} \eeq Here $k_{in}= \sqrt{2 T M}$, $I_1 \approx
\bar \alpha^2 M/2$ is the ionization potential and $M$ has the
meaning of the reduced mass for a pair of recombining particles. The
constant $\bar \alpha$ for recombining particles with charges
$Z_1$ and $Z_2$ is related with the fine structure constant $\alpha$
by $\bar \alpha= Z_1 Z_2 \alpha$. The approximation
Eq.(\ref{recdisc}) followed from the known result for
electron-proton recombination
\beq \sigma_{rec}=\sigma_r
  =\sum_i \frac{8\pi}{3^{3/2}} \bar \alpha^3 \frac{e^4}{Mv^2i^3} \frac{1}{(Mv^2/2+I_i)},
\label{recep} \eeq
where $M$ and $v$ are the reduced mass and velocity of the particles;
 $I_i$ - ionization potential  ($I_i=I_1/i^2$).

 To sum approximately over 'i' it was noted in  \cite{4had} that $\sigma_r\propto 1/i$
 for $I_i >> Mv^2/2=T_{eff}$ while at $I_i<T_{eff}$ the cross section
 $\sigma_i\propto 1/i^3$ falls down rapidly.

 The "Coulomb-like" attraction of $y$-charges can lead to their
radiative recombination. It can be described in the analogy to the
process of free monopole-antimonopole annihilation considered in
\cite{ZK}. The potential energy of the Coulomb-like interaction
between $A$ and $C$ exceeds their thermal energy $T$ at the
distance
$$ d_0 = \frac{\alpha_y}{T}.$$
Following the classical solution of energy loss due to radiation,
converting infinite motion to finite, free $y$-charges form bound
systems at the impact parameter \cite{ZK,4had} \beq a \approx
(T/m)^{3/10} \cdot d_0. \label{impact} \eeq The rate of such a
binding is then given by \cite{4had} \beq \sv = \pi a^2 v \approx
\pi \cdot (m/T)^{9/10} \cdot (\frac{\alpha_y}{m})^2 \approx
\label{sigimpact} \eeq
$$\approx 2 \cdot 10^{-12}
(\frac{\alpha_y}{1/30})^{2}(\frac{300K}{T})^{9/10}S_2^{-11/10}
\frac{cm^3}{s}$$
The successive evolution of this highly excited atom-like bound
system is determined by the loss of angular momentum owing to the
$y$-radiation. The time scale for the fall into the center in this
bound system, resulting in $AC$ recombination, was estimated
according to classical formula in \cite{DFK,4had}
\beq \tau = \frac{a^3}{64 \pi} \cdot (\frac{m}{\alpha_y})^2 =
\frac{\alpha_y}{64 \pi} \cdot (\frac{m}{T})^{21/10} \cdot
\frac{1}{m} \label{recomb} \eeq
$$\approx 2 \cdot 10^{-4} (\frac{\alpha_y}{1/30})(\frac{300K}{T})^{21/10}S_2^{11/10} \s.$$
As is easily seen from Eq.(\ref{recomb}) this time scale of $AC$
recombination $\tau \ll m/T^2 \ll m_{Pl}/T^2$ turns out to be much
less than the cosmological time at which the bound system was
formed.

The above classical description assumes $a=
\alpha_y/(m^{3/10}T^{7/10}) \gg  1/(\alpha_y m)$ and is valid at
$T \ll T_{rc}=m  \alpha_y^{20/7} \approx 60 \MeV
S_2(\frac{\alpha_y}{1/30})^{20/7}$ \cite{4had,Fargion:2005xz}.
Since $T_{rc} \gg I_{OHe}$ effects of radiative recombination can
also contribute $AC$-binding due to OLe-helium catalysis. However,
the rate of this binding is dominated by (\ref{catETa}) at \beq T
\le T_a (\frac{\alpha_y \bar
\alpha^{6/5}}{\pi^2})^{10/19}(\frac{m_{He}}{m})^{11/19} \approx
100 \eV (\frac{\alpha_y}{1/30})^{10/19}S_2^{-30/19} \label{sigrc}
\eeq and the radiative recombination becomes important only at a
temperature much less, than in the period of cosmological
OLe-helium catalysis.

\section*{\label{radiative} Appendix 8. Problems of "minimal" AC-cosmology}
The minimal realization of cosmological scenario, based on the
AC-model \cite{5} does not go beyond its Lagrangian, which does
not contain mechanisms for the generation of an AC-particle
asymmetry and does not involve additional $U(1)$ symmetry.
Therefore, the minimal AC-cosmological scenario involves an
AC-particle symmetric cosmology, in which abundance of particles
and antiparticles is equal and $AC$ leptons do not possess
$y$-interaction. In such a scenario modern dark matter is a
mixture of equal amounts of $(AC)$ atoms and $(\bar A \bar C)$
"anti-atoms", as well as the "products of incomplete cosmological
combustion" OLe-helium $(HeA)$, C-OLe-helium $(\bar C He)$ and
anomalous helium ($(\bar Aee)$ and $(Cee)$) should be present in
the Universe.
This case is described by the results of the present paper in the
limit $\alpha_y=0$ and $\kappa_{A} =\kappa_{C} = 0.$ For
generality we consider $m_A \neq m_C$ and take for definiteness
$m_C > m_A$.

The thermal history of symmetric AC-matter looks as follows in
chronological order for $m_C = 100 S_{2C}{\GeV} > m_A = 100
S_{2A}{\GeV}$

1) $10^{-10}S_{2C}^{-2}\s \le t \le 6 \cdot10^{-8}S_{2C}^{-2}\s$
at $m_C \ge T \ge T_{fC}=m_C/30 \approx 3 S_{2C} \GeV$ AC-lepton
pair $C \bar C$ annihilation and freezing out (Subsection
\ref{Efreezing} and Appendix 6). The abundance of frozen out $C
\bar C$ pairs is given by $r_{C}=r_{\bar C}= 1.5 \cdot
10^{-13}S_{2C}\cdot (1 - \ln{(S_{2C})}/30)$.

2) $10^{-10}S_{2A}^{-2}\s \le t \le 6 \cdot10^{-8}S_{2A}^{-2}\s$
at $m_A \ge T \ge T_{fA}=m_A/30 \approx 3 S_{2A} \GeV$. Pairs $A
\bar A$ annihilate and freeze out (Subsection \ref{Efreezing} with
Appendix 2) with abundance $r_{A}=r_{\bar A}= 1.5 \cdot
10^{-13}S_{2A}\cdot (1 - \ln{(S_{2A})}/30)$.

3) $ t \sim 2.5 \cdot 10^{-3}S_{2C}^{-2}\s$  at $T \sim I_{C} = 20
S_{2C} \MeV.$ The temperature corresponds to the binding energy
$I_{C} = Z_C^4 \alpha^2 m_C/4 \approx 20 S_{2C} \MeV$ ($Z_C=2$) of
twin C-positronium "atoms" $(C^{++} \bar C^{--})$ , in which $\bar
C^{--}$ annihilate with $(C^{++}$. After this annihilation the $C
\bar C$ pairs abundance is given by (compare with subsection
\ref{anE}) $r_C= r_{\bar C} \approx 1.25 \cdot 10^{-15} S_{2A}$.
These pairs are eliminated in the course of a successive evolution
of AC-matter owing to OLe-helium catalysis.

4) $6 \cdot 10^{-4}S_{2}^{-2}\s \le t \le 0.5 S_{2}^{-2}\s$ at
$I_{AC}\ge T \ge I_{AC}/30 \approx 40 S_{2} \MeV$, where
$S_{2}=S_{2A}S_{2C}/(S_{2A}+S_{2C})$. In this period recombination
of negatively charged AC-leptons $A^{--}$ and $\bar C^{--}$ with
positively charged $C^{++}$ and $\bar A^{++}$ can lead to a
formation of AC- lepton "atoms" $(AC)$ and anti-atoms $(\bar A
\bar C)$ with potential energy $I_{AC}= Z_A^2 Z_C^2\alpha^2 m/2
\approx 40 S_2 \MeV$ ($Z_A=Z_C=2$, $m$ is the reduced mass)
(compare with subsection \ref{recE}). Together with neutral $(AC)$
atoms and $(\bar A \bar C)$ "anti-atoms" a significant fraction
of free charged $A^{--}$, $\bar C^{--}$, $C^{++}$ $\bar A^{++}$ is
also left even for $S_{2A} < 6$. At $S_{2A} > 6$ recombination in
$(AC)$ and $(\bar A \bar C)$ states is not effective.

5) $ t \sim 2.5 \cdot 10^{-3}S_{2A}^{-2}\s$  at $T \sim I_{A} = 20
S_{2A} \MeV.$ The temperature corresponds to the binding energy
$I_{A} = Z_A^4 \alpha^2 m_A/4 \approx 20 S_{2A} \MeV$ ($Z_A=2$) of
twin E-positronium "atoms" $(A^{--} \bar A^{++})$, in which $\bar
A^{++}$ annihilate with $A^{--}$. This annihilation results in the
frozen out $A \bar A$ pairs abundance $r_A= r_{\bar A} \approx
1.25 \cdot 10^{-15} S_{2A}$ (compare with subsection \ref{anE}).
For $S_{2A} \ll S_{2C}$ the abundance of these pairs is much less
than the abundance of  $(C^{++} \bar C^{--})$ pairs. Under this
condition in the course of a successive evolution of AC-matter
these frozen out $A^{--}$ and $\bar A^{++})$ become dominantly
 bound in primordial $(AC)$ and $(\bar A \bar C)$ "atoms".

6) $100\s \le t \le 300\s$  at $100 \keV\ge T \ge I_{o}/27 \approx
60 \keV,$ where $I_{\bar C He} = I_{o} = Z_{He}^2 Z_{A}^2 \alpha^2
m_{He}/2 = Z_{He}^2 Z_{C}^2 \alpha^2 m_{He}/2 = 1.6 \MeV$ is in
the limit $m_C \gg m_A\gg m_{He}$ the ionization potential of both
$(^4He^{++}A^{--})$ and $(^4He^{++} \bar C^{--})$ "atoms". Helium
$^4$He is formed in the result of the Standard Big Bang
Nucleosynthesis and virtually all free $A^{--}$ and $\bar C^{--}$
are trapped by $^4$He in OLe-helium $(^4He^{++}A^{--})$ and
C-OLe-helium $(^4He^{++} \bar C^{--})$ (section 4). Being formed,
OLe-helium catalyzes binding of free $C^{++}$ with its constituent
$A^{--}$ into $(AC)$ atoms and annihilation of free $\bar
A^{++})$ with $A^{--}$, while C-OLe-helium $(^4He^{++} \bar
C^{--})$ catalyzes $(\bar A \bar C)$ binding and $(C^{++} \bar
C^{--})$ annihilation.

For $S_{2A} \ll S_{2C}$ virtually all $A^{--}$ and $\bar A^{++}$
form $(AC)$ and $(\bar A \bar C)$, while the dominant part of
$(C^{++} \bar C^{--})$ pairs annihilate. Effects of this $(C^{++}
\bar C^{--})$ annihilation, catalyzed by C-OLe-helium at $T \sim
60 \keV$, do not cause any contradictions with observations. In
the result the abundance of $(AC)$ and $(\bar A \bar C)$ is given
by $r_{(AC)}= r_{(\bar A \bar C)} \approx 1.25 \cdot 10^{-15}
S_{2A}$. The density of these AC-atoms should not exceed the CDM
density, corresponding at $S_{2A} \ll S_{2C}$ to $r_{(AC)}=
r_{(\bar A \bar C)} \approx 2 \cdot 10^{-12}/ S_{2C}$. It leads to
the condition
$$S_{2A}S_{2C} \le 1600.$$ For $S_{2A} \ll S_{2C}$ the remaining
abundance of free $\bar A^{++}$ and OLe-helium $(HeA)$ is
exponentially small, but the surviving abundance of $C^{++}$ and
C-OLe-helium $(^4He^{++} \bar C^{--})$ is significant, being of
the order of $r_C=r_{\bar HeC} \approx \frac{1}{J_{o}} \approx 7
\cdot 10^{-18},$ or relative to baryons $f_C=r_C/r_b=f_{\bar
CHe}=r_{\bar CHe}/r_b \approx 10^{-6}$.

Here the main problem for symmetric AC-cosmology arises. By
construction AC-matter in this cosmological scenario contains an
equal amount of particles and antiparticles. This explosive
material should be present on the successive stages of
cosmological evolution:

7) $t \sim 2.5 \cdot 10^{11}\s$  at $T \sim I_{He}/30 \approx 2
eV.$ Here $I_{He}= Z^2 \alpha^2 m_{e}/2 = 54.4 \eV$ is the
potential energy of an ordinary He atom. Formation of anomalous
helium atoms. Free $C^{++}$ with charge $Z=+2$ recombine with
$e^-$ and form anomalous helium atoms $(eeC^{++})$.

8) $t \sim 10^{12}\s$  at $T \sim T_{RM} \approx 1 \eV.$ AC-matter
dominance starts with $(AC)$ and $(\bar A \bar C)$  "atoms",
playing the role of CDM in the formation of  Large Scale
structures.

9) $z \sim 20.$ Formation of galaxies starts, triggering effects
of AC-matter annihilation.

Such effects even from a sparse component of $(eeC^{++})$
anomalous helium and C-OLe-helium $(^4He^{++} \bar C^{--})$ are
dramatic for the considered model. As it was revealed for hadrons
of the 4th generation in \cite{4had}, a decrease of anomalous
helium is accompanied by $\gamma$ radiation, which at
$f_C=r_C/r_b=f_{\bar CHe}=r_{\bar CHe}/r_b > 10^{-9}$ exceeds the
observed $\gamma$ ray background flux. The set of problems
\cite{4had}, related with neutrino and gamma radiation from
$(C^{++} \bar C^{--})$ C-OLe-helium catalyzed annihilation in
terrestrial matter and with effects of such annihilation in large
volume detectors, add troubles to the considered model.

Moreover, the dominant form of dark matter is also explosive in
the symmetric AC-cosmology and $(AC)$ and $(\bar A \bar C)$
annihilation in the Galaxy should create a gamma flux, which
exceeds the observed gamma background at $$S_{2A}S_{2C} \le
3600.$$

The above list of troubles seems to be unavoidable for symmetric
AC-cosmology, in which the explosive character of AC-matter is hardly
compatible with its presence in the modern Universe. This is the
reason, why we consider in the main part of our paper an asymmetric
AC-cosmology with AC-particle excess.

Note that at large $S_2>10^5$ asymmetric AC-cosmology coincides
with the symmetric case with all its troubles. Strictly speaking
this case corresponds to the mass of AC-leptons which might highly
exceed reheating temperature and the creation of AC-lepton pairs
is related with the physics of the preheating stage. However it is
hardly probable to avoid this cosmological problem for super heavy
AC-leptons what can be considered as a cosmological upper limit,
of about $10^7$GeV, for the AC-lepton mass.

\section*{\label{EpMatter} Appendix 9. Suppression of $A$ and $C$ species in dense matter bodies without $y$-interaction}
Cross sections of OLe-helium and anomalous helium interaction with
matter differ by about ten orders of magnitude. In the absence of
$y$ interaction, holding these two species together by the
condition of $y$-neutrality, the corresponding difference in their
mobilities inevitably leads to fractionating of these two
components. The problem of suppression of anomalous helium in
dense matter under the condition of fractionating is discussed in
the present Appendix.

To catalyze the processes, reducing the abundance of anomalous
helium, first of all $(HeA)$ should be effectively captured by a
matter body.

A matter object with size $R$ with a number density $n_{at}=n_b/A$
of atoms with atomic weight $A$ is opaque for $(HeA)$ and can
effectively capture it, if \beq n_{at} R \sigma_{trAb}= n_b R
\sigma_{trAb} > 1. \label{Epcap} \eeq Here the transport cross
section for the $(HeA)$ energy transfer to an atom $\sigma_{trA} =
A\sigma_{trAp}$ is expressed through such a cross section per
nucleon \beq \sigma_{trAb} =  \pi R_{o}^2 \frac{m_{p}}{m_A}
\approx 10^{-27}/S_2 \cm^{2}, \label{sigpEpcap} \eeq where $R_{o}
= 1/(Z_E Z_{He} \alpha m_{He}) \approx 2 \cdot 10^{-13} \cm$ is
given by Eq.(\ref{REHe}). The condition Eq.(\ref{Epcap}) for an
effective capture of $(HeA)$ reads as
$$n_{b} R  > 10^{27} S_2 \cm^{-2}.$$
A similar condition Eq.(\ref{Epcap}) for the effective capture of
$(eeC^{++})$, having atomic cross sections for the interaction
with matter $\sigma_a \sim 10^{-16}\cm^2$, is given by
$$n_{b} R  > 10^{18} S_2 \cm^{-2},$$
where for $(eeC^{++})$ the transport cross section per nucleon is
equal to \beq \sigma_{tra}= \sigma_a (m_p/m_C) \approx 10^{-18}
S_2 \cm^2. \label{sigpUEcap} \eeq Matter objects, satisfying the
condition Eq.(\ref{Epcap}) for all these species, can, in
principle, provide the conditions for $(HeA)$ catalysis, but,
being captured, $(HeA)$ should survive in the object and the
catalysis should be effective.

The temperature in hot stellar interiors normally does not reach
the value of $I_{o}=1.6 \MeV$ (which might be reached and exceeded
only at Supernova explosions). Therefore an $(HeA)$ "atom" cannot
be ionized by thermal radiation. On the other hand, at $T < T_a
\approx 64/S_2 \keV$ the rate of catalysis is given by
Eq.(\ref{catETa}) and in planets (in particular, in the Earth)
this rate is much less than Eq.(\ref{Epsv}). Remember that we
consider here the case without $y$-interaction and without
$y$-radiative recombination, induced by this interaction.

Note that at such a low rate of thermal catalysis, even if $(HeA)$
atoms were contained in the primordial terrestrial matter in the
pre-galactic proportion with $(eeC^{++})$, they could not
effectively reduce the abundance of this form of anomalous helium
below the experimental limits. Indeed, taking in Eq.(\ref{supsym})
$n \approx 2 \cdot 10^{24} \cm^{-3}$, $\sv^{A}_{cat} \approx
10^{-24} S_2^{3/2}\cm^3/\s$ from Eq.(\ref{catETa}) at $T\sim 300K$
and $t_A \approx 10^{17}\s$, we get the residual abundance of
anomalous helium $\xi_C \sim 5 \cdot 10^{-18}/S_2^{3/2}$, which
exceeds the experimental upper limit at $S_2 < 9$.

A possible enhancement of the catalysis may be related with the
$(HeA)$ excess in the Solar system, and, in particular the Earth.
In such a mechanism the terrestrial abundance of anomalous helium
$(eeC^{++})$ is suppressed due to $(HeA)$ catalyzed $(AC)$ binding
of most of the incoming flux $I_{C}$ reaching the Earth.

In the framework of our consideration, interstellar gas contains a
significant ($\sim f_g \xi_{C}$) fraction of $(eeC^{++})$. When
the interstellar gas approaches Solar System, it is stopped by the
Solar wind in the heliopause at a distance $R_h \sim 10^{15} \cm$
from the Sun. In the absence of detailed experimental information
about the properties of this region we assume for our estimation
that all the incoming ordinary interstellar gas, containing
dominantly ordinary hydrogen, is concentrated in the heliopause
and the fraction of this gas, penetrating this region towards the
Sun, is pushed back to the heliopause by the Solar wind. In the
result, to the present time during the age of the Solar system
$t_E$ a semisphere of width $L\sim R_h$ is formed in the distance
$R_h$, being filled by a gas with density $n_{hel} \sim (2 \pi
R_h^2 v_g t_E n_g)/(2 \pi R_h^2 L) \sim 10^8 \cm^{-3}.$ The above
estimations show that this region is transparent for $(HeA)$, but
opaque for atomic size remnants, in particular, for $(eeC^{++})$.
Though the Solar wind cannot directly stop heavy $(eeC^{++})$, the
gas shield in the heliopause slows down their income to Earth and
suppresses the incoming flux $I_{C}$ by a factor $S_h \sim
1/(n_{hel} R_h \sigma_{tra})$, where according to
Eq.(\ref{sigpUEcap}) $\sigma_{tra} \approx 10^{-18} S_2 \cm^2$. So
the incoming flux, reaching the Earth, can be estimated as
\cite{4had} \beq I_{C} = \frac{\xi_{C} f_g n_g v_g}{8\pi}S_h
\approx 10^{-8}\frac{S_h}{5 \cdot 10^{-5}}(cm^2 \cdot s \cdot
ster)^{-1} \label{IincP} \eeq This suppression provides an
additional reason for the generation of an $(HeA)$ excess in the
Earth.

For the $(HeA)$ excess $k=\Delta n_{HeA}/n$ the residual abundance
of anomalous helium can be suppressed exponentially \beq \xi_{i} =
\xi_{i0} \exp(- k n \sv^{A}_{cat} t_b). \label{matEpexs} \eeq Such
an excess can appear in the Solar system since the matter of the
disc (with $n_b R \sim 3 \cdot 10^{20}\cm^{-2}$) is transparent
for nuclear interacting $(HeA)$ and opaque for atomic
$(eeC^{++})$. Then the incoming flux of $(HeA)$ is proportional to
the halo velocity dispersion ($v_h \sim 300 \km/\s$), while the
flux of $(eeC^{++})$ comes to Solar system with a velocity
dispersion in the disc ($v_d \sim 20 \km/\s$) and it can be
additionally suppressed while crossing the heliopause (see
discussion in \cite{4had}). For an equal initial abundance of
$(HeA)$ and $(eeC^{++})$ this difference in  the incoming fluxes
results in an excessive amount of $(HeA)$ in Solar system.

For the $(HeA)$ excess, linearly growing with time $k n_b = j_A
\cdot t$ (here $j_A$ is the increase of excessive $(HeA)$ per unit
time in unit volume) at $t > \tau$, where
$$ \tau = \frac{1}{\sqrt{j_A \sv_{cat}^A}},$$ the
abundance of primordial anomalous helium decreases as \beq \xi_{i}
= \xi_{i0} \exp(- \frac{t_b^2}{2 \tau^2}) \label{matEpexsgr} \eeq
and falls down below the experimental upper limit  at $\tau < 0.1
t_b$, corresponding to $4 \cdot 10^{-9}S_2^{-3/2}\cm^{-3}\s^{-1}$.
Note that the income of $(HeA)$, captured and homogeneously
distributed in  the Earth, gives rise to  \cite{4had} \beq j_{A} =
\frac{n_O v_h}{4 R_E} \approx 2.5\cdot 10^{-9}\cm^{-3}\s^{-1},
\label{income} \eeq where the Earth's radius $R_E \approx 6 \cdot
10^8 \cm$ and the number density $n_O$ of OLe-helium is determined
by the local halo density $\rho_{0h} \approx 0.3
\frac{GeV}{\cm^3}$ as
$$n_O = \xi_{A}\frac{\rho_{0h}}{m_p}.$$
For the income Eq.(\ref{income}) $\tau \sim 2\cdot
10^{16}S_2^{-3/4}\s.$

We see that the income of $(HeA)$ can create in the Earth an
$(HeA)$ excess, which provides an effective suppression of
primordial anomalous helium in the terrestrial matter. If we take
into account the income of interstellar anomalous helium, $j_i$,
on the timescale $t > \tau$ the solution reads
$$\xi_{i} = \frac{j_{i} \tau}{n_b} \exp(- \frac{t_b^2}{2 \tau^2}).$$

The above consideration assumes a timescale, exceeding the
diffusion timescale of $(HeA)$ in the Earth ($t_{dif} \sim (n_b
\sigma_{trAb} R_E) \cdot R_E/v_T \sim 4\cdot 10^{9}S_2^{-1/2}\s$,
where the thermal velocity of $(HeA)$ in terrestrial matter is
$v_T \sim 2 \cdot 10^4 S_2^{-1/2}\cm$) or the timescale of water
circulation in the Earth ($t_w \approx 3\cdot 10^{10}\s$, see the
discussion in \cite{4had}). Since $\tau  \gg t_w \geq t_{dif}$, on
the timescale $t_{dif}$ we can neglect the increase of $(HeA)$
excess with time and take its value, gained to the present time,
as constant $k n_b \approx j_A t_E$, which for $j_A$, given by
Eq.(\ref{income}) is equal to $k n_b \approx 2.5\cdot 10^{8}
\cm^{-3}$. The timescale of the $(HeA)$ catalysis is then
$$t_{cat} = \frac{1}{k n_b \sv_{cat}^A} \approx 4\cdot 10^{15}S_2^{-3/2} \s,$$
being for $S_2 <10^3$ much larger than the relaxation timescales
$t_{dif}$ and $t_w$. It means that $(HeA)$ catalysis for most of
incoming anomalous helium atoms takes place after they acquire a
stationary distribution. However, $(HeA)$ in the infalling flux
have momentum $k \sim m v_h/c \sim 1/R_{o}$ and thus they have a
correspondingly larger cross section of catalysis. Therefore the
problem of anomalous helium can involve considerations of
non-thermal catalysis, taking place before the slowing down of
this flux.

All these complications, induced by fractionating of OLe-helium
and anomalous helium are naturally avoided in AC model with
$y$-interaction, considered in the present paper.
\bibliography{noncom,kosmo}

\end{document}